%% file: paper.tex
\UseRawInputEncoding
\documentclass[10pt,journal,compsoc]{IEEEtran}

\ifCLASSOPTIONcompsoc
  \usepackage[nocompress]{cite}
\else
  \usepackage{cite}
\fi
\usepackage{rotating}
\usepackage{pdflscape}
\usepackage{adjustbox}
\usepackage{blindtext}
\newsavebox{\myimage}

\ifCLASSINFOpdf
\else
\fi
\usepackage{balance}  
\usepackage{graphics} 
\usepackage{txfonts}
\usepackage{times}    
\usepackage[pdftex]{hyperref}
\usepackage{color}
\usepackage{textcomp}
\usepackage{booktabs}
\usepackage{ccicons}

\newcommand{\emt}[1]{\emph{``#1''}}
\newcommand{\info}[1]{$^{#1}$}

\usepackage{cite}
\usepackage{url}
\usepackage{fancybox}
\usepackage{multirow}
\usepackage{flushend}
\usepackage{booktabs}
\usepackage{tabularx}
\usepackage{comment}
\usepackage{array}
\usepackage[flushleft]{threeparttable}
\usepackage{mdframed}
\graphicspath{{}{images/}{dia/}}
\DeclareGraphicsExtensions{.pdf,.png}

\usepackage{tcolorbox}
\usepackage{listings}
\usepackage{courier}
\usepackage{hyperref}

\usepackage{tikz}
\definecolor{1c1}{RGB}{188,162,6}
\definecolor{1c2}{RGB}{137,129,80}
\definecolor{1c3}{RGB}{239,167,31}
\definecolor{1c4}{RGB}{88,194,241}
\definecolor{1c5}{RGB}{6,180,188}

\tikzset{mynode/.style={draw=white,solid,circle,fill=green,inner sep=1pt, thick,
text=black}}
\tikzset{arrow line/.style={dashed, line width= 2.5pt, color=#1}}

\def\bf{\textbf}

\def\fig {Figure~}

\def\tbl {Table~}
\def\tbls {Tables~}

\def\sec {Section~}
\def\secs {Sections~}

\def\app {Appendix~}
\def\it{\textit}

\def\tt{\mct}
\newcommand{\ib}[1]{{\textbf {\textit { #1}}}}

\newcommand{\co}[1]{{\textsc {\small{ #1}}}}

\newcommand{\mct}[1]{{\footnotesize {\texttt {#1}}}}

\newcommand{\api}[1]{{\sf{\texttt\small{#1}}}}

\usepackage{paralist}

\usepackage{enumitem}

\newcommand{\nd}{\vspace{1mm}\noindent}
\newcommand\q[2]{\textsc{\small{#1}$^{#2}$}}

\usepackage[small,bf]{caption}

\usepackage{sparklines}
\definecolor{sparkspikecolor}{named}{red}
\usepackage{wrapfig}
\usepackage{bchart}

\usepackage{scalerel}

\usepackage{tikz}
\definecolor{1c1}{RGB}{188,162,6}
\definecolor{1c2}{RGB}{137,129,80}
\definecolor{1c3}{RGB}{239,167,31}
\definecolor{1c4}{RGB}{88,194,241}
\definecolor{1c5}{RGB}{6,180,188}
\tikzset{mynode/.style={draw=white,solid,circle,fill=green,inner sep=1pt, thick,
text=black}}
\tikzset{arrow line/.style={dashed, line width= 2.5pt, color=#1}}

\newcommand*\circled[1]{\tikz[baseline=(char.base)]{
            \node[shape=circle,draw,inner sep=1pt] (char) {#1};}}

\newcommand*\circledc[1]{\tikz[baseline=(char.base)]{
            \node[shape=circle,draw,inner sep=1pt,fill=lime] (char)
            {#1};}}
\newcommand{\bnd}[1]{\noindent\bf{\small\circledc{#1}} }

\newcommand{\rev}[1]{\textcolor{black}{ #1}}

\newcounter{o}
\begin{document}

\title{Understanding How and Why Developers Seek and Analyze API-related Opinions}

\author{Gias Uddin,
        Olga Baysal,
        Latifa Guerrouj, and
        Foutse Khomh
        }

%


\IEEEtitleabstractindextext{%
\begin{abstract}
With the advent and proliferation of online developer forums as informal
documentation, developers often share their opinions about the APIs they use.
Thus, opinions of others often shape the developer's perception and decisions related to software
development. For example, the choice of an API or how to \it{reuse} the
functionality the API offers are, to a considerable degree, conditioned upon
what other developers think about the API.
While many developers refer to and rely on such opinion-rich information about
APIs, we found little research that investigates the use and benefits of public opinions.
To understand how developers seek and evaluate API
opinions, we conducted two surveys involving a total of 178 software
developers. We analyzed the data in two dimensions, each corresponding to
specific needs related to API reviews:
\begin{inparaenum}[(1)]
\item Needs for seeking API reviews, and
\item Needs for automated tool support to assess the reviews.
\end{inparaenum}
We observed that developers seek API reviews and often have to summarize
those for diverse development needs (e.g., API suitability). Developers also
make conscious efforts to judge the trustworthiness of the provided opinions
and believe that automated tool support for API reviews analysis can assist in
diverse development scenarios, including, for example, saving time in API selection as well as making
informed decisions on a particular API features.
\end{abstract}

}

\maketitle

\IEEEdisplaynontitleabstractindextext

%
\IEEEpeerreviewmaketitle

\input{intro.tex}

\input{related-work.tex}

\input{motivation.tex}

\input{pilot-survey.tex}

\input{methodology.tex}

\input{results-opinion.tex}
\input{discussions.tex}

\input{research-journey.tex}
\input{threats.tex}

\input{summary.tex}


\section*{Acknowledgements}
We thank all developers who took part in our survey for their time, participation, and
feedback, and Lalit Azad who participated in the open-coding of the primary
survey.

\bibliographystyle{IEEEtran}
\bibliography{consolidated}
\input{appendix.tex}
\end{document}

%% file: intro.tex
\IEEEraisesectionheading{\section{Introduction}\label{sec:introduction}}

APIs (Application Programming Interfaces)
offer interfaces to reusable software
components. Modern-day rapid software development is often facilitated
by the plethora of open-source APIs available for any given development
task. The online development portal GitHub~\cite{website:github} now hosts more
than 67 million public repositories. We can observe a radical increase from the 2.2
million active repositories hosted in GitHub in 2014. While many of the public repositories in
GitHub may not be code-based or are personal projects, we observed similar growth in many
other online API package managers. For example, we observed an increase in the number of open-source APIs
shared in all the following package managers (as of July 2018) : \begin{inparaenum}
\item 11,782\% for Javascript APIs in {\em npm} package manager (from 5,646 in December 2011),
\item 334\% for Java APIs in online {\em maven} central (from 55,785 in March 2013),
\item 2,447\% for C\# APIs in online {\em nuget} repository (from 4,799 in February 2012),
\item 1,477\% for Python APIs in {\em PyPI} (from 9,362 in March 2010).
\end{inparaenum} Javascript, Java, C\# and Python are among the top five most popular programming
languages in Stack Overflow\footnote{We used the GitHub API and modulecounts.com to
 collect the statistics}. Developers can share their APIs in other package managers as well, such as {\em Bower} for Javascript,
 {\em Rubygems.org} for Ruby, etc.
%
%



\begin{figure*}
\centering
\includegraphics[scale=.55]{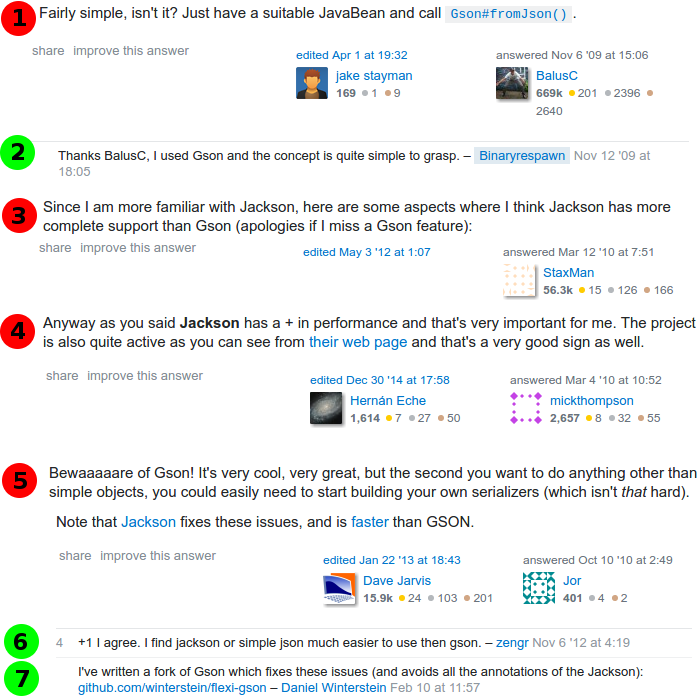}
	\caption{Example of a Stack Overflow discussion about two APIs. Green-coloured
	circles are comments, and red-coloured ones are answers.}
 \label{fig:soSentimentPosts}
\end{figure*}

With a myriad of APIs being available, developers now face a new challenge --- how to choose the right API.
For any given task, we now expect to see multiple competing APIs. For example, in the mapping community,
developers can choose from multiple web APIs, such as Google Maps APIs, Bing Maps APIs, Apple Maps APIs, MapBox, OpenLayer, etc.
The selection and adoption of an API depends on a number of factors~\cite{Robillard-APIsHardtoLearn-IEEESoftware2009a}, such
as the availability of learning resources, the design and usability of the API~\cite{Myers-ImprovingAPIUsability-CACM2016,Stylos-UsabilityObjectConstructor-ICSE2007a,Stylos-MakingAPIsmoreUsable-PhDThesis2009}.
Developers can learn APIs by using the API official documentation. However, the official documentation can be
often incomplete, obsolete, and/or  incorrect~\cite{Uddin-HowAPIDocumentationFails-IEEESW2015,Robillard-FieldStudyAPILearningObstacles-SpringerEmpirical2011a}.
In our previous study of more than 300 developers at IBM, we found that such problems in an API official documentation can motivate a developer to select other competing APIs.

To overcome the challenge of selecting an API among available choices and properly learning it, many developers seek help and insights from other
developers in online developer forums. \fig\ref{fig:soSentimentPosts} presents the screenshot of seven
Stack Overflow posts. In \fig\ref{fig:soSentimentPosts}, the four red-coloured circles are
answers \circled{1}, \circled{3}-\circled{5} and three green-coloured circles are
comments \circled{2}, \circled{6}, \circled{7}.
The oldest post (at the top) is dated from November 06, 2009, while the most recent
one (at the bottom) is from February 10, 2016. These posts express developers' opinions about two Java APIs
(Jackson~\cite{website:jackson} and Gson~\cite{website:gson}) offering JSON
parsing features for Java. None of the posts contain any code snippets.
The first answer \circled{1} representing a positive opinion about the Gson API
motivates the developer `binaryrespawn' to use it \circled{2}. In the next
answer \circled{3}, the user `StaxMan' compares Gson with Jackson, favoring
Jackson for offering better support, and based on this feedback, `mickthomson'
\circled{4} decides to use Jackson instead of Gson. Three out of the four
answers \circled{3}--\circled{5} imply a positive
sentiment towards Jackson but a negative one about Gson.
Later, the developer `Daniel Winterstein' develops a new version of Gson
fixing existing issues, and shares his API \circled{7}. This example illustrates
how developers share their experiences and insights, as well as how they
influence and are influenced by other developers' opinions. A developer looking for only code
examples for Gson would have missed the important insights about the API's
limitations, which may have affected his development activities. Thus, opinions
extracted from the informal discussions can drive developers' decision making.

\begin{figure*}
\centering
\hspace*{-.5cm}%
\includegraphics[scale=.70]{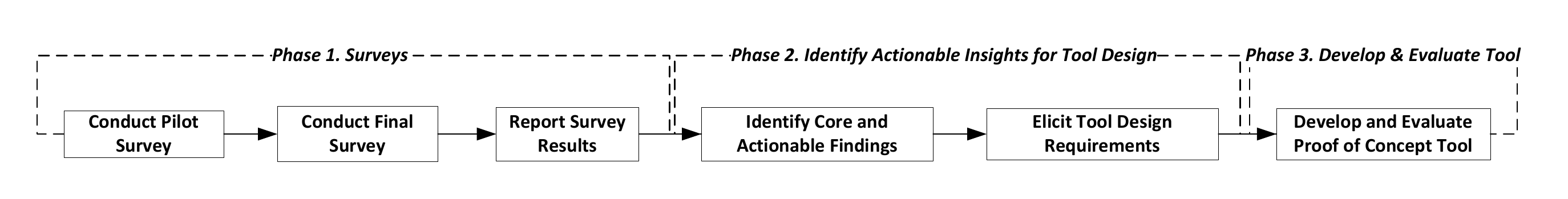}
	\caption{The major research steps undertaken during and after the two surveys reported in this paper}
 \label{fig:researchJourneySteps}
\end{figure*}

Indeed, opinions are key determinants to many of the activities
related to software development, such as developers' productivity
analysis~\cite{Ortu-AreBulliesMoreProductive-MSR205}, determining developers
burnout~\cite{Mika-MiningValenceBurnout-MSR2016}, improving software
applications~\cite{Guzman-SentimentAnalysisAppReviews-RE2014}, and
developing awareness tools for software development
teams~\cite{Guzman-EmotionalAwareness-FSE2013,Murgia-DoDevelopersFeelEmotion-MSR2014,
Novielli-ChallengesSentimentDetectionProgrammerEcosystem-SSE2015}.
Research on APIs has produced important contributions, such as automatic usage
inference mining~\cite{ZimmermannZeller-eRose-TSE2006a,Michail-DataMiningLibraryReusePatternsAssociation-ICSE2000a},
automatic traceability recovery between API elements and
learning
resources~\cite{Dagenais-RecoDocPaper-ICSE2012a,Rigby-CodeElementInformalDocument-ICSE2013,Bacchelli-LinkEmailSourceCode-ICSE2010a}, as well as recommendation systems to facilitate code
reuse~\cite{Holmes-Strathcona-TSE2006a,Ekwa-APIExplorer-ECOOP2011a} (see
\sec\ref{sec:background}).
However, to the best of our knowledge, there is no research that focuses on the
analysis of developers' perception about API reviews and how such
opinions affect their API-related decisions. As illustrated in \fig\ref{fig:soSentimentPosts},
while developer forums serve as communication channels for discussing the
implementation of the API features, they also enable the
exchange of opinions or sentiments expressed on numerous APIs,
their features and aspects.
Given the presence of sentiments in
the forum posts and opinions about APIs, such insights can be leveraged to develop
techniques to automatically analyze API reviews in forum posts. Such
insights can also contribute not only to the development of an empirical body of
knowledge on the topic, but also to the design of tools that analyze opinion-rich information.

To fill this gap in the literature, we conducted two surveys involving a total
of 178  software developers. The \it{goals} of our study are to \it{understand}
 \begin{inparaenum}[(1)]
\item how  software developers seek and value opinions about APIs,
and
\item what tools can better support their analysis and evaluation of the
API reviews.
\end{inparaenum}
The \it{subjects}
are the surveys' participants and the \it{objects}
are the API reviews that the developers encounter in their daily development
activities from diverse resources. The \it{context} consists of the various development
activities that can be influenced by the API reviews.
Through an
exploratory analysis of the surveys' responses, we answer the following research
questions:

\nd\bf{RQ1: How do developers seek and value opinions about APIs in developer
forums?}

The developers reported that they seek opinions about APIs in forum posts to
support diverse development needs, such as API selection, documentation,
learning how to use an API, etc.
The developers valued the API reviews. However, they were also cautious while
making informed decisions based on those reviews due to a number of factors related
to the quality of the provided opinions, such as, the lack of insights
into the prevalence of the issue reported in the opinion, the trustworthiness of
the provided opinion (e.g., marketing initiatives vs subjective opinion), etc. The developers wished for the support of different mechanisms to
aggregate opinions about APIs and to assess the quality of the provided opinions
about APIs in the developer forums.


\nd\bf{RQ2: What tool support is needed to help developers with analyzing
and assessing API reviews in developer forums?}

The developers consider that automated tools of diverse nature can be developed
and consulted to properly analyze API reviews in forum posts, e.g.,
visualizations of the aggregated sentiment about APIs to determine their
popularity, understanding the various aspects (e.g., performance) about
APIs and how other developers rate those aspects about
APIs based on their usage of the APIs, etc.
The developers mentioned that the huge volume of available information and
opinions about APIs in the forum posts can hinder their desire to get quick, digestible, and actionable insights about APIs. The developers also mentioned that in the
absence of any automated summarization technique to help them, they leverage different
features in forum posts to get summarized viewpoints of the APIs, e.g., skimming
through highly-ranked posts, using tags to find similar APIs, etc. Developers also
envision diverse potential summarization approaches to help them address
such needs, e.g., a dedicated portal to show aggregated opinions about APIs.

In \fig\ref{fig:researchJourneySteps}, we show the three major research phases we
undertook during and after conducting the two surveys.
\begin{description}[style=nextline]
\item[Phase 1. Design, Conduct, and Report Surveys] We conducted two surveys. We analyze the survey responses using both statistical and qualitative analyses.
This paper primarily focuses on the design, analysis, and reporting of the two surveys (see \secs\ref{sec:motivation} - \ref{sec:discussion}).

\item[Phase 2. Identify Actionable Insights for Tool Design] We identify requirements from the survey results to develop techniques and tools to
assist developers in their exploration of opinions about APIs from developer forums. In \sec\ref{sec:research-journey}, we discuss
the actionable findings from the survey results that could be used for future tool designs.
\item[Phase 3. Develop and Evaluate Techniques and Tools]  We develop techniques based on the findings from Phase 2. We incorporate the techniques in our prototype tool, called  Opiner.
Opiner is a search engine~\cite{Uddin-OpinerReviewToolDemo-ASE2017,Uddin-OpinerReviewToolDemo-ASE2017}. Using Opiner, developers can search for an API by name and explore the opinions and usage scenarios related to APIs. The opinions and
usage scenarios are automatically mined from Stack Overflow. 
In  \sec\ref{sec:research-journey}, we briefly describe Opiner.
\end{description}

\nd In this paper, we make the following main  contributions:
\begin{enumerate}

\item \bf{Surveys.} The design of two surveys and the collected data
involving the responses of 178 software engineers.

\item \bf{Analysis.} A detailed analysis of the survey responses that provides
insights into:
\begin{inparaenum}[(1)]
\item How  developers seek and analyze opinion-rich API information.
\item Needs for automated tool supports
to make informed, proactive, and efficient decisions based on analysis of API reviews. 
\end{inparaenum} \rev{In \tbl\ref{tab:comparison-related-work}, we compare the major findings of this paper against the state of the art research on APIs. In \sec\ref{sec:background}, 
we discuss the related work in details.}

%
\end{enumerate}
\begin{table*}[t]
  \centering
  \caption{Comparison between our findings and prior findings}
    \begin{tabular}{p{1.8cm}p{4cm}|p{5cm}|p{6.3cm}}\toprule
    \textbf{Theme} & \multicolumn{1}{l}{\textbf{Our Study}} & \multicolumn{1}{l}{\textbf{Prior Study}} & \textbf{Comparison} \\
\midrule
        \multirow{2}{*}{\parbox{1.8cm}{\bf{How developers learn to select and use APIs}}} 
    & 
     \multirow{2}{*}{\parbox{4cm}{
    \begin{inparaenum}[(1)]
    \item Developers seek opinions to support diverse development needs, e.g., API selection, feature improvement, etc.
    \item Developers seek opinions to compensate for the shortcomings in API official documentation (e.g., incorrectness, etc.).
    \item Developers consider the combination of code examples and opinions about an API as a form of API documentation.  
    \end{inparaenum}
     }}

    &  Robillard and DeLine~\cite{Robillard-FieldStudyAPILearningObstacles-SpringerEmpirical2011a} identified in a series of survey and interviews that the most severe obstacles developers faced while
     learning new APIs were related to the official documentation of the APIs. 
    & Our study confirms the findings of~\cite{Robillard-FieldStudyAPILearningObstacles-SpringerEmpirical2011a} that developers find API official documentation can be incomplete. 
    In addition, we find that developers leverage API reviews in forums to compensate for those shortcomings. \\
    \cmidrule{3-4}
    &   
    & Carroll et al.~\cite{Carroll-MinimalManual-JournalHCI1987a} designed ``minimal manual'' to support task-based documentation after observing  that
    the learning of the developers was often interrupted by their self-initiated problem-solving tasks 
    while using API official documentation.   
    & While Carroll et al.~\cite{Carroll-MinimalManual-JournalHCI1987a} created ``minimal manual'' manually, our 
    results show that we can leverage developer forums to develop ``minimal manual'' by combining code examples and API reviews, because developers consider both as a form of documentation. \\
    \midrule
    {\bf{Sentiment analysis of software artifacts}} 
    & 
    
    (1) Developers use positive and negative opinions of other developers as an indicator of quality of the discussed API and code examples.
    (2) Developers face challenges to determine the quality of provided opinions (e.g., trustworthiness, recency, real-world facts, etc.). 
    & The attributes in developer forums (e.g., upvotes, downvotes, etc.) are used to analyze the quality of posts
and their roles in the Q\&A process~\cite{Calefato-SOSuccessfulAnswers-MSR2014,Bajaj-MiningQuestionsSO-MSR2014,
Lal-MigratedQuestionsSO-APSEC2014,Correa-DeletedQuestionSO-WWW2014,Vasilescu-SocialQAKnowledgeSharing-CSCW2014,Kavaler-APIsUsedinAndroidMarket-SOCINFO2013},
to analyze developer profiles (e.g., personality traits of the most and low
reputed users)~\cite{Bazelli-SOPersonalityTraits-ICSM2013,
Ginsca-UserProfiling-DUBMOD2013}, or to determine the influence of badges
in Stack Overflow~\cite{Anderson-SOBadge-WWW2013}.
    & Our findings show that more insights can be derived 
    by analyzing the opinions of developers towards APIs, in addition to~\cite{Calefato-SOSuccessfulAnswers-MSR2014,Bajaj-MiningQuestionsSO-MSR2014,
Lal-MigratedQuestionsSO-APSEC2014,Correa-DeletedQuestionSO-WWW2014,Vasilescu-SocialQAKnowledgeSharing-CSCW2014,Kavaler-APIsUsedinAndroidMarket-SOCINFO2013}. 
    Our findings highlight the needs for research into the analysis of opinion quality, e.g., 
    by gaining deeper insights into developer reputations and badges following~\cite{Bazelli-SOPersonalityTraits-ICSM2013,
Ginsca-UserProfiling-DUBMOD2013,Anderson-SOBadge-WWW2013}.  \\
    \midrule
    \bf{Analysis of APIs in developer forums} 
    & Developers prefer to learn about different API aspects from the opinions of other developers in forum posts using automated analysis (e.g., find all opinions discussing about the 
    performance of an API).
    & Zhang and Hou~\cite{Zhang-ProblematicAPIFeatures-ICPC2013} identified problematic API features in the discussion Stack Overflow posts, by detecting 
    sentences with negative sentiments. Treude and Robillard~\cite{Treude-APIInsight-ICSE2016} 
    mined important insights about an API type from the textual contents of Stack Overflow. 
    & 
    Unlike Zhang and Hou~\cite{Zhang-ProblematicAPIFeatures-ICPC2013}, our study shows that both positive and negative opinions about API features need to be identified.
    The insights gained for each API type by~\cite{Treude-APIInsight-ICSE2016} can be enhanced by also including the diverse opinions about API aspects. \\
    \midrule
    \multirow{2}{*}{\parbox{1.8cm}{\bf{Summarization of software artifacts}}} 
    & Developers mostly rely on search engines to explore opinions about APIs. They  
    are frequently overwhelmed with the huge volume of opinions about APIs in forums. They wished for an automatic 
    summarizer by mining those opinions. 
    & 
    Topic modeling has been used to find dominant discussion topics in Stack Overflow~\cite{Barua-StackoverflowTopics-ESE2012,Rosen-MobileDeveloperSO-ESE2015}, and 
    to find recurrent themes in API learning obstacles~\cite{Wang-DetectingAPIUsageObstacles-MSR2013}. Both natural language and structural summaries of 
    source code have been studied extensively~\cite{Rastkar:2014,murphy:context,Holmes-Strathcona-TSE2006a, anvik:context,Sridhara-MethodSummary-ASE2010,Moreno13,McBurney14,Guerrouj15-nier,Murphy:summarization}.
    &  
    Unlike~\cite{Barua-StackoverflowTopics-ESE2012,Rosen-MobileDeveloperSO-ESE2015}, our study 
    shows that a finer grained topic-based summarization is required (e.g., using only opinions). 
    Given the needs to analyze opinions by diverse API aspects (e.g., performance, usability), an aspect-based~\cite{Kim-OpinionSummarizerSurvey-UIUC2011} opinion summarization for APIs can be useful, which is different 
    from source code summarization~\cite{Rastkar:2014,murphy:context,Holmes-Strathcona-TSE2006a, anvik:context,Sridhara-MethodSummary-ASE2010,Moreno13,McBurney14,Guerrouj15-nier,Murphy:summarization}.\\
    \cmidrule{2-4}
    & Besides an opinion summarizer, developers also asked for tool support to assist in their development tasks 
    by leveraging opinions about APIs, such as, API comparator, trend analyzer, API opinion miner, etc.  
    & Several tools have been developed to harness knowledge about APIs from developer forums, such as automatically generating comments to explain a 
    code example~\cite{Wong-AutoCommentSO-ASE2013}, recommending experts to answer a question in Stack Overflow~\cite{Chang-RoutingQuestionsSO-SNAM2013}, etc.  
    & Unlike~\cite{Wong-AutoCommentSO-ASE2013,Chang-RoutingQuestionsSO-SNAM2013}, we offer insights on the usage of opinions in tool development to support 
    development tasks. Our findings offer possible extensions to existing research, e.g., include reactions towards a code example of~\cite{Wong-AutoCommentSO-ASE2013} 
    to show its quality. \\
          \bottomrule
    \end{tabular}%
  \label{tab:comparison-related-work}%
\end{table*}%


%

%% file: related-work.tex
\section{Related Work} \label{sec:background}
As noted in \sec\ref{sec:introduction}, the findings of this paper motivated us to pursue a research journey that
contributed to the development of our proof-of-concept tool, Opiner (see  \sec\ref{sec:research-journey}). Specifically, our subsequent
research projects focused on the design, development, and evaluation of techniques and tools to mine
and summarize opinions and usage scenarios about APIs from developer forums. The research journey is captured in the following manuscripts:
\begin{enumerate}
  \item In \cite{Uddin-OpinionMining-TSE2018}, we present a benchmark dataset of 4,522 sentences from Stack Overflow, each labelled as API aspects, such as performance, usability, etc.
  The catalog of API aspects is derived from the two surveys of this paper (Q11 and Q15 in Final Survey and Q15 from pilot survey). We leverage
  the benchmark dataset to develop machine learning supervised classifiers to automatically detect API aspects discussed in opinionated sentences. 
  We then present a suite of algorithms to automatically mine opinions about APIs from Stack Overflow. We report the evaluation of each technique.
  \item In \cite{Uddin-OpinerReviewAlgo-ASE2017}, we present 
  two algorithms to summarize opinions about APIs from Stack Overflow. The design and development
  of the algorithms were motivated by the findings from this paper, e.g., 
  developers prefer to seek opinions about API aspects, such as performance, etc. We compare the
  algorithms against six off-the-shelf summarization algorithms.
  \item In \cite{Uddin-OpinerReviewToolDemo-ASE2017}, we present 
  the Opiner architecture that supports the mining and summarization of opinions from Stack Overflow.
  \item In \cite{Uddin-MiningUsageScenarios-TechReport2018}, we present a framework to automatically
   mine usage scenarios about APIs from Stack Overflow. We present an empirical study
  to investigate the value of the framework.
  \item In \cite{Uddin-UsageSummarization-TSE2018}, we present four algorithms to automatically 
  summarize usage scenarios about APIs and their evaluation using four user studies.
\end{enumerate} As noted above, the findings from this paper have 
formed the cornerstone towards the development and evaluation of the techniques and tools presented in the above papers.

Other related work can be categorized into four areas.
\begin{inparaenum}[(1)]
\item Studies conducted to understand how developers learn to select and use APIs,
\item Analysis of APIs in developer forums,
\item Sentiment analysis in software engineering, and
\item Summarization of software artifacts.
\end{inparaenum} We discuss the related work below.

\subsection{How developers learn to select and use APIs}\label{sec:related-work-howDevLearnSelectApis}
While our surveys focused on the role of opinions to support development tasks, previous studies mainly focused on interviews  or empirical studies
to understand the role of API official documentation to support the development tasks~\cite{Sohan-UsabilityRESTAPI-VLHCC2017,Uddin-HowAPIDocumentationFails-IEEESW2015,Wang-DetectingAPIUsageObstacles-MSR2013,Barzillay-StackOverflow-Springer2013,Ekwa-StudyUnfamiliarAPIs-ICSE2012,Robillard-FieldStudyAPILearningObstacles-SpringerEmpirical2011a,Treude-HowProgrammersAskQuestionsInWeb-ICSE2011}.
Developers in our survey seek opinions about APIs to support diverse development needs (e.g., API selection) as well 
as to compensate for shortcomings in API documentation, e.g., when the documentation is incomplete or ambiguous. The problems in API official 
documentation are previously reported in multiple studies, such as~\cite{Robillard-FieldStudyAPILearningObstacles-SpringerEmpirical2011a,Carroll-MinimalManual-JournalHCI1987a,Uddin-HowAPIDocumentationFails-IEEESW2015,Shull-InvestigatingReadingTechniquesForOOFramework-TSE2000}, etc.

Robillard and DeLine~\cite{Robillard-FieldStudyAPILearningObstacles-SpringerEmpirical2011a}
conducted a survey and a series of qualitative interviews of software developers
at Microsoft to understand how developers learn APIs. The study identified
that the most severe obstacles developers faced while learning new APIs were
related to the official documentation of the APIs. With API
documentation, the developers cited the lack of code examples and the absence of task-oriented
description of the API usage as some major blockers to use
APIs. 
The benefit of task-based documentation over traditional hierarchy-based documentation (e.g., Javadoc) was previously 
also reported by Carroll et al.~\cite{Carroll-MinimalManual-JournalHCI1987a} who observed developers
while using traditional documentation to learn an API (e.g., Javadoc, manual).
They found that the learning of the developers was often interrupted by
their self-initiated problem-solving tasks that they
undertook during their navigation of the documentation. During such unsupervised
exploration, they observed that developers ignored groups and entire sections of
a documentation that they deemed not necessary for their development task at hand. Unsurprisingly, such unsupervised exploration often led to mistakes. They
conjectured that traditional API documentation is not designed to support such
active way of developers' learning. To support developers' learning of APIs from
documentation, they designed a new type of API documentation, called as the
\it{minimal manual}, that is task-oriented and that helps the users resolve
errors~\cite{Cai-FrameworkDocumentation-PhDThesis2000,Carroll-MinimalManual-JournalHCI1987a,
Rossen-SmallTalkMinimalistInstruction-CHI1990a,Meij-AssessmentMinimalistApproachDocumentation-SIGDOC1992}. In a subsequent study of 43 participants, 
Shull et al.~\cite{Shull-InvestigatingReadingTechniquesForOOFramework-TSE2000} also confirmed the effectiveness of example-based documentation 
over hierarchy-based documentation.

The advent of cloud-based software development has popularized the adoption
of Web APIs, such as, the Google Map APIs, API mashups in the ProgrammableWeb, and
so on. Tian et al.~\cite{Tian-ExploratoryStudyWebApi-EASE2017} conducted an
exploratory study to learn about the features offered by the Web APIs
via the ProgrammableWeb portal and the type of contents supported in the
documentation of those APIs. They observed that such Web APIs offer diverse
development scenarios, e.g., text mining, business analysis, etc. They
found that the documentation of the APIs offer insights into different
knowledge types, e.g., the underlying intent of the API features, the
step-by-step guides, etc. Sohan et al.~\cite{Sohan-UsabilityRESTAPI-VLHCC2017} observed
that REST API client developers face problems while using an API without usage examples, such
as correct data types, formats, required HTTP headers, etc. Intuitively, the developer forums
can be used for missing code examples for an API, because developers in our surveys mentioned that they rely
on the API usage discussions in developer forums when the API official documentation can be missing.

The generation of such task-based documentation can be challenging. However, the developers in our survey reported that they consider 
the combinations of code examples and reactions towards the examples about an API in the forum as a form of API documentation. Intuitively, the Q\&A format 
of online developer forums (e.g., Stack Overflow) follow task-based documentation format, e.g., the task is described in the question and the solution with code examples 
and opinions in the answer posts. Leveraging usage scenarios about APIs posted in the developer forums can be
necessary, when API official documentation does not include those scenarios and
can often be
outdated~\cite{
Dagenais-DeveloperDocumentation-FSE2010a}. Indeed, documentation that does not meet the expectations of its
readers can lead to frustration and a major loss of
time,
or even in an API being
abandoned~\cite{Robillard-FieldStudyAPILearningObstacles-SpringerEmpirical2011a}.
To address the shortcomings in API official documentation, research
efforts focused on the
linking of API types in a formal documentation (e.g., Javadoc) to code examples
in forum posts where the types are
discussed~\cite{Subramanian-LiveAPIDocumentation-ICSE2014}, presenting
interesting textual contents from Stack Overflow about an API type in the formal
documentation~\cite{Treude-DocumentationInsightsSO-ICSE2016}, etc.
However, our study in this paper shows that the plethora of usage discussions
available for an API in the forum posts can pose challenges to the developers
to get quick and actionable insights into how an API can be used for a given
development task. One possible way to assist developers is to generate
on-demand developer documentation from forum
posts~\cite{Robillard-OndemandDeveloperDoc-ICSME2017}. However, to be able to do
that, we first need to understand what specific problems persist in the API
official documentation, that should be addressed through such documentation
efforts.
If we know of the common documentation problems, we can then prioritize those problems
and investigate techniques to leverage API usage
scenarios posted in the developer documents to address those problems.
In a recent study~\cite{Uddin-HowAPIDocumentationFails-IEEESW2015}, we conducted surveys of more than 300 developers
at IBM to understand the problems developers faced while using API official documentation. We observed 10
common documentation problems, such as documentation incompleteness, incorrectness, ambiguities, etc.
Therefore, we can leverage the findings from our study to design API documentation resources that can address
the problems developers that are commonly observed in API documentation.
%

\subsection{Sentiment Analysis of Software Artifacts}\label{sec:related-work-sentiment}
Developers in our surveys reported that they use the positive and negative opinions towards an API as an indicator of quality of the features offered by the API. 
While our findings shed light on developers' perceptions of API quality by leveraging opinions, recent research of sentiment analysis of 
software artifacts focused on the mining of sentiments and emotions from software repositories. 
Ortu et al.~\cite{Ortu-AreBulliesMoreProductive-MSR205} observed weak
correlation between the politeness of the developers in
the comments and the time to fix the issue in Jira,
i.e., bullies are not more productive than others in a software development
team. M\"{a}ntyl\"{a}  et al.~\cite{Mika-MiningValenceBurnout-MSR2016} correlated VAD (Valence, Arousal, Dominance) scores~\cite{Warriner-VadLexicons-BRM2013} in
Jira issues with the loss of productivity and burn-out in software engineering
teams. They found that the increases in issue's priority correlate with increases in Arousal.
Pletea et al.~\cite{Pletea-SecurityEmotionSE-MSR2014} found that
security-related discussions in GitHub contained more negative comments. Guzman et
al.~\cite{Guzman-SentimentAnalysisGithub-MSR2014} found that GitHub
projects written in Java have more negative comments as well as the comments
posted on Monday, while the developers in a distributed team are more positive.
Guzman and Bruegge~\cite{Guzman-EmotionalAwareness-FSE2013} summarized emotions
expressed across collaboration artifacts in a
software team (bug reports, etc.) using
LDA~\cite{Blei-LDA-JournalMachineLearning2003} and sentiment analysis. The team
leads found the summaries to be useful, but
less informative.

We observed in our surveys that while developers analyze opinions to learn about APIs, they also find it challenging to assess the quality (e.g., trustworthiness) of the 
provided opinions due to the presence of such opinions scattered across multiple unrelated forum posts.  Related research has focused on the 
assessment of post quality using post attributes (e.g., post score) and their roles in the Q\&A process~\cite{Calefato-SOSuccessfulAnswers-MSR2014,Bajaj-MiningQuestionsSO-MSR2014,
Lal-MigratedQuestionsSO-APSEC2014,Correa-DeletedQuestionSO-WWW2014,Vasilescu-SocialQAKnowledgeSharing-CSCW2014,Kavaler-APIsUsedinAndroidMarket-SOCINFO2013},
to analyze developer profiles (e.g., personality traits of the most and low
reputed users)~\cite{Bazelli-SOPersonalityTraits-ICSM2013,
Ginsca-UserProfiling-DUBMOD2013}, or to determine the influence of badges
in Stack Overflow~\cite{Anderson-SOBadge-WWW2013}. In contrast to the above work, our findings motivate the needs to develop 
opinion quality assessment models, tools and techniques for forum posts.


\subsection{Analysis of APIs in Developer Forums}\label{sec:related-work-analysisOfApiDeveloperForums}
Developers in our surveys prefer to seek opinions about different API aspects and wish to leverage automated tools to support such analysis (e.g., get all 
the opinions discussing about the performance of an API, etc.). A closely related work is the identification of problematic 
API features in Stack Overflow by Zhang and Hou~\cite{Zhang-ProblematicAPIFeatures-ICPC2013}, who considered sentences with negative sentences 
as indicators of problematic API features. In a parallel study, Wang and Godfrey~\cite{Wang-DetectingAPIUsageObstacles-MSR2013}  
hypothesize that the more discussions an API class generate in the forum posts, the more
likely that it is problematic to use. By applying topic modeling on the posts, they observed several
recurrent themes of API usage obstacles, such as learning the interactions among components. Unlike both \cite{Zhang-ProblematicAPIFeatures-ICPC2013} and \cite{Wang-DetectingAPIUsageObstacles-MSR2013} 
our findings motivate the needs to study both positive and negative opinions about APIs to obtain finer-grained insights about API aspects.
Intuitively, such fine-grained insights can also be useful to compare competing APIs for a given task. 

A large volume of API research has devoted into the automatic mining of insights about APIs from Stack Overflow. In contrast to our study, they 
the studies are mainly empirical (i.e., automatic mining techniques) and they do not consider opinions about APIs. For example, Treude and Robillard~\cite{Treude-APIInsight-ICSE2016} 
developed machine learning tools to detect insightful sentences about API types from Stack Overflow. Parnin et al.~\cite{Parnin2012} have investigated API classes discussed in
Stack Overflow using heuristics based on exact matching
of classes names with words in posts (title, body, snippets, etc.).
Using a similar approach, Kavaler et al.~\cite{Kavaler-APIsUsedinAndroidMarket-SOCINFO2013} analyzed the
relationship between API usage and their related Stack Overflow discussions. Both studies found
a positive relationship between API class usage and the volume of Stack Overflow discussions.
More recent research \cite{Linares-Vasquez-2014,Guerrouj15-saner} investigated the relationship between API changes and developer discussions. 
Treude et al.~\cite{Treude-HowProgrammersAskQuestionsInWeb-ICSE2011} categorized
questions and posts from Stack Overflow to understand the types of questions asked in Stack Overflow. They observed that
developer forums are the most effective for code reviews and conceptual questions. The findings corroborate to our study participants
who mentioned that they seek expertise in API usage in forum posts.

\subsection{Summarization of Software Artifacts}\label{sec:related-work-summarization}
Developers in our surveys wish for tools to automatically summarize the huge volume of opinions posted about APIs from developer forums.
Such summaries can be different from the research conducted to produce Natural language and structural summaries of source code~\cite{Rastkar:2014,
murphy:context,
Holmes-Strathcona-TSE2006a,
anvik:context,
Sridhara-MethodSummary-ASE2010,
Moreno13,
McBurney14,
Guerrouj15-nier,
Haiduc:Summarization,
Murphy:summarization,
Ying-SelectionPresentationCodeExample-FSE2014,Rodeghero-CodeSummarizationEyeTracking-ICSE2014}.

One potential technique in text-based summarization is topic modeling, which can be applicable to summarize opinions (e.g., in other domains~\cite{Kim-OpinionSummarizerSurvey-UIUC2011}).
Previously, topic modeling has been used to study dominant discussion
 topics in developer forums~\cite{Barua-StackoverflowTopics-ESE2012,Rosen-MobileDeveloperSO-ESE2015}. Our survey findings motivate for a finer grained 
 analysis using topic modeling, such as analysis of only opinions instead of all textual contents. Given that 
 developers in our survey prefer to analyze opinions by API aspects (e.g., performance, usability), another 
 potential summarization approach could be aspect-based summarization~\cite{Kim-OpinionSummarizerSurvey-UIUC2011} of API opinions, as we 
 developed in Opiner~\cite{Uddin-OpinerReviewAlgo-ASE2017} based on the findings from this study.

%% file: motivation.tex
\begin{sidewaysfigure*}
\includegraphics[width=25cm,height=35cm,keepaspectratio]{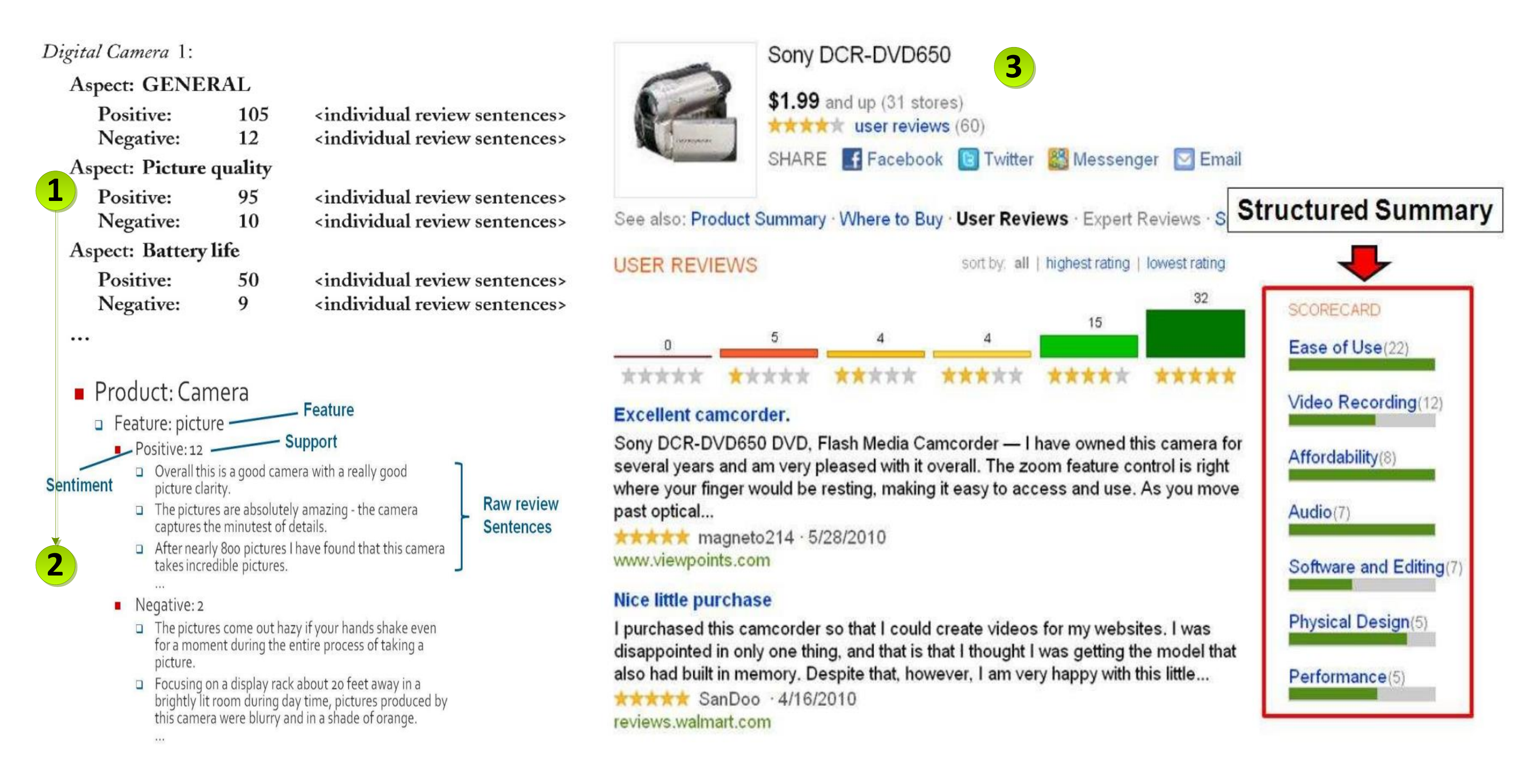}%
\caption{Screenshots of opinion summarization engine for camera reviews. The
     different aspects (e.g., picture quality) are used to present summarized
     viewpoints about the camera.
     The circle 3 shows an incarnation of the camera product reviews in the now defunct Bing product
     search. The screenshots are taken from \cite{Kim-OpinionSummarizerSurvey-UIUC2011}. }
\label{fig:opiniontools-camera}
\end{sidewaysfigure*}
\section{Research Context} \label{sec:motivation}
We further motivate the needs for better understanding of the
impact of opinions about APIs on the developers first by taking cues from other domains (\sec\ref{subsec:motivation-other-domains}) and 
then by demonstrating the prevalence of opinions about APIs in Stack Overflow based on small study in \sec\ref{subsec:study-prevalence-opinions}. 
We discuss the rationale behind each research question, along with its
sub-questions in \secs\ref{subsec:rq1} and \ref{subsec:rq2}. We then discuss the motivation behind the two surveys we conducted 
to answer the research questions (\sec\ref{subsec:motivation-surveys}).

\subsection{Opinion Analysis in Other Domains}\label{subsec:motivation-other-domains}

Our research on
API reviews was motivated by similar research in the other domains.
Automated sentiment analysis and opinion mining about entities (e.g., cars,
camera products, hotels, restaurants) have been challenging but a practical
research area due to their benefits for consumers (e.g.,  guiding them in choosing a hotel or selecting a camera product).  

In \fig\ref{fig:opiniontools-camera}, we show the screenshot of three
separate initiatives on the automatic collection and aggregation of
reviews about camera products. The first two tools are developed in the academia and the third tool
was developed as part of Microsoft's Bing Product Search. The first
two circles \circled{1} and \circled{2}, show two preliminary outlines
of such a tool presented by Liu et
al.~\cite{liu-sentimentanalysis-handbookchapter-2010}.
The third circle \circled{3} shows a similar tool in the Microsoft Bing Product Search. In all
of the three tools, positive and negative opinions about a camera product are
collected and aggregated under a number of aspects (e.g., picture quality,
battery life, etc.). When no predefined aspect is found, the opinions are
categorized under a generic aspect ``GENERAL'' (aspect is named as ``feature'' in
\circled{2}). The opinions about the camera can be collected from diverse
sources, e.g., online product reviews, sites selling the product, etc. For
example, Google collects reviews about hotels and restaurants through a separate
gadget in its online search engine.

%

\subsection{API Reviews in Developer Forums}\label{subsec:study-prevalence-opinions}
%
%
\rev{Similar to the camera reviews in
\fig\ref{fig:opiniontools-camera}, API reviews can be found in forum posts. As we demonstrated in \fig\ref{fig:soSentimentPosts}, opinions about APIs 
can be prevalent in developer forums. In fact, we observed that more than 66\% of posts that are tagged
``Java'' and ``JSON'' in the Stack Overflow data contain at least
one positive or negative sentiment\footnote{We used sentiment detection algorithm we developed in~\cite{Uddin-OpinionValue-TSE2018} on the Stack Overflow 2014 data dump}. 
In \tbl\ref{tbl:datasets-overview} we show
descriptive statistics of the dataset.
There were 22,733 posts from 3,048 threads with scores greater than zero. We did
not consider any post with a negative score because such posts are considered
as not helpful by the developers in Stack Overflow. The last column ``Users''
show the total number of distinct users that posted at least one
answer/comment/question in those threads. To identify uniqueness of a user, we
used the user\_id as found in the Stack Overflow database. On average, around
four users participated in one thread, and more than one user participated in
2,940 threads (96.4\%), and a maximum of 56 distinct users participated in one
thread~\cite{website:stackoverflow-338586}. From this corpus, we identified
the Java APIs that were mentioned in the posts. To identify the Java APIs, we
used our API database.}

\rev{Our API database consists of the Java official APIs and
the Java APIs listed in the two software portals Ohloh~\cite{website:ohloh}
and Maven central.~\cite{website:maven-central}\footnote{\scriptsize We crawled
Maven in March 2014 and Ohloh in December 2013.} We crawled the Javadocs of
five official Java APIs (SE 6-8, and EE 6,7)
and collected information about 875 packages and 15,663 types.
 We consider an official Java package as an API in the
 absence of any guidelines available to consider otherwise. In total, our API database
 contains 62,444 distinct Java
 APIs. All of the APIs (11,576) hosted in Maven central are for Java. From
 Ohloh, we only included the Java APIs (50,863) out of
 the total crawled (712,663). We considered a project in Ohloh as a Java API
 if its main programming language was
 Java.}

\rev{We collected the opinionated sentences about APIs using a technique
we developed in~\cite{Uddin-OpinionMining-TSE2018}. The
technique works as follows:
\begin{enumerate}[leftmargin=10pt]
  \item \bf{Load and preprocess} Stack Overflow posts.
  \item \bf{Detect opinionated sentences} using a rule-based algorithm. The algorithm is an adaptation of 
  the Domain Sentiment Orientation (DSO)~\cite{Hu-MiningSummarizingCustomerReviews-KDD2004} 
  algorithm for software engineering. Similar adaptation of the algorithm was previously reported by Blair-Goldensohn et al.~\cite{BlairGoldensohn-SentimentSummarizerLocalReviews-NLPIX2008} for 
  Google local product reviews. The algorithm computes sentiment score of each sentence by detecting positive and negative sentiment words in the sentence.  
  \item \bf{Detection of API names} in the forum texts and hyperlinks based a set of heuristics (e.g., exact and fuzzy matching) and
  \item \bf{Association of APIs to opinionated sentences} using
  heuristics, e.g., proximity between API names and opinionated sentences.
\end{enumerate}}

\begin{table}[tbp]
\centering
\scriptsize
\caption{Statistics of the dataset (A = Answers, C = Comments).}
\centering
\begin{tabular}{rrrr|rr|r}\toprule
\textbf{Threads} & \textbf{Posts} & \textbf{A} & \textbf{C}
& \textbf{Sentences} & \textbf{Words} & \textbf{Users} \\
\midrule
3048 & 22.7K & 5.9K & 13.8K & 87K & 1.08M & 7.5K \\ \midrule
\textbf{Average} & 7.46 & 1.93 & 4.53 & 28.55 & 353.36 & 3.92 \\
\bottomrule
\end{tabular}
\label{tbl:datasets-overview}
\end{table}

\rev{In \tbl\ref{tbl:opinionated-sentence-dataset}, we present summary statistics of
the opinionated sentences detected in the dataset. Overall 415 distinct APIs
were found. While the average number of opinionated sentences per API was
37.66, it was 2,066 for the top five most reviewed APIs. In fact, the top
five APIs contained 66.1\% of all the opinionated sentences in the
posts. These APIs are \api{jackson},
\api{Google Gson}, \api{spring framework}, \api{jersey}, and
\api{org.json}.} 

\begin{table}[tbp]
\centering
\scriptsize
\caption{Distribution of opinionated sentences across APIs.}
\begin{tabular}{rrrr|rrr}\toprule
\multicolumn{ 4}{c}{\textbf{Overall}} & \multicolumn{ 3}{c}{\textbf{Top Five}} \\
\midrule
\textbf{API} & \textbf{Total} & \textbf{+Pos} &
\textbf{--Neg} & \textbf{Total} & \textbf{+Pos} & \textbf{--Neg} \\
\midrule
415 & 15,627 & 10,055 & 5,572 & 10,330  & 6,687  & 3,643  \\ \midrule
\multicolumn{1}{l}{\textbf{Average}} & 37.66 & 24.23 & 13.43 & 2,066 & 1,337.40 & 728.60 \\
\bottomrule
\end{tabular}
\label{tbl:opinionated-sentence-dataset}
\end{table}

\begin{table*}[t]
  \centering
  \caption{The formulation of the research questions in the study.}
    \begin{tabular}{r|r|l}\toprule
    \multicolumn{1}{l}{\textbf{RQ1}} & \multicolumn{2}{l}{\textbf{How do developers seek and value opinions about APIs?}} \\
    \midrule
    \textbf{1.1} & \multicolumn{2}{l}{\textit{\textbf{ Theme - Opinion Needs:
    How do developers seek opinions about APIs?}}} \\
    \cmidrule{2-3}
          & \textbf{1.1.a} & \textit{Where do developers seek opinions?} \\
          & \textbf{1.1.b} & \textit{What motivates developers to seek opinions about APIs?} \\
          & \textbf{1.1.c} & \textit{What challenges do developers face while seeking API reviews?} \\
    \cmidrule{2-3}
    \textbf{1.2} & \multicolumn{2}{l}{\textit{\textbf{Theme - Opinion Quality:
    How do developers assess the quality of the provided opinions about APIs?}}} \\
    \midrule
    \multicolumn{1}{l}{\textbf{RQ2}} & \multicolumn{2}{l}{\textbf{How can
    the support for automated processing of opinions assist developers to
    analyze API reviews?}}
    \\
    \midrule
    \textbf{2.1} & \multicolumn{2}{p{44.55em}}{\textit{\textbf{Theme - Tool
    Support:
    What tool support is needed to help developers to assess API reviews?}}}
    \\
    \cmidrule{2-3}
    \textbf{2.2} & \multicolumn{2}{p{44.55em}}{\textit{\textbf{Theme -
    Summarization Needs: What are the developers' needs for summarization of API
    reviews?}}}
    \\
    \cmidrule{2-3}
          & \textbf{2.2.a} & \textit{What problems in API reviews motivate the need for summarization?} \\
          & \textbf{2.2.b} & \textit{How summarization of API reviews can support developer decision making?} \\
          & \textbf{2.2.c} & \textit{How do developers expect API reviews to be summarized?} \\
    \bottomrule
    \end{tabular}%

  \label{tab:rqs}%
\end{table*}%

\subsection{Reasons for Seeking Opinions About APIs (RQ1)}\label{subsec:rq1}
We aim to understand how and why developers seek and analyze
such opinions about APIs and how such information can shape their perception and
usage of APIs. The first goal of our study is to learn how
developers seek and value the opinions of other developers.
\subsubsection{Motivation}
By analyzing how and
why developers seek opinions about APIs, we can gain insights into the role of API reviews in the daily
 development activities of software developers. The first step towards understanding
 the developers needs, is to learn
about the resources they currently leverage to seek information and opinions
about APIs. The developers may use diverse resources to seek opinions about
APIs. If a
certain resource is used more than other resources, the analysis of the resource can be given more priority over others.

\rev{As we noted in \sec\ref{subsec:study-prevalence-opinions}, our observation of the API reviews shows that opinions about APIs are shared in
the developer forums.} Therefore, an understanding of how the different development
activities can be influenced and supported through the API reviews can provide us with insights about the factors that motivate developers to seek opinions
as well as the challenges that they may face during this process. By learning about these challenges while
seeking reviews about APIs, we can gain insights into
the complexity of the problem that needs to be addressed to
assist developers in their exploration of API reviews.

Finally, opinions are by themselves \it{subjective}, i.e., opinions about APIs
stem from the personal belief or experience of the developers who use the APIs.
Therefore, developers may face challenges while assessing the validity of claims by other developers. By analyzing what factors can hinder and support the
developers' assessment of the quality of the provided API reviews,
we can gain insights into the challenges developers face while leveraging the API reviews.

\subsubsection{Approach}
We examine developer needs for API reviews via the following questions:
\begin{itemize}
  \item RQ1.1: How do developers seek opinions about APIs?
  \item RQ1.2: How do developers assess the quality of the provided
  opinion?
\end{itemize}

To exhaustively answer these questions, we further divide
  RQ1.1 into three sub-questions:
\begin{itemize}
\item {\it{RQ1.1.a: Where do developers seek opinions about APIs?}}
\item {\it{RQ1.1.b: What motivates developers to seek opinions?}}
\item {\it{RQ1.1.c: What challenges do developers face while seeking for
opinions?}}
\end{itemize}

\subsection{Tool Support to Analyze API Reviews (RQ2)} \label{subsec:rq2}
The second goal of our study is to understand
 the needs for tool support to facilitate automated analysis of API reviews.

\subsubsection{Motivation}
To understand whether developer needs any tools to analyze API reviews, we first need to understand what tools developers may be using currently to
analyze the reviews and what problems they may be facing. Such analysis can offer
  insights into how research in this direction can offer benefits
  to the developers through future prototypes and tool supports.
A predominant direction in the automated processing of reviews
  in other domains (e.g., cars, cameras, products) is to summarize the reviews.
  For the domain of API reviews, it can also help to know how developers
  determine the needs for opinion summarization about APIs. The first step is to determine the feasibility of the
  existing cross-domain opinion summarization techniques to the domain of API
  reviews. Such analysis can provide insights into how the
  summarization approaches adopted in other domains can be applicable to the domain of API reviews.
  By learning about the specific development needs that can be
  better supported though the summarization of API reviews, we can gain
  insights into the potential use cases API review summaries can support. By understanding how developers expect to see summaries of
  API reviews, we can gain insights into whether and how different
  summarization techniques can be designed and developed for the domain of API reviews.
   It is, thus, necessary to know what specific problems in the API reviews should be
  summarized and whether priority should be given to one API aspect over another.

\subsubsection{Approach}
We pose two research questions to understand the needs for tool support to
analyze API reviews:
\begin{itemize}
  \item RQ2.1: What tool support is needed to help developers with analyzing
  and assessing API-related opinions?
  \item RQ2.2: What are the developers' needs for summarization of
  opinion-rich information?
\end{itemize}
To answer these two questions exhaustively, we further divide RQ2.2
  into three sub-questions:
\begin{itemize}
  \item {\it{RQ2.2.a: What problems in API reviews motivate the needs for
  summarization? }}
  \item {\it{RQ2.2.b: How can summarization of API reviews  support the developer's
  decision making processes? }}
  \item {\it{RQ2.2.c: How do developers expect API reviews to be
  summarized? }}
\end{itemize}

\subsection{The Surveys}\label{subsec:motivation-surveys}
We learn about the developer needs for API reviews and tool support for analyzing the
reviews through two surveys. We conducted the first survey as a pilot survey and the second as the primary one.
The purpose of the pilot survey is to identify and correct potential
ambiguities in the design of the primary survey.
Both the pilot and the primary surveys share the same goals. However, the
questions of the primary survey are refined and made more focused based on the
findings of the pilot survey. For example, in the pilot survey, we mainly
focused on the GitHub developers. \rev{We picked GitHub for our pilot survey, because previous research shows that GitHub developers use third-party APIs 
(e.g., open source APIs) and would like to stay aware of changes in APIs in their development tasks to become remain productive~\cite{Treude-SummarizingDevelopmentActivity-FSE2015}. 
We focused on open source community because open source development is embraced by both individual developers as well as big and small 
companies (both as contributors to API development as well API usage). In addition, due to the openness, 
the community may also be more reliant on developer forums to inform choices. Such interactions 
can be visible to any other developers for further analysis (e.g., during their decision making about an API).  
This offers us access to diverse viewpoints (i.e., opinions) of developers on APIs that may be used in diverse development needs and contexts.} 

We found that most of the respondents in our
pilot survey considered the developer forums (e.g., Stack Overflow) as the primary source of opinions
about APIs. Therefore, in the primary survey, our focus was to understand how
developers seek and analyze opinions in developer forums. Stack Overflow is arguably the most popular online forums to
share and discuss code and opinions about open-source APIs. Therefore, in our primary survey, we picked developers who are
actively involved in the discussions of Stack Overflow posts.

In \sec\ref{sec:pilot-survey}, we discuss the design and the summary of results
of the pilot survey. In \secs\ref{sec:methodology} and \ref{sec:results}, we
discuss the design and detailed results of the primary survey.

%% file: pilot-survey.tex
\section{The Pilot Survey}\label{sec:pilot-survey}
The pilot survey consisted of 24 questions:
three demographic questions, eight multiple-choice,
five Likert-scale questions, and eight open-ended questions. In \tbl\ref{tbl:s1questions}, we show all the questions of the pilot
survey (except the demographic questions) in the order they appeared in the
survey questionnaires. The
demographic questions concern the participants role (e.g., software developer or
engineer, project manager or lead, QA or testing engineer,  other),
whether they are actively involved in software development or not,
and their experience in software development (e.g., less than 1 year, more than
10 years, etc.). The survey was hosted in Google forms and can be viewed at
\url{https://goo.gl/forms/8X5jKDKilkfWZT372}.

\subsection{Pilot Survey Participants} We sent the pilot survey invitations to
2,500 GitHub users. The 2,500 users were randomly sampled from 4,500 users from GitHub.
{The 4,500 users were collected using the GitHub API. The GitHub API returns GitHub users
starting with an ID of 1. We stopped calling the API after it returned the first 4,500 users.}
From the number of emails sent to GitHub users, 70
emails were bounced back for various reasons, e.g., invalid (domain expired) or non-existent email addresses,
making it 2,430 emails being actually delivered. A few users emailed us saying that they were not interested in
participating due to the lack of any incentives. Finally, a total of 55
developers responded.
In addition, we sent the invitation to 11 developers in a software development
company in Ottawa, Canada. The company was selected based on a personal contact we had within the company.
The company was involved in multiple different software
projects involving open-source APIs. Out of the 11, nine responded. Among the
GitHub participants: \begin{enumerate}
\item 78\% of respondents said that they are software
developers (11\% are project managers and 11\% belong to ``other'' category),
\item 92\% are actively involved in software development, and
\item 64\% have more than 10 years of software development experience, 13\% of
them have between 7 and 10 years of experience, 9\% between 3
to 6 years, 8\% between 1 to 2 years and around 6\% less than 1 year of
experience.
\end{enumerate} Among the nine Industrial participants, three were team leads
and six were professional developers. All nine participants were actively involved
in software development. The nine participants had professional development
experience between five to more than 10 years.

\begin{table*}[h]
\vspace{-3mm}
  \centering
  \caption{Pilot survey questions with key highlights from responses. Subscript with a question number shows number of responses.}
  \vspace{-2mm}
   \resizebox{\textwidth}{!}{%
    \begin{tabular}{r|p{17cm}}\toprule
    \textbf{RQ1.1} & \textbf{Opinion Needs} \\ \midrule
    1$_{60}$    & Do you value opinion of other developer when deciding on what API to
    use? \it{(yes / no)} \\
          & \ib{\begin{inparaenum}
          \item Yes 95\%,
          \item No 5\%
          \end{inparaenum} } \\
          \cmidrule{2-2}
    2$_{60}$    & Where do you seek help/opinions about APIs? \it{(five sources and
    others)} \\
          & \ib{\begin{inparaenum}
          \item Developer Forums 83.3\%,
          \item Co-worker 83.3\%,
          \item Mailing List 33.3\%,
          \item IRC 26.7\%,
          \item Others 35\%
          \end{inparaenum}} \\
 \cmidrule{2-2}
    3$_{60}$   & How often do you refer to online forums (e.g, Stack Overflow) to get
    information about APIs? \it{(five options)} \\
          & \ib{\begin{inparaenum}
          \item Every day 23.3\%,
          \item Two/three times a week 36.7\%,
          \item Once a month 25\%,
          \item Once a week 10\%,
          \item Never 5\%
          \end{inparaenum}} \\
 \cmidrule{2-2}
    4$_{60}$    & When do you seek opinions about APIs? \it{(5-point Likert scale for
    each option)} \\
          & \ib{
          \begin{inparaenum}
          \item Select among choices 83.3\%,
          \item Select API version 18.3\%,
          \item Improve a feature 61.7\%,
          \item Fix bug 63.3\%,
          \item Determine a replacement 73.3\%,
          \item Validate a selection 45\%,
          \item Develop a competing API 58.3\%,
          \item Replace an API feature 51.7\%
          \end{inparaenum}
          } \\
\cmidrule{2-2}
    5$_{21}$   & What are other reasons of you referring to the opinions about
    APIs from other developers? \it{(text box)} \\
          &
          \ib{
          \begin{inparaenum}
          \item Opinion trustworthiness analysis 18.8\%
          \item API usage 15.6\%,
          \item API suitability 15.6\%,
          \item API maturity 6.3\%,
          \item Usability analysis 6.3\%
          \end{inparaenum}
          }
           \\
\cmidrule{2-2}
    6$_{28}$    & What is your biggest challenge while seeking opinions about an API?
    \it{(text box)} \\
          &
          \ib{
          \begin{inparaenum}
          \item Trustworthiness analysis 19.4\%
          \item Too much info 16.1\%,
          \item Biased opinion 12.9\%,
          \item Documentation 9.7\%,
          \item Staying aware 6.5\%
          \end{inparaenum}
          }
          \\
\cmidrule{2-2}
    \bf{RQ1.2}    & \bf{Opinion Quality} \\
\midrule
    20$_{60}$   & How do you determine the quality of a provided opinion in a forum
    post? (e.g., Stack Overflow) \it{(seven options)} \\
          & \ib{\begin{inparaenum}
          \item Date 55\%,
          \item Votes 66.7\%,
          \item Supporting links 70\%,
          \item User profile 45\%,
          \item Code presence 73.3\%,
          \item Post length 16.7\%,
          \item Others 1.7\%
          \end{inparaenum}} \\
\cmidrule{2-2}

    21$_{15}$  & What other factors in a forum post can help you determine the
    quality of a provided opinion? \it{(text box)} \\
          &
          \ib{
       \begin{inparaenum}
       \item Documentation 63.3\%,
       \item Expertise 10.5\%,
       \item Trustworthiness 10.5\%,
       \item Situational relevance 10.5\%,
       \item Biased opinion 5.3\%
       \end{inparaenum}
          }
          \\

\cmidrule{2-2}
    \textbf{RQ2.1} & \textbf{Tool Support} \\
\midrule
    7$_{60}$    & What tools can better support your understanding of opinions about
    APIs in online forum discussions? \it{(5-point Likert scale for each
    option)}
    \\
          & \ib{
          \begin{inparaenum}
          \item Opinion mine \& summarize 68.3\%,
          \item Sentiment mine 45\%,
          \item Comparator analyze 73.3\%,
          \item Trends 56.7\%,
          \item Co-mentioned APIs 76.7\%
          \end{inparaenum}
          } \\
\cmidrule{2-2}
    8$_{17}$  & What other tools can help your understanding of opinions about APIs
    in online forum discussions? \it{(text box)}\\
          &
          \ib{
          \begin{inparaenum}
          \item Reasoning 18.2\%,
          \item Extractive summaries 18.2\%,
          \item Dependency info 12.1\%,
          \item Documentation 9.1\%,
          \item Expertise find 9.1\%
          \end{inparaenum}
          }
           \\
\cmidrule{2-2}
    \textbf{RQ2.2} & \textbf{Summarization Needs} \\
\midrule
    9$_{60}$   & How often do you feel overwhelmed due to the abundance of opinions
    about an API? \it{(four options)} \\
          & \ib{\begin{inparaenum}
          \item Every time 5\%,
          \item Some time 48.3\%,
          \item Rarely 33.3\%,
          \item Never 13.3\%
          \end{inparaenum}} \\
\cmidrule{2-2}
    10$_{60}$  & Would summarization of opinions about APIs help you to make a better decision
    on which one to use? \it{(yes/no)} \\
          & \ib{
          \begin{inparaenum}
          \item Yes 83.3\%,
          \item No 16.7\%
          \end{inparaenum}
          } \\
\cmidrule{2-2}
    11$_{59}$ & Opinions about APIs need to be summarized because? \it{(5-point
    Likert scale for each option)} \\
          & \ib{
          \begin{inparaenum}
          \item Too many posts with opinions 60\%,
          \item Interesting opinion in another post 66.7\%,
          \item Contrastive viewpoints missed 55\%,
          \item Not enough time to look for all opinions 56.7\%
          \end{inparaenum}
          } \\
\cmidrule{2-2}
    12$_{60}$   & Opinion summarization can improve the following decision making
    processes. \it{(5-point Likert scale for each option)} \\
          &
          \ib{
          \begin{inparaenum}
          \item Select among choices 85\%,
          \item Select API version 45\%,
          \item Improve a feature 48.3\%,
          \item Fix bug 40\%,
          \item Determine a replacement 78.3\%,
          \item Validate a selection 53.3\%,
          \item Develop a competing API 48.3\%,
          \item Replace an API feature 46.7\%
          \end{inparaenum}
          }
          \\
\cmidrule{2-2}
    13$_{9}$   & What other areas can be positively affected having support for
    opinion summarization? \it{(text box)} \\
          &
          \ib{\begin{inparaenum}
          \item API usage 21.4\%,
          \item Trustworthiness analysis 21.4\%,
          \item Documentation 9.5\%,
          \item Maturity analysis 14.3\%,
          \item Expertise 7.1\%
          \end{inparaenum}}
          \\
\cmidrule{2-2}
    14$_{9}$  & What other areas can be negatively affected having support for
    opinion summarization? \it{(text box)} \\
          &
          \ib{\begin{inparaenum}
          \item Trustworthiness analysis 33.3\%,
          \item Info overload 33.3\%,
          \item API usage 8.3\%,
          \item Reasoning 8.3\%
          \end{inparaenum}}
          \\
\cmidrule{2-2}
    15$_{60}$   & An opinion is important if it contains discussion about one or more
    of the following API aspects? \it{(5-point Likert scale for each option)} \\
          &\ib{\begin{inparaenum}
          \item Performance 86.7\%,
          \item Security 85\%,
          \item Usability 85\%,
          \item Documentation 88.3\%,
          \item Compatibility 75\%,
          \item Community 73.3\%,
          \item Bug 86.7\%,
          \item Legal 43.3\%,
          \item Portability 50\%,
          \item General Features 48.3\%,
          \item Only sentiment 15\%
          \end{inparaenum}} \\
\cmidrule{2-2}
    16$_{16}$   & What are the other factors you look for in API opinions? \it{(text
    box)}
    \\
          &
          \ib{\begin{inparaenum}
          \item API usage 28.6\%,
          \item Usability 19\%,
          \item Expertise 14.3\%,
          \item Standards 9.5\%,
          \item Reasoning 9.5\%
          \end{inparaenum}}
          \\
\cmidrule{2-2}
    17$_{60}$  & What type of opinion summarization would you find most useful?
    \it{(five types)}
    \\
          & \ib{
          \begin{inparaenum}
          \item In a paragraph 83.3\%,
          \item Divided into aspects 78.3\%,
          \item Top N by most recent 36.7\%,
          \item Divided into topics 35\%,
          \item Others 1.7\%
          \end{inparaenum}
          } \\
\cmidrule{2-2}
    18$_{60}$  & How many keywords do you find to be sufficient for topic
    description? \it{(five options)} \\
          & \ib{\begin{inparaenum}
          \item Five words 61.7\%,
          \item 5-10 words 28.3\%,
          \item 10-15 words 5\%,
          \item 15-20 words 3.3\%,
          \item Not helpful at all 1.7\%
          \end{inparaenum}} \\
\cmidrule{2-2}
    19$_{7}$   & What are the other ways opinions about APIs should be summarized?
    \it{(text box)}
    \\
          &
          \ib{
          \begin{inparaenum}
          \item API Usage 25\%,
          \item Expertise 8.3\%,
          \item Documentation 8.3\%,
          \item Popularity 8.3\%,
          \item Testing 8.3\%
          \end{inparaenum}
          }
          \\
\bottomrule
    \end{tabular}%
    }
  \label{tbl:s1questions}
    \vspace{-2mm}

\end{table*}%

\subsection{Pilot Survey Data Analysis}
We analyzed the survey data using statistical and
qualitative approaches.
For the open-ended questions, we applied an open coding
approach~\cite{miles_1994}. {Open coding includes labelling of concepts/categories
in textual contents based on the properties and dimensions of the entities (e.g., an API) about which
the contents are provided. In our open coding, we followed the card sorting approach~\cite{Fincher-MakingSenseofSurveyData-JES2005}.
In card sorting, the textual contents are divided into \it{cards}, where each card denotes a \it{conceptually coherent
quote}. For example,
consider the following sentence in \fig\ref{fig:soSentimentPosts} (from answer circled as \circled{5}): \emt{Note that
Jackson fixes these issues, and is faster than GSON.} The sentence has two different conceptual coherent quotes, ``Note that Jackson fixes
these issues'', and ``and is faster than GSON''. The first quote refers to fixes to issues (i.e., bugs) by the API. The second quote
refers to the ``performance'' aspect of the API. In our analysis, as we analyzed the quotes,
themes and categories emerged and evolved during the open coding process.}

We created all of the ``cards'', splitting the responses for eight
open-ended questions. This resulted into 173 individual quotes; each generally
corresponded to individual cohesive statements. In further analysis, the
first two authors acted as coders to group cards
into themes, merging those into categories.
We analyzed the responses to each open-ended question in three steps:

\begin{enumerate}
\item The two coders independently performed card sorts on the 20\% of the
cards extracted from the survey responses to identify initial card groups. The
coders then met to compare and discuss their identified groups.

\item The two coders performed another independent round, sorting another 20\%
of the quotes into the groups that were agreed-upon in the previous step. We
then calculated and report the coder reliability to ensure the integrity of the
card sort. We selected two popular reliability coefficients for
nominal data: percent agreement and Cohen's
Kappa~\cite{Cohen-Kappa-EducationalPsy1960}.

Coder reliability is a measure of agreement among multiple coders for how they
apply codes to text data. To calculate agreement, we counted the number of
cards for each emerged group for both coders and used ReCal2~\cite{recal} for
calculations. The coders achieved the \it{almost perfect} degree of agreement; on
average two coders agreed on the coding of the content in 96\% of the time (the
average percent agreement varies across the questions and is within the range
of 92--100\%; while the average Cohen's Kappa score is 0.84 ranging between
0.63--1 across the questions).

\item The rest of the card sort (for each open-ended question), i.e., 60\% of
the quotes, was performed by both coders together.
\end{enumerate}

\subsection{Summary of Results from the Pilot Survey}
In this section, we briefly discuss the major findings from the pilot survey,
that influenced the design of the primary survey. A detail report of the findings is in
our online appendix~\cite{website:opinionsurvey-online-appendix}. Out of the 64 respondents who completed the
demographic questions, maximum 60 respondents answered the other questions.

In \tbl\ref{tbl:s1questions}, the type of each question (e.g.,
open or closed-ended) is indicated beside the question (in \it{italic}
format). The \it{key findings} for each question
are provided under the question  in \tbl\ref{tbl:s1questions} (in \ib{bold
italic} format). The key findings are determined as follows:
\begin{description}
\item[Responses to open-ended questions.] The response to an open-ended question
is provided in a text box (e.g., Q5). For each such question, we show
the top five categories (i.e., themes) that emerged from the responses of the
question. The frequency of categories under a question
is used to find the top five categories for
the question.
\item[Likert-scale questions] Each such question can have multiple options for a
user to choose from (e.g., Q7). Each option can be rated across five scales:
Strongly Agree, Agree, Neutral, Disagree, and Strongly Disagree. For
each question, we calculate the percentage of respondents that agreed (=
Total Agreed + Total Strongly Agreed) with the provided option.
\item[Multiple choice questions.] There were two types of multiple choice
questions. \begin{inparaenum}
\item Only Select One: Each such question can have more than one choice
and a participant can pick one of the the options (e.g., Q1).
\item Can Select Multiple. The participant can pick one,
more than one, or all the choices, when the options are of type multiple
select (e.g., Q17).
\end{inparaenum}
For each option, we calculate the percentage of respondents that picked the
option.
\end{description}

\nd\bf{Needs for API Reviews (RQ1.1).} 95\% of the participants reported that
they consider opinions of other developers while making a selection of an API.
The primary sources of such opinions for them are both developer forums and
co-workers (83.3\%). 69.3\% of the participants reported to have consulted
developers forums to gain information about APIs. Most of the
participants (83.3\%) reported that they consider opinions about
APIs while making a selection of an API among multiple choices.
They also seek opinions to determine a replacement of an API (73.3\%), as well
as fixing a bug (63.3\%), etc. Among the other reasons for the
participants to seek opinion were the learning of API usage and
the analysis of validity of a provided API usage (e.g, if it
actually works). The participants reported that making an informed
decision among too much opinion, and
judging the trustworthiness of the provided opinions are the major challenges
they face while seeking and analyzing opinions.

\nd\bf{Needs for API Review Quality Analysis (RQ1.2).} Most of the participants (73.3\%) considered an opinion, accompanied by a code
example, to be of high quality (Q20).
70\% of the participants considered the presence of links to supporting
documents as a good indicator of the quality of the provided opinion, while 55\%
believed that the posting date is an important factor.
Stack Overflow uses upvotes and downvotes of a post as a gauge of
the public opinion on the content. 66.7\% of developers
attributed a higher count of votes on the
post to their assumption of the higher quality of the corresponding opinion in
the post.
The user profile,
another Stack Overflow feature, was found as useful by 45\% of
participants to evaluate the quality of the provided opinion. The length of the
post where the opinion is provided was not considered as a contributing factor
during the assessment of the opinion quality (agreed by only 16.7\% of the
responders). One participant did not agree with any of the choices, while two
participants mentioned two additional factors:
\begin{inparaenum}[\bf(1)]
\item real example, i.e., the provided code example should correspond to a real
world scenario; and
\item reasoning of the provided opinion.
\end{inparaenum}

\nd\bf{Tools for API Review Analysis (RQ2.1).} More than 83\% of the respondents
agreed that opinion summarization is much needed for several reasons. Since
opinions change over time or across different posts, developers wanted to be
able to track changes in API opinions and to follow how APIs evolve. {In our surveys, we
sought to explain each option in the closed questions with a short description or examples. The description for 
each question can be found in the links to the surveys (see \url{https://goo.gl/forms/8X5jKDKilkfWZT372} for the pilot survey). For example, in
Q11 of the pilot survey (\tbl\ref{tbl:s1questions}), the two options ``Opinions can change over time'' and
``Opinions can evolve over time'' are differentiated as follows: Opinions about an API about a feature can change from bad to good, e.g., developers
did not like it before, but like it now (i.e., we need a contrastive viewpoint). In contrast, \it{overall}
opinions about an API can evolve, e.g., it is more positive now due to
the increase in adoption by users (e.g., by becoming more usable), etc.}

\nd\bf{Tools for API Review Summarization (RQ2.2).} The vast majority of responders also thought that an
interesting opinion about an API might be expressed in a
post that they have missed or have not looked at.
The need for opinion summarization was also motivated
by the myriad of opinions expressed via various developer forums.
Developers believe that the lack of time to search for
all possible opinions and the possibility of missing an important
opinion are strong incentives for having opinions organized in a better way.

While developers were interested in tool support for mining and summarizing
opinions about APIs, they also wanted to link such opinions to other related
dimensions, such as, usage, maturity, documentation, and user expertise. 85\% of
the respondents believed that opinion summarization could help them select an
API among multiple choices. More than 78\% believed that opinion summaries can
also help them find a replacement to an API.
Developers agreed that summaries can also help them develop a new
API to address the needs that are currently not supported, improve a software
feature, replace an API feature, validate the choice of an API,
select the right version of an API, and fix a bug (48.3\%, 48.3\%, 46.7\%,
53.3\%, 45\%, and 40\%, respectively).

\subsection{Needs for the Primary Survey}\label{sec:needs-for-primary-survey}
The responses in our pilot survey
showed that opinions about APIs are important in diverse development needs.
However, the survey had the following limitations:
\begin{description}[style=nextline]
\item[Design.] A number of similar questions were asked in pairs. For example,
in Q4, we asked the developers about the reasons to seek opinions.
The developers were provided eight options to choose from. In Q5, we asked the
developers to write about the other reasons that could also motivate them to
seek opinions. Q5 is an open-ended question. Both Q4 and Q5
explore similar themes (i.e., needs for opinion seeking). Therefore, the
responses in Q5 could potentially be biased due to the respondents already being
presented the eight possible options in Q4. A review of the manuscript based on
the pilot survey (available in our online appendix~\cite{website:opinionsurvey-online-appendix})
both by the colleagues
and the reviewers in Transaction of Software Engineering pointed out that a better approach would have been to ask
Q5 before Q4, or only ask Q5. Moreover, the respondent should not be given an
option to modify his response to an open-ended question,
if a similar themed closed-ended question is asked afterwards.
\item[Sampling.] We sampled 2,500 GitHub developers out of the first 4,500
GitHub IDs as returned by the GitHub API. We did not investigate the
relevant information of the developers, e.g., are they still active in software
development? Do they show expertise in a particular programming languages?, and
so on. This lack of background information on the survey population can also
prevent us from making a formal conclusion out of the responses of the survey.
\item[Response Rate.] While previous studies involving GitHub developers also
reported low response rate (e.g., 7.8\% by Treude et
al.~\cite{Treude-SummarizingDevelopmentActivity-FSE2015}), the response rate in
our pilot survey was still considerably lower
(only 2.62\%). There can be many reasons for such low response rate. For
example, unlike Treude et al.~\cite{Treude-SummarizingDevelopmentActivity-FSE2015}, we did not
offer any award/incentives to participate in the survey. However, our lack of
enough knowledge of the GitHub survey population prevents us from making a
definitive connection between the low response rate and the lack of incentives.
\end{description}
We designed the primary survey to address the above limitations. Specifically,
we took the following steps to avoid the above problems  in our primary survey:
\begin{enumerate}
  \item We asked all the open-ended questions before the corresponding
  closed-ended questions. The respondents only saw the closed-ended questions
  after they completed their responses to all the open-ended questions. The
  respondents were not allowed to change their response to any open-ended
  question, once they were asked the closed-ended questions.
  \item We conducted the primary survey with a different group of software
  developers, all collected from Stack Overflow. To pick the survey
  participants, we applied a systematic and exhaustive sampling process
  (discussed in the next section).
  \item We achieved a much higher response rate (15.8\%) in our primary survey.
\end{enumerate}  In the next
section, we discuss in details about the design of the primary
survey. In \sec\ref{sec:discussion}, we
briefly compare the results of the pilot and primary survey on the similar-themed question pairs.

%% file: methodology.tex
\begin{table*}[htbp]
 \caption{Questions asked in the primary survey. The horizontal lines
 between questions denote sections} \begin{tabular}{r|p{14cm}|l|r}\toprule
    {\textbf{No}} & \textbf{Question} & \textbf{Theme} &
    \multicolumn{1}{l}{\textbf{Map (Pilot)}} \\
    \midrule
    1     & Do you visit developer forums to seek info about APIs? \it{(yes/no)}
    & RQ1.1.a    &  \\
    \midrule
    2     & List top developer forums you visited in the last two years \it{(text
    box)} &
    RQ1.1.a    &  \\
    3     & Do you value the opinion of other developers in the developer forums when deciding on what API to use? \it{(yes/no)} &
    RQ1.2 & 1 \\
	\midrule
    4     & What are your reasons for referring to opinions of other developers
    about APIs in online developer forums? \it{(text
    box)} & RQ1.1.b    & 5 \\
    5     & How do you seek information about APIs in a developer forum? How do
    you navigate the multiple posts?  & RQ1.1.c    &  \\
    6     & What are your biggest challenges when seeking for opinions about an
    API in an online developer forum? \it{(text box)} &
    RQ1.1.c & 6 \\
    7     & What factors in a forum post can help you determine the quality of a
    provided opinion about an API? \it{(text box)} & RQ1.2    & 21 \\
    8     & Do you rely on tools to help you understand opinions about APIs in
    online forum discussions? \it{(yes/no)} & RQ2.1    &  \\
    9     & If you don't use a tool currently to explore the diverse opinions about APIs in developer
    forums, do you believe there is a need of such a tool to help you find the right
    viewpoints about an API quickly? \it{(yes/no)} & RQ2.1    &  \\
    10    & You said yes to the previous question on using a tool to navigate forum posts.
    Please provide the name of the tool. \it{(text box)} & RQ2.1    & 8 \\
	\midrule
    11    & What are the important factors of an API that play a role in
    your decision to choose an API? \it{(text box)} & RQ2.2.c    & 16 \\
	\midrule
    12    & Considering that opinions and diverse viewpoints about an API
    can be scattered in different posts and threads of a developer forum,
    what are the different ways opinions about APIs can be summarized from
    developer forums? \it{(text box)} & RQ2.2.c    & 19 \\
    13    & What areas can be positively affected by the
    summarization of reviews about APIs from developer forums? \it{(text box)} &
    RQ2.2.b
    & 13 \\
    14    & What areas can be negatively affected by the
    summarization of reviews about APIs from developer forums? \it{(text box)} &
    RQ2.2.b
    & 14 \\
	\midrule
    15    & An opinion is important if it contains discussion about
    the following API aspects? \it{(5-point Likert scale for each option)} &
    RQ2.2.c & 15 \\
    16    & Where do you seek help/opinions about APIs? \it{(5 opinions + None +
    Text box to write other sources)} & RQ1.1.a    & 2 \\
    17    & How often do you refer to online forums (e.g., Stack Overflow) to
    get information about APIs? \it{(five options)} & RQ1.1.a    & 3 \\
    18    & When do you seek opinions about APIs? \it{(5-point Likert scale for
    each option)} & RQ1.1.b    & 4 \\
    19    & What tools can better support your understanding of API reviews in
    developer forums? \it{(5-point Likert scale for each option)} &
    RQ2.1 & 7 \\
    20    & How often do you feel overwhelmed due to the abundance of opinions
    about an API? \it{(four options)} & RQ2.2.a    & 9 \\
    21    & Would a summarization of opinions about APIs help you to make a better decision on which
    one to use? \it{(yes/no)} & RQ2.2.a    & 10 \\
    22    & Opinions about an API need to be summarized because \it{(5-point
    Likert scale for each option)} & RQ2.2.a    & 11 \\
 	\midrule
    23    & Opinion summarization can improve the following decision making processes
    \it{(5-point Likert scale for each option)} & RQ2.2.b    & 12 \\
    \midrule
    24    & Please explain why you don't value the opinion of other developers in
    the developer forums. \it{(text box)} & RQ1.2
    &
    \\
    \bottomrule
    \end{tabular}%
  \label{tab:s2questions}%
\end{table*}%
\section{Primary Survey Design} \label{sec:methodology}
We conducted the primary survey with a different group of  software
developers. Besides the three demographic questions, the primary survey contained 24
questions. In \tbl\ref{tab:s2questions}, we show the questions in the order they
were asked. We show how the questions from the pilot survey were asked in the
primary survey in the last column of \tbl\ref{tab:s2questions}. The survey was
conducted using Google forms and is available for view at:
\url{https://goo.gl/forms/2nPVUgBoqCcAabwj1}.

%

{As we noted in \sec\ref{subsec:motivation-surveys}, in our primary survey, we focused on understanding how and whether
developers seek and analyze API reviews in developer forums. This decision was based on
the observations from our pilot survey. The participants in our pilot survey reported
the developer forums as their primary sources for seeking opinions about APIs (along with co-workers).
One of our major goals from the surveys is to elicit requirements for tool designs to facilitate API analysis using API reviews.
Developer forums, such as Stack Overflow, can be a sharing place for co-workers as well. Moreover,
the design and deployment of such tools can be better facilitated if the data is already available and shared in the forum posts.}


In \tbl\ref{tab:s2questions}, the horizontal
lines between questions denote sections. For example, there is only one question
in the first section (Q1). Once a developers responds to all the questions in a
section, he is navigated to the next section. Depending on the type of the
question, the next section is determined. For example, if the answer to first
question (``Do you visit developer forums to seek info about APIs?'') is a
`No', we did not ask him any further questions.
If the answer is a `Yes', the respondent is navigated to the second question. The
navigation between the sections was designed to ensure two things:
\begin{inparaenum}
\item that we do not
ask a respondent irrelevant questions. For example, if a developer does not value the
opinions of other developers, it is probably of no use asking him about his
motivation for seeking opinions about APIs anyway, and
\item that the response to a question is not biased by another relevant
question. For example, the first question in the third section of
\tbl\ref{tab:s2questions} is Q4 (``What are your reasons for referring to
opinions of other developers about APIs in online developer forums?''). This was
an open-ended question, where the developers were asked to write their responses
in a text box. A relevant question was Q18 in the sixth section (``When do you
seek opinions about APIs?''). The developers were given eight options in a
Likert scale (e.g., selection of an API among choices). The developers were
able to answer Q18 only after they answered Q4. The developers were not allowed
to return to Q4 from Q18. We adopted similar strategy for all such
question pairs in the primary survey. In this way, we avoided the problem of
potential bias in the developers' responses in our primary survey.
\end{inparaenum}

The second
last column in \tbl\ref{tab:s2questions} shows how the questions are
mapped to the two research questions (and the sub-questions)
that we intend to answer. The pilot and the primary surveys contained
similar questions. The last column of \tbl\ref{tab:s2questions} shows how 18 of
the 24 questions in the primary survey were similar to 18 questions in the
Pilot survey. While the two sets of questions are similar, in the primary survey
the questions focused specifically on developer forums. For example, Q4 in the
primary survey (\tbl\ref{tab:s2questions}) was ``What are your reasons for referring to
opinions of other developers about APIs in online developer forums?'' The
similar question in the pilot survey was Q5 (\tbl\ref{tbl:s1questions}): ``What
are other reasons for you to refer to the opinions about APIs from other
developers?''

\begin{figure}
\centering
\includegraphics[scale=0.90]{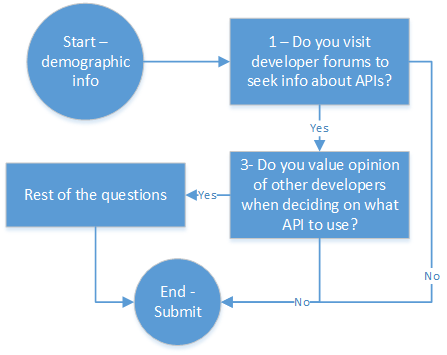}
\caption{Navigation between sections in the primary survey.}
\label{fig:survey2navigation}
\end{figure}

To ensure that we capture developers' experience about API
reviews properly in the primary survey, we also asked six questions that were
not part of the pilot survey. The first two questions (discussed below) in the
primary survey were asked to ensure that we get responses from developers who
indeed seek and value opinions about APIs.
The first question was ``Do you visit developer forums to seek info about
APIs?''. If a developer responded with a `No' to this question,
we did not ask him any further questions. We did this because
\begin{inparaenum}
\item developer forums
are considered as the primary resource in our pilot survey and
\item automated review analysis
techniques can be developed to leverage the developer forums.
\end{inparaenum} The second new question was about asking the participants
about the top developer forums they recently visited. Such information can be
useful to know which developer forums can be leveraged for such analysis. The third new
question was ``Do you value the opinion of other developers in the forum posts
while deciding on what API to use?''. If the response was a `no', we asked the
participant only one question (24): ``Please explain why you don't value the
opinion of developers in the forum posts''. We asked this question to understand
the potential problems in the opinions that may be preventing them from
leveraging those opinions. In
\fig\ref{fig:survey2navigation}, we show how the above
two questions are used to either navigate into the rest of the survey questions
or to complete the survey without asking the respondents further questions.

%

%

%
\begin{figure}
\centering
\includegraphics[scale=0.65]{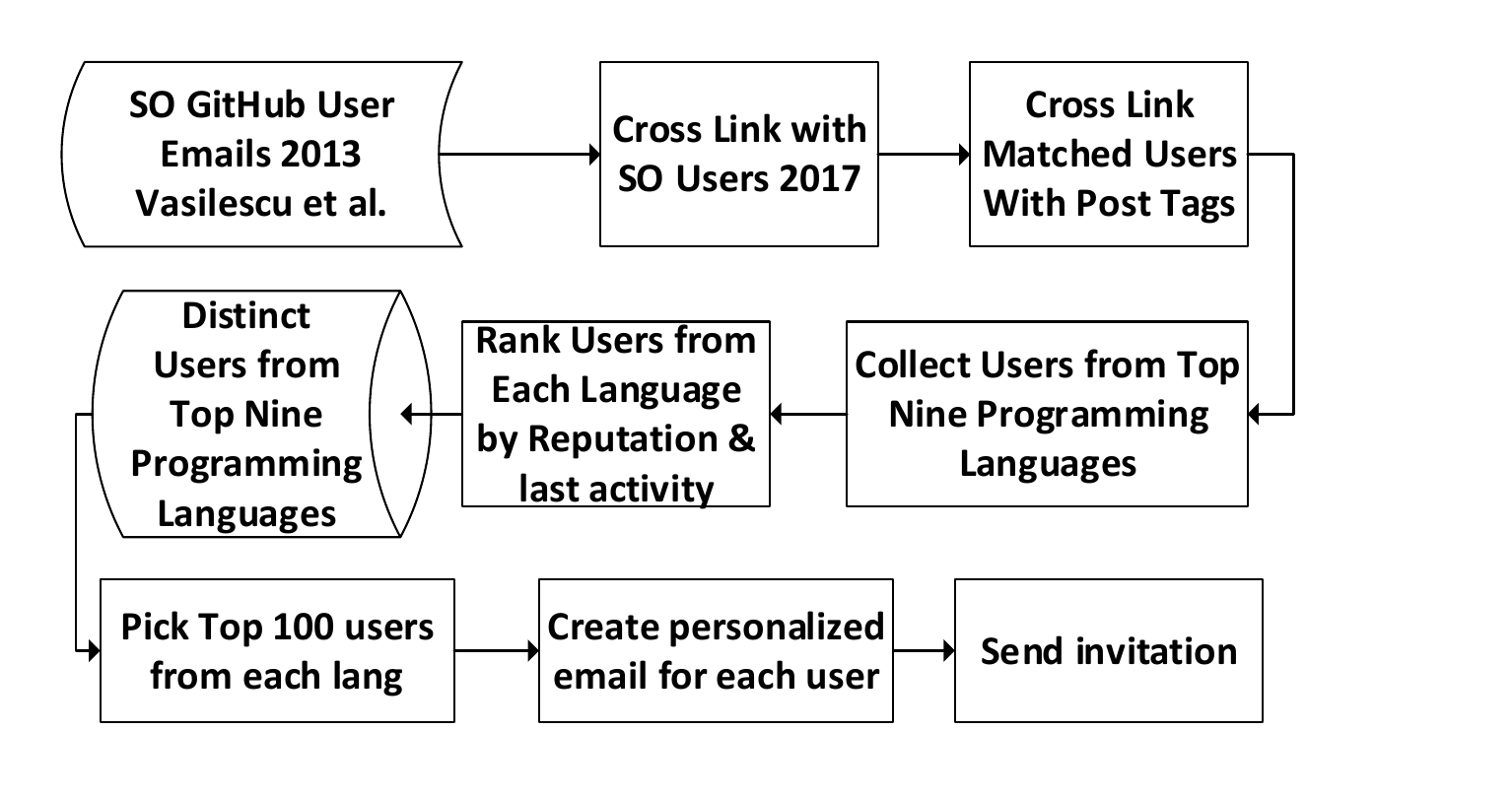}
\caption{Sampling of participants for the primary survey.}
\label{fig:sampling-survey2}
\end{figure}

\begin{table*}[t]
  \centering
  \caption{Summary statistics of the primary survey population and sampled users
  from the population.}
  \resizebox{\textwidth}{!}{%
    \begin{tabular}{lr|r|rrr|rrr|rrrrr}\toprule
          & {\textbf{Total}} & {\textbf{Threads}} & \multicolumn{3}{c}{\textbf{Created}} & \multicolumn{3}{c}{\textbf{Last Active}} & \multicolumn{5}{c}{\textbf{Reputation}} \\
          \cmidrule{4-14}
          & {\textbf{Users}} & {\textbf{Contributed}} & \textbf{2012}
          & \textbf{2011} & \textbf{$<=$ 2010} & \textbf{2017} & \textbf{2016} &
          \textbf{$<=$ 2015} & \textbf{Min} & \multicolumn{1}{r}{\textbf{Max}} &
          \multicolumn{1}{r}{\textbf{Median}} & \multicolumn{1}{r}{\textbf{Avg}} & \multicolumn{1}{r}{\textbf{Std}} \\
          \midrule
    \textbf{Population} & 88,021 & 3,233,131 & 16,080 & 26,257 & 45,684 & 40,780
    & 10,409 & 36,832 & 1     & 627,850 & 159 & 1,866.5 & 10,169.1 \\
    \textbf{Sample} & 900   & 772,760 & 47    & 151   & 702   & 900   & 0     &
    0     & 5,471  & 627,850 & 18,433.5 & 43,311.22 & 66,738 \\
    \bottomrule
    \end{tabular}%
    }
  \label{tbl:s2summarystats}%
\end{table*}%

\subsection{Participants}
We targeted developers that participated in the Stack Overflow forum
posts (e.g., asked a question or provided an answer to a question in Stack
Overflow). Out of a total of 720 invitations sent, we received 114 responses
(response rate 15.8\%).
Among those, 72.8\% responded that they visit developers forums and they
value the opinion of other developers in the forums.
The distribution of the profession of those participants is:
\begin{inparaenum}
\item Software developers 79.5\%,
\item Tech/Team Lead 13.3\%,
\item Research Engineer 3.6\%,
\item Student 3.6\%, and
\item Other 1.2\%
\end{inparaenum}. The distribution of experience of the participants is:
\begin{inparaenum}
\item 10+ years 56.6\%,
\item Seven to 10 years 28.9\%, and
\item Three to six years 14.5\%.
\end{inparaenum} To report the experience, the developers were given five
options to choose from (following Treude et
al.~\cite{Treude-SummarizingDevelopmentActivity-FSE2015}): \begin{inparaenum}
\item less than 1 year,
\item 1 to 2 years,
\item 3 to 6 years,
\item 7 to 10 years, and
\item 10+ years.
\end{inparaenum} None of the respondents reported to have software
development experience of less than three years. Therefore, we received
responses from experienced developers in our primary survey. 97.6\% of them were
actively involved in software development.

\subsubsection{Sampling Strategy}
To recruit the
participants for an empirical study in software engineering, Kitchenham et
al.~\cite{Kitchenham-GuidelinesEmpiricalResearchSE-TSE2002} offered two
recommendations:
\begin{description}
  \item[Population] \emt{Identify the population from which the subjects and
  objects are drawn}, and
  \item[Process]\emt{Define the process by which the subjects and objects were
  selected}.
\end{description} We followed both of the two suggestions to recruit the
participants for our primary survey (discussed below).

 The contact
information of users in Stack Overflow is kept hidden from the public to ensure that the users are not spammed.
Vasilescu et
al.~\cite{Vasilescu-AssociationSOGithub-SocialCom2013} correlated the user email
hash from Stack Overflow to those in GitHub. To ensure that the mining of such
personal information is not criticized by the Stack Overflow community in
general, Bogdan Vasilescu started a question in Stack Overflow with the theme, which
attracted a considerable number of developers from the Stack Overflow community
in 2012~\cite{website:stackoverflow-q135104}. The purpose of Vasilescu et
al.~\cite{Vasilescu-AssociationSOGithub-SocialCom2013} was to see how
many of the Stack Overflow users are also active in GitHub.
In August of 2012, they found 1,295,623 Stack Overflow users.
They cross-matched 93,772 of those users in GitHub. For each of those matched users
in GitHub, they confirmed their email addresses by mapping their
email hash in Stack Overflow to their email addresses as they shared in GitHub.
Each user record in their dataset contains three fields: \begin{inparaenum}[(1)]
\item UnifiedId: a unique identifier for each record (an auto-incremental
integer),
\item GitHubEmail: the email address of the user as posted in GitHub,
\item SOUserId: the ID of the user in Stack Overflow.
\end{inparaenum} We used this list of 93,772 users as the
\it{potential population source} of the primary survey. In our
survey invitation, we were careful not to spam the developers. For example, we
only sent emails to them twice (the second time as a reminder). In addition,
we followed the Canadian Anti-Spam rules~\cite{website:canad-antispam}
while sending the emails, by offering each email recipient the option to `opt-out'
from our invitation. We did not send the
second, i.e., reminder email to the users who decided to opt-out.

\nd We sampled 900
users from the 93,772 users as follows (\fig\ref{fig:sampling-survey2}):

\begin{enumerate}[leftmargin=10pt]
\item \bf{Match:} Out of the 93,772 users in the list of Vasilescu et
al.~\cite{Vasilescu-AssociationSOGithub-SocialCom2013}, we found 92,276 of them
in the Stack Overflow dataset of March 2017. We cross-linked the user `ID' in
the Stack Overflow dataset to the `SOUserId' field in the dataset of Vasilescu et
al.~\cite{Vasilescu-AssociationSOGithub-SocialCom2013} to find the users.
For each user, we collected the following information: \begin{inparaenum}[(1)]
\item Name,
\item Reputation\footnote{The reputation of a user in Stack Overflow is
based on the votes from other developers},
\item User creation date, and
\item Last accessed date in Stack Overflow.
\end{inparaenum}

\item \bf{Discard:} Out of the 92,276 users, we discarded 4,255 users
whose email addresses were also found in the list of 4,500 GitHub users that we used to
sample the 2,500 GitHub users for the pilot survey. Thus, the size
of the target population in our primary survey was 88,021.

\item \bf{Tag:} For each user out of the population (88,021), we then
searched for the posts in the Stack Overflow data dump of 2017 where he has
contributed by either asking the question or answering the question.
For each user, we created a user tag frequency table as follows:
\begin{inparaenum}
\item For each post contributed by the user, we collected the id of the
thread of the post.
\item For each thread, we collected the list of tags assigned to it. We put all
those tags in the tag frequency table of the user.
\item We computed the occurrence of each tag in the tag frequency table of the
user
\item We ranked the tags based on frequency, i.e., the tag with the highest
occurrence in the table was put at the top.
\end{inparaenum}

\item \bf{Programming Languages:} We assigned each user to one of
the following nine programming languages:\begin{inparaenum}[(1)]
\item Javascript,
\item Java,
\item C\#,
\item Python,
\item C++, 
\item Ruby, 
\item Objective-C, 
\item C, and 
\item R. 
\end{inparaenum} {The nine programming languages are among the top 10 programming
languages in Stack Overflow, based on the number of questions tagged as languages. In our survey population, we observed that more than 95\% of the users
tagged at least one of the languages in the Stack Overflow posts.} We assigned a user to a language, if the
language had the highest occurrence among the nine languages in the user
frequency table of the user. If a user did not have any tag resembling any of
the nine languages, we did not include him/her in the sample.

\item \bf{Ranking of Users.} For each language, we ranked the users by
reputation and activity date, i.e., first by reputation and then if two users had the same
reputation, we put the one at the top between the two, who was active more
recently.

\item \bf{Create sample} For each language, we picked the top 100 users. For
each user, we created a personalized email and sent him/her the survey invite.
\end{enumerate}

In \tbl\ref{tbl:s2summarystats}, we show the summary statistics of the
primary survey population and the sampled users. The 88,021 users contributed to
more than 3.2M threads. As of September 2017, Stack Overflow hosts 14.6M threads and 7.8M
users. Thus, the users in the population corresponds to 0.012\% of all the users
in Stack Overflow, but they contributed to 22.1\% of all threads in Stack
Overflow. \rev{All of the 900 users were active in Stack
Overflow as early as 2017, even though most of them first created their account
in Stack Overflow on or before 2010. Moreover, each of the sampled users was highly regarded in the
Stack Overflow community, if we take their reputation in the forum as a metric
for that -- the minimum reputation was 5,471 and the maximum reputation was 627,850, with a
median of 18,434. Therefore, we could expect that the answers from these
users would ensure informed insights about the needs for opinions in reviews about APIs posted in Stack Overflow.}
%

\begin{figure*}[t]
\centering
\includegraphics[scale=0.50]{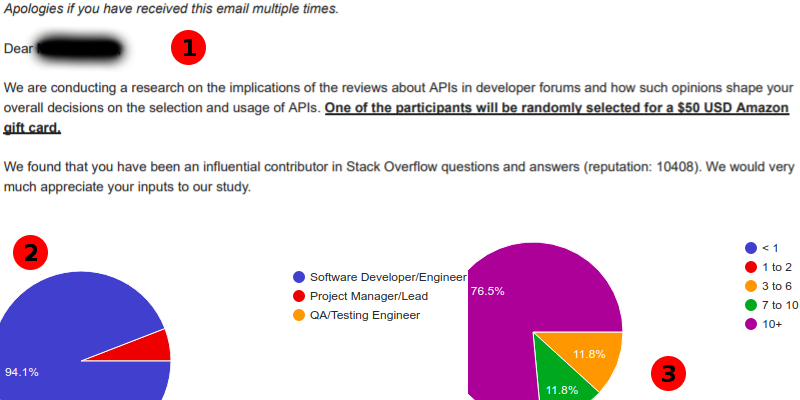}
\caption{Snapshots of the primary survey email and responses}
\label{fig:response-snapshot-survey2}
\end{figure*}

\subsubsection{Participation Engagement}
While targeting the right population is paramount for a survey, convincing the
population to respond to the survey is a non-trivial task. As Kitchenham et
al.~\cite{Kitchenham-GuidelinesEmpiricalResearchSE-TSE2002} and Smith et
al.~\cite{Smith-DeveloperParticipationSurvey-Chase2013} noted, it is necessary to
consider the contextual nature of the daily activities of software developers
while inviting them to the survey. While sending the survey invitations, we
followed the suggestions of Smith et
al.~\cite{Smith-DeveloperParticipationSurvey-Chase2013} who reported their
experience on three surveys conducted at Microsoft. Specifically, we followed
five suggestions, two related to \it{persuasion} (Liking, Authority and
credibility) and three related to \it{social factors} (Social benefit,
compensation value, and timing).

\begin{description}
\item[Liking.] Nisbett and Wilson~\cite{Nisbett-HaloEffect-PSP1977} explained
the cognitive bias \it{halo effect} that people are \it{more
likely to comply with a request from a person they have positive affect
afterwards}~\cite{Smith-DeveloperParticipationSurvey-Chase2013}. Their advice to
leverage the positive affection is to communicate with the people in the study
by their name. For our survey, we created a personalized email for each user by
\begin{inparaenum}[(i)]
\item addressing the developer by his name in the email, and
\item including the reputation of the developer in the email.
\end{inparaenum} In \fig\ref{fig:response-snapshot-survey2}, we show a
screenshot of an email sent to one of the users \circled{1}.

%
\item[Authority and credibility.] Smith et
al.~\cite{Smith-DeveloperParticipationSurvey-Chase2013} pointed out that the
\emph{``compliance rates rise with the authority and credibility of the
persuader''}. To ensure that the developers considered our survey as
authentic and credible, we highlighted the nature of the research in both the
email and the survey front page. We also cited that the survey was governed by
the formal Ethics Approval from McGill University and that
the reporting to the survey would be anonymized.

\item [Social Benefit.] Edwards found that participants are
more likely to respond to survey requests from
universities~\cite{Edwards-IncreasingResponseRatesPostal-BMJ2002}. The reason
is that, participants may be more likely to respond to a survey if they know
that their response would not be used for commercial
benefits, rather it will be used to benefit the society (e.g., through
research).
We contacted the survey participants using the academic emails of all the authors of the paper.
We also mentioned in the email that the survey is conducted as a PhD research of
the first author, and stressed on the fact that the survey was not used for any
commercial purpose.

\item [Compensation value.] Smith et
al.~\cite{Smith-DeveloperParticipationSurvey-Chase2013} observed in the
surveys conducted at Microsoft that people will likely comply with a
survey request if they owe the requester a favor. Such reciprocity can be
induced by providing an incentive, such as a gift card.
Previous research showed that this technique alone can double the
participation rate~\cite{James-SurveyMonetaryIncentives-PublicOpinion1992}. In our
primary survey, we offered a 50 USD Amazon Gift card to one of the randomly
selected participants.

\item [Timing.] As Smith et
al.~\cite{Smith-DeveloperParticipationSurvey-Chase2013} experienced, the time
when a survey invitation is sent to the developers, can directly impact their
likelihood of responding to the email. They advise to avoid sending the survey
invitation emails during the following times: \begin{inparaenum}
\item Monday mornings (when developers just quickly want to skim through their
emails)
\item Most likely out of office days (Mondays, Fridays, December month).
\end{inparaenum} We sent the survey invitations only on the three other weekdays
(Tuesday-Thursday). We conducted the survey in August 2017.
\end{description}

 Out of the 900 emails we sent to the sampled users, around 150 bounced back for various reasons (e.g., old or unreachable email). Around 30 users requested by
email that they would not want to participate in the survey. Therefore, the
final number of emails sent \it{successfully} was 720. We note that the email
list compiled by Vasilescu et
al.~\cite{Vasilescu-AssociationSOGithub-SocialCom2013} is from 2013. Therefore,
it may happen that not all of the 720 email recipients may have been using those
email addresses any more. Previously, Treude et
al.~\cite{Treude-SummarizingDevelopmentActivity-FSE2015} reported a response
rate of 7.8\% for a survey conducted with the Github developers.
Our response rate (15.8\%) is more than the response rate of Treude et
al.~\cite{Treude-SummarizingDevelopmentActivity-FSE2015}.

\subsection{Survey Data Analysis}
We analyzed the primary survey data using the same statistical and
qualitative approaches we applied in the pilot survey. We created a total
of 947 quotes from the responses to the open-ended questions. We created quotes from each response as follows:
\begin{inparaenum}
\item We divided it into sentences.
\item We further divided each sentence  into individual clauses. We
used semi-colon as a separator between two clauses. Each clause was considered a quote.
\end{inparaenum}

Two coders analyzed the quotes. The first author was
the first coder. The second coder was selected
as a senior software engineer working in the Industry in Ottawa, Canada.
The second coder is not an author of this manuscript.

The two coders together coded
eight of the nine open-ended questions (Q4-Q7, Q11-Q14 in
\tbl\ref{tab:s2questions}). The other open-ended question was Q24 (``explain why you don't value the
opinion of other developers \ldots''). Only 10 participants responded to that
question, resulting in 17 quotes. The first coder labelled all of those. In
\tbl\ref{tab:agreementc1c2Fors2}, we show the agreement level between the two
coders for the eight open-ended questions. The last row in
\tbl\ref{tab:agreementc1c2Fors2} shows the number of quotes for each of the
questions. To compute the agreement between the coders, we used the online
recal2 calculator~\cite{recal}. The calculator reports
the agreement using four measures:
\begin{inparaenum}
\item Percent agreement,
\item Cohen $\kappa$~\cite{Cohen-Kappa-EducationalPsy1960},
\item Scott's Pi~\cite{Scott-Pi-PublicOpinion1955}, and
\item Krippendorff's $\alpha$~\cite{Krippendorff-Alpha-HumanCommunications2004}
\end{inparaenum} The Scott's pi is extended to more than two coders in Fleiss'
$\kappa$. Unlike Cohen $\kappa$, in Scott's Pi the coders have the
same distribution of responses. The Krippendorff's $\alpha$ is more sensitive to
bias introduced by a coder,
and is recommended over Cohen $\kappa$~\cite{website:joyce-pickingbeststats-2013}.
The agreement (Cohen $\kappa$) between the coders is above 0.75 for all
questions except Q7. For Q7, the agreement is above the substantial
level~\cite{Viera-KappaInterpretation-FamilyMed2005}.
\begin{table}[t]
  \centering
  \caption{The agreement level between the coders in the open coding.}
    \begin{tabular}{lrrrrrrrr}\toprule
          & \multicolumn{1}{l}{\textbf{Q4}} & \multicolumn{1}{l}{\textbf{Q5}} & \multicolumn{1}{l}{\textbf{Q6}} & \multicolumn{1}{l}{\textbf{Q7}} & \multicolumn{1}{l}{\textbf{Q11}} & \multicolumn{1}{l}{\textbf{Q12}} & \multicolumn{1}{l}{\textbf{Q13}} & \multicolumn{1}{l}{\textbf{Q14}} \\
          \midrule
    Percent & 98.9  & 98.6  & 99.0  & 96.4  & 98.9  & 98.4  & 100   & 100 \\
    Cohen $\kappa$ & 0.90  & 0.76  & 0.80  & 0.66  & 0.80  & 0.75  & 1     & 1
    \\
    Scott $\pi$ & 0.90  & 0.77  & 0.80  & 0.67  & 0.81  & 0.76  & 1     & 1 \\
    Krippen $\alpha$ & 0.90  & 0.77  & 0.80  & 0.67  & 0.81  & 0.76  & 1     & 1
    \\ \midrule
    \bf{Quotes} & 112   & 104   & 112   & 121   & 150   & 122   & 101   & 108 \\

    \bottomrule
    \end{tabular}%
  \label{tab:agreementc1c2Fors2}%
\end{table}%

%% file: results-opinion.tex
\section{Primary Survey Results} \label{sec:results}
\begin{figure}[t]
\centering
\hspace*{-.5cm}%
\includegraphics[scale=0.78]{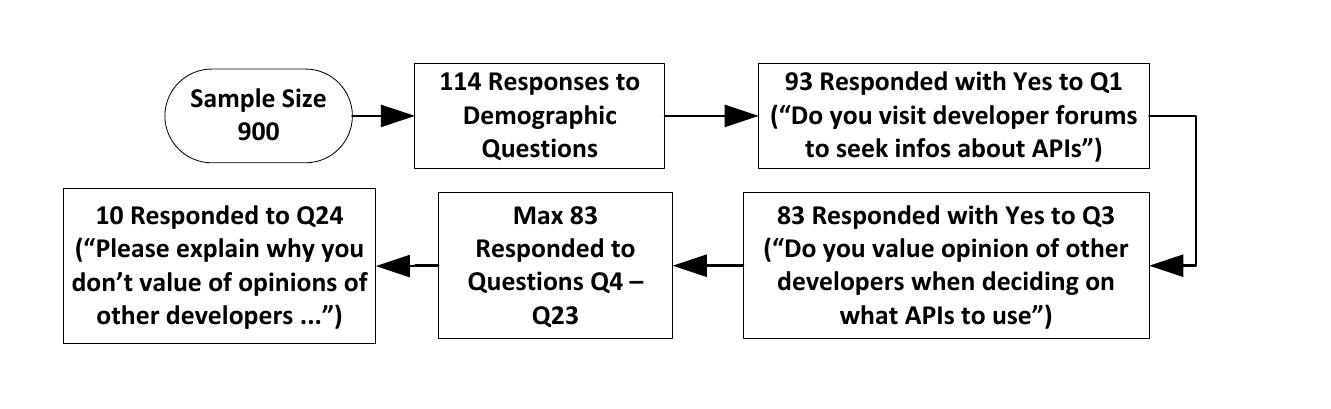}
\caption{The number of responses to the questions in the primary survey}
\label{fig:ResponseNumsS2}
\end{figure}

During the open-coding process of the nine open-ended questions, 37 categories
emerged (excluding `irrelevant' and responses such as `Not sure'). We labelled a
quote as `Not Sure' when the respondent in the quote mentioned that he was
not sure of the specific answer or did not understand the question. For
example, the following quotes were all labelled as `Not sure': \begin{inparaenum}
\item \emt{Don't know}
\item \emt{I am not sure}
\item \emt{This is a hard problem.}
\item \emt{I'm not sure what this question is asking.}
\end{inparaenum} The 37 categories were observed a total of 1,019 times in the
quotes. 46 quotes have more than one category. For example, the following quote
has three categories (Documentation, API usage, and Trustworthiness): \emt{Many
responses are only minimally informed, have made errors in their code samples, or directly contradict first-party documentation.}
In \tbl\ref{tab:s2Categories} (\app\ref{sec:app-open-coding-results}), we show the distribution of the
categories by the questions. In \fig\ref{fig:ResponseNumsS2}, we show the number of participants that answered
the different questions in the primary survey. We discuss the results below.
\subsection{Reasons for Seeking Opinions about APIs (RQ1.1)}
We report the responses of the developers along the three
sub-questions: \begin{inparaenum}
\item Sources for development and opinion needs,
\item Factors motivating developers to seek opinion about APIs from
developer forums, and
\item Challenges developers face while seeking for opinions about APIs from
developer forums.
\end{inparaenum}

\subsubsection{Sources for Opinions about APIs (RQ1.1.a)}
We asked the developers three questions (Q2, 16, 17 in
\tbl\ref{tab:s2questions}). 


\bnd{Q2} \it{We asked them to give a list of the top developer forums
they visited in the last two years}. The respondents reported 40 different developer
forums.
Stack Overflow and its companion sites in Stack Exchange were reported the most
(94 times), at least once by each respondent.
The other major forums listed were (in order of frequency):
\begin{inparaenum}
\item GitHub issue tracker (9),
\item Google developer forum (8),
\item Apple developer forum (6),
\item Twitter/Quora/Reddit/Slack (3)
\end{inparaenum} The other mentions were blog lists, internal mailing lists,
XDA, Android forums, etc.

\bnd{Q16} \it{We further asked them which sources they use to
seek opinions about APIs}. We gave them six options to choose
from:
\begin{inparaenum}
\item Developer forums, e.g., Stack Overflow (Picked by all 83 developers who
mentioned that they valued the opinions of other developers and that they
visit developer forums).
\item Co-workers (63 developers),
\item IRC chats (22 developers),
\item Internal mailing lists (24 developers),
\item None (0 developers),
\item Others. For `Others', we gave them a text box to write the name of the
forums. Among the other sources, developers picked a variety of online
resources, such as, Google search engine, Hacker news, blogs, Slack (for
meetups), GitHub, Twitter, etc.
\end{inparaenum}

The Google and Apple developer forums were present in the list of forums
that the developers visited in the last two years, but they were absent in the
list of forums where developers visit to seek opinions.
Stack Overflow was picked in both questions as the most visited site by the developers both as a general purpose forum to find
information about APIs and as a forum to seek opinions about APIs. {We observed the
presence of Twitter in both lists (Q2 and Q16). Therefore, besides Stack Overflow, Twitter can be another
resource to support developer's learning needs, as was previously observed by Sharma et al.~\cite{Sharma-TwitterSerendipitous-SANER2017}.}

\bnd{Q17} \it{We asked the developers about their frequency of
visiting the online developer forums (e.g., Stack Overflow) to get information about APIs (Q17)}.
There were six options to choose from:
\begin{inparaenum}
\item Every day (picked by 36.1\% of the developers),
\item Two/three times a week (32.5\%),
\item Once a week (14.5\%),
\item Once a month (16.9\%),
\item Once a year (0\%),
\item Never (0\%).
\end{inparaenum} Therefore, most of the developers (83.1\%) reported that they
visit developer forums at least once a week, with the majority (36.1\%) visiting
the forums every day. Each developer reported to visit the developer forums to
seek opinions about APIs at least once a month.

\begin{tcolorbox}[title=Reasons for Seeking Opinions about APIs (RQ1.1) \hrule
\it{Sources for Opinions about APIs (RQ1.1.a)},
opacityback=0, standard jigsaw,]
Stack Overflow was considered as the major source to seek information and
opinions about APIs. Developers also
use diverse informal documentation resources, e.g., blogs, Twitter, GitHub issue tracking system, etc. 
\end{tcolorbox}
\subsubsection{Factors Motivating Opinion Seeking (RQ1.1.b)}
We asked
the developers two questions (Q4,18  in
\tbl\ref{tab:s2questions}). For each category as we
report below (e.g., \co{Expertise}\info{31,31} below), the superscript $(n,m)$
is interpreted as follows:
$n$ for the number of quotes found in the responses of the
questions, and $m$ is the number of total distinct participants provided those
responses. A similar format was
previously used by Treude et
al.~\cite{Treude-SummarizingDevelopmentActivity-FSE2015} (except $m$, i.e.,
number of respondents).

\bnd{Q4} \it{We asked the developers to write about the reasons for them to
refer to opinions about APIs from other developers}. The developers seek
opinions for a variety of reasons:  

\begin{enumerate}[leftmargin=10pt]

\item To gain overall \co{Expertise}\info{31,31} about an API by learning
from experience of others. Developers consider the opinions from other expert
developers as indications of real-word experience and hope that such experience
can lead them to the right direction, \emt{Getting experience of others can save a lot of time if you end up
using a better API or end up skipping a bad one.}.

\item To learn about the \co{Usage}\info{17,16} of an API by analyzing the
opinion of other developers posted in response to the API usage scenarios in
the posts. Developers consider that certain edge cases of an API usage can only
be learned by looking at the opinions of other developers, \emt{It's especially helpful to learn about problems others' faced that might only be evident after consider time is spent experimenting with or 
using the API in production.} They expect to learn
about the potential issues about an API feature from the opinions, before they
start using an API, \emt{Because there is always a trick and even best practice which can only be provided by other developers.} 

\item To be able to \co{Select}\info{13,12} an API among multiple choices for
a given development. The developers think that the quality of APIs can vary
\it{considerably}. Moreover, not all APIs may suit the needs at hand. Therefore,
they expect to see the pros and cons of the APIs before making a selection. \emt{If I don't know an API and I need to pick one, getting the opinion of a few others helps make the decision.}
To make a
selection or to meet the specific development needs at hand, the developers
leverage the knowledge about the API aspects expressed in the opinions about the competing APIs.
\begin{enumerate}[leftmargin=10pt]
\item the \co{Documentation}\info{11,11} support, e.g., when the official
documentation is not sufficient enough, \emt{possibility to get an answer to a specific question (which may not be explained in other sources like the API's documentation)}.
\item the \co{Community}\info{11,9} engagement in the forum posts and mailing
lists, \emt{for instance, Firebase team members are active on Stack Overflow.}
\item the \co{Usability}\info{1,1} and design principles of the API and the \co{Performance}\info{8,8} of the API to assess API quality, 
\emt{It allows me to see common gotchas, so I can estimate the quality of the API based on those opinions, not just my own.}
\end{enumerate} 
\item To improve \co{productivity}\info{9,8} by saving time in the decision
making process, \emt{Getting experience of others can save a lot of time if you end up
using a better API or skipping a bad one.}

\item To \co{trust}\info{13,11} and to \it{validate} the
claims about a specific API usage or feature (e.g., if a posted code example is
good/safe to use), because \emt{They represent hands-on information from fellow devs, free of marketing and business.}
\end{enumerate}

\begin{figure*}
\centering
\includegraphics[scale=.86]{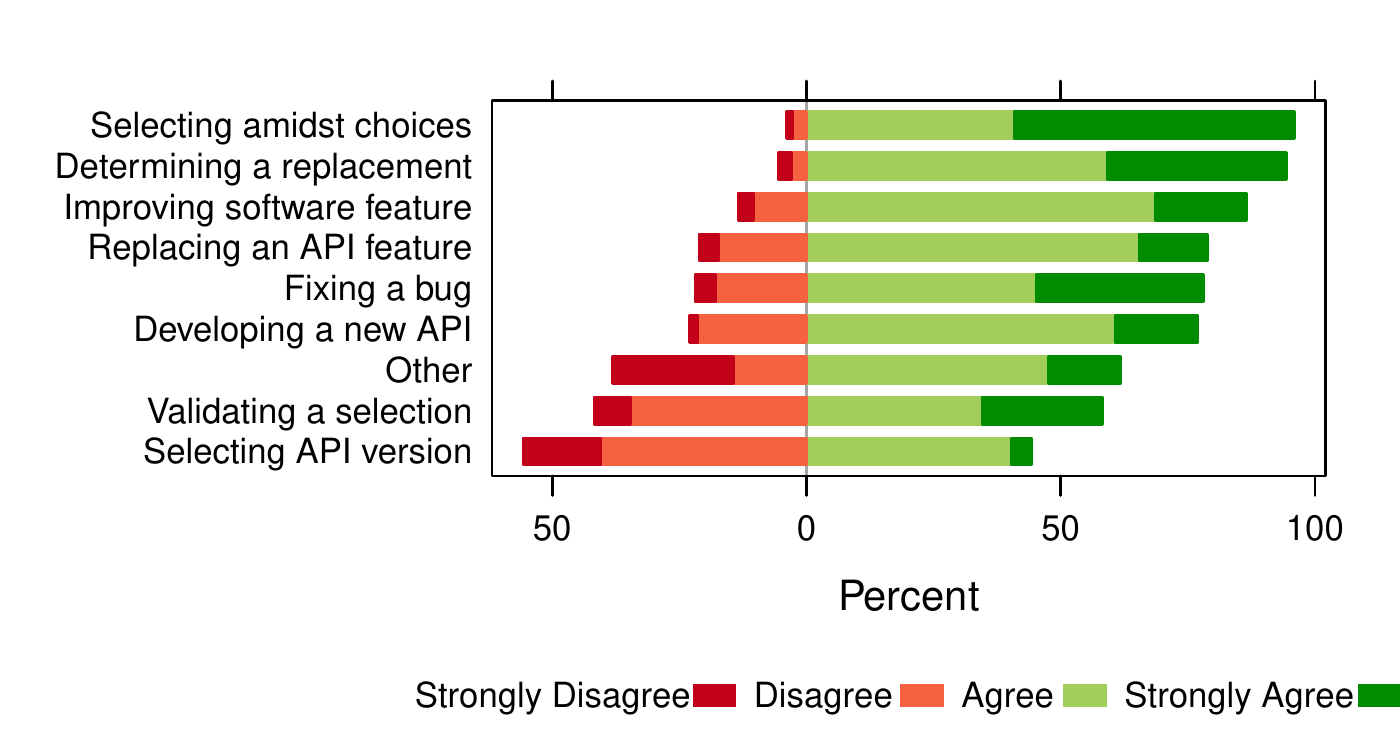}
\caption{Developer motivation for seeking opinions about APIs.}
\label{fig:WhenDoYouSeekOpinionsAboutAPIs}
\end{figure*}
\bnd{Q18}\it{We asked the developers about the specific development needs that
may motivate them to seek opinions about APIs.} We
solicited opinions using a five-point Likert-scale question. The analysis of the
responses (\fig\ref{fig:WhenDoYouSeekOpinionsAboutAPIs})
shows that developers find selection-related factors (i.e., determining
replacement of an API and selection of an API among choices)
to be the most influential for seeking for opinions (88\% and 80.7\%
of all respondents agreed/strongly agreed, respectively). Developers also
seek opinions when they need to improve a software feature or to fix a bug (69.9\% and 68.7\% agreement, respectively). 
56.6\% of the respondents seek help during the development of a new
API, such as, addressing the shortcomings of the existing API. Developers are
less enthusiastic to rely on opinions for validating their choice of an API to
others (38.6\%), or replacing one version of an API with another version
(27.7\%). Only 15.7\% of the respondents mentioned that they seek opinions
for reasons not covered by the above options. Note that while the `Neutral'
responses are not shown in \fig\ref{fig:WhenDoYouSeekOpinionsAboutAPIs}, we
include the neutral counts in our calculation of percent agreements above.
Our charts to show the results of Likert-scale questions (e.g.,
\fig\ref{fig:WhenDoYouSeekOpinionsAboutAPIs}) follow similar formats as adopted in previous
research~\cite{Kononenko-CodeReviewQuality-ICSE2016,Bird-BranchWhatIf-FSE2012}.

\begin{tcolorbox}[title=Reasons for Seeking Opinions about APIs (RQ1.1) \hrule
\it{Factors Motivating Opinion Seeking (RQ1.1.b)},
opacityback=0, standard jigsaw,]
Developers seek opinions about APIs in developer forums to support diverse
development needs, such as building \it{expertise} about API aspects,
learning nuances about specific API usage, making a selection of
an API among choices, etc.
\end{tcolorbox}

\subsubsection{Challenges in Opinion Seeking (RQ1.1.c)}
We asked developers two questions (Q5, 6 in \tbl\ref{tab:s2questions}) to
determine the way developers seek information and opinions about APIs and the
challenges they face while seeking and analyzing the opinions.

\bnd{Q5}\it{We asked the developers to write about how they seek information
about APIs in a developer forums and how they navigate through multiple forums posts
to form an informed insight about an API.} More than 83\% of the respondents
include some form of \co{searching}\info{94,79} using a general purpose search
engine (e.g., Google) or using the provided search capabilities in the forums. 
In the absence of a suitable alternative, the developers learn to trust such resources, \emt{I use Google and trust Stack Overflow's mechanisms for getting the best posts into the top search results}
However, they also cite that they learn to be patient to get the \it{right} results from the search engines, \emt{Google and patience}.  
Because such
search can still return many results, developers employ different
mechanisms to navigate through the results and to find the information
that they can consider as the \it{right} result. Such mechanisms include, the
ranking (e.g., top hits) of the results, the manual \co{Similarity}\info{3,3}
assessment of the search results by focusing on the \it{recency} of the opinion,
the analysis of the presence of \co{Sentiments}\info{4,4} in the post about the
API,  \emt{I usually google keyword and then look for the positive and negative response}.

\bnd{Q6} \it{We asked developers to write about the challenges they face, while
seeking for opinions about APIs from developer forums.} The major challenge
developers reported was to get the \co{situational relevancy}\info{26,22} of the
opinions within their usage needs. Such difficulties can stem from diverse
sources, such as, finding the right question for their problem, or realizing
that the post may have only the fraction of the answers that the developer was
looking for, \emt{It's hard to know what issues people have with APIs so it's difficult to come up with something to search for.}
Another relevant category was \co{Recency}\info{13,13}
of the provided opinion, as developers tend to struggle whether the provided
opinion is still valid within the features offered by an API, \emt{Information may be outdated, applying to years-old API versions}. The
assessment of the \co{Trustworthiness}\info{21,20} of the provided opinion as
well as the opinion provider were considered as challenging due
to various reasons, such as, lack of insight into the expertise
level of the opinion provider, bias of opinion provider, etc. According to 
one developer, \emt{Sometimes it's hard to determine a developer's experience with
certain technologies, and thus, it may be hard to judge the API based on that dev's opinion alone}.
%

The \q{information overload}{3,3} in the forum posts while navigating through
the forum posts was mentioned as a challenge, \emt{They provide many different opinions, and I have
to aggregate all information and make my own decision}. The absence of answers to
questions,  as well as the presence of low quality questions and answers, and
the \q{Lack of Information}{1,1} in the answers were frustrating to the
developers during the analysis of the opinions. The developers mention about
their extensive reliance on the \q{Search}{15,11} features offered by both Stack
Overflow and general purpose search engines to find the right information.
Finding the right keywords was a challenge during the searching. Developers
wished for better search support to easily find duplicate and unanswered
questions. Lack of knowledge on how to search can prove a blocker in such situations, \emt{If I'm inexperienced or lacking information in a particular area,
I may not be using the right terminology to hit the answers I'm looking for.}

Developers expressed their difficulties during their \q{selection}{6,6} of an
API by leveraging the opinions. The lack of opinions and information for newer
APIs is a challenge during the selection of an API, \emt{With newer APIs, there's often very little information.}. Developers can have specific
requirements (e.g., performance) for their \q{Usage}{8,8} of an API, and finding
the opinion corresponding to the requirement can be non-trivial (e.g., is the
feature offered by this API scalable?).
The necessity to analyze opinions based on specific API aspects was highlighted
by the developers, such as, 

\begin{enumerate}[leftmargin=10pt]
  \item link to the \q{Documentation}{4,4} support in the
opinions,\emt{Ensuring that the information is up-to-date and accurate requires
going back-and-forth between forums and software project's official docs to
ensure accuracy (in this case official docs would be lacking, hence using
forums is the first place)}
\item \q{compatibility}{4,4} of an API feature in different versions, \emt{filtering for a particular (current) API version is difficult.}
\item \q{Usability}{2,2} of the API, and the activity, and
engagement of the supporting \q{community}{9,7}, \emt{But looking at several questions in a particular tag will help give a feeling for the library and sometimes its community too.} 
\end{enumerate} The lack of a proper mechanism
to support such analysis within the current search engine or developer forums
make such analysis difficult for the developers.
%
Finally, getting an instant insight into the \q{expertise}{3,3} of the opinion
provider is considered as important by the developers during their analysis of
the opinions. The developers consider that getting such insight can be
challenging, \emt{Need to figure out if the person knows what they are talking about and is in a situation comparable to ours.}

\begin{tcolorbox}[title=Reasons for Seeking Opinions about APIs (RQ1.1) \hrule
\it{Challenges in Opinion Seeking (RQ1.1.c)},
opacityback=0, standard jigsaw,]
Majority of the developers leverage search engines to look for opinions.
They manually analyze the presence of sentiments in the posts to pick the
important information. The developers face challenges in such exploration for a variety of
reasons, such as the difficulty in associating such opinions within the contexts of
their development needs, the lack of enough evidence to validate
the trustworthiness of a given claim in an opinion, etc.
\end{tcolorbox}

\subsection{Needs for API Review Quality Assessment (RQ1.2)}
We asked three questions (Q3,7,24 in
\tbl\ref{tab:s2questions}, two open-ended).

\bnd{Q3} \it{We asked the developers whether they value the opinion about APIs
of the other developers in the developer forums.} 89.2\% of the respondents
mentioned that they value the opinion of other developers in the
online developer forums, such as, Stack Overflow. 10.8\%
reported that they do not value such opinions.

\bnd{Q24} \it{We asked the 10.8\% participants who do not value the opinion of
other developers to provide us the reasons why they do not value such opinions.} The
developers cited their concern about \co{trustworthiness}\info{4,3} as the main
reason, followed by the lack of \co{Situational relevancy}\info{2,2} of
the opinion to suit specific development needs. The concerns related to the
trustworthiness of the opinion can stem from diverse factors, such as, the
\it{credibility} and the \it{experience} of the poster, \emt{They are usually biased and sometimes
blatant advertisements.}. Lack of trust can also stem from the inherent
bias someone may possess towards a specific computing platform (e.g.,
Windows vs Linux developers). 
We note that both of these two categories were mentioned
also by the developers who value opinions of other developers. However, they
mentioned that by looking for diverse opinions about an entity, they can form
a better insight into the trustworthiness of the provided claim and code
example.

We now report
the results from responses of the 89.2\% developers who reported that they
value the opinion of other developers.

\bnd{Q7}\it{The open-ended question asked developers to write about the factors
that determine the quality of the provided opinion.} The following factors are used to assess the quality 
of the opinions:
\begin{enumerate}[leftmargin=10pt,topsep=0pt]
\item The quality of the provided opinions as comparable to API official
 \q{documentation}{54,49}. Forum posts are considered as informal
 documentation. The developers expected the provided opinions to be
 considered as an alternative source of API documentation. Clarity of the provided opinion with proper
writing style and enough details with links to support each claim are considered
as important when opinions can be considered as of good quality.
In particular, the following metrics are cited to assess opinions as a form of API documentation.
 
\begin{enumerate}[leftmargin=10pt,topsep=0pt]
   \item \bf{Completeness} of the provided opinion based on the 
   \emt{Exhaustiveness of the answer and  comparison to others}
   \item \bf{Acceptable writing quality} by using \emt{proper language} and \emt{Spelling, punctuation and grammar.} 
   \item \bf{Accompanying with facts}, such as using \it{screenshots or similar
   suggestions}, \it{examples from real world applications}, or 
   \emt{Cross-referencing other answers, comments on the post, alternate answers to the same question.}
   \item \bf{Brevity of the opinion} by being \emt{brief and to the point}
   \item \bf{Providing context} by discussing \emt{identified usage
   patterns and recognized API intent}
	\item \bf{Accompanying code examples with reaction}, \emt{Code samples, references to ISOs and other standards, clear writing that demonstrates a grasp of the subjects at hand}
\end{enumerate}
\item The \q{Reputation}{35,33} of the opinion provider and upvote to the forum post
where the opinion is found,  \emt{If the poster has a high Stack Overflow score and a couple of users comment positively about it, that weighs pretty heavily on my decision.}
\item The perceived \q{Expertise}{13,12} of the opinion provider within the
context of the provided opinion, \emt{It's easy to read a few lines of technical commentary and identify immediately whether the writer is a precise thinker who has an opinion worth reading.}
\item The \q{Trustworthiness}{8,5} of the opinion provider who demonstrates
   \it{apparent fairness and weighing pros/cons}.
   \item The \q{situational relevance}{12,9} of the provided opinion within a given
development needs, and
\item The \q{recency}{6,6} of the opinion, \emt{posts more than a few years old are likely to be out of date and counterproductive.}
\end{enumerate}

\begin{tcolorbox}[title=Needs for API Review Quality Assessment (RQ1.2),
opacityback=0, standard jigsaw,]
Developer analyze the quality of the opinions about APIs in the forum posts by
considering the opinions as a source of \it{API documentation}. They employ a
number of \it{metrics} to judge the quality of the provided opinions, e.g.,
the clarity and completeness of the provided opinion, the presence of code examples, the presence of detailed links supporting the claims, etc.
\end{tcolorbox}

\subsection{Tool Support for API Review Analysis (RQ2.1)}
We asked four questions (Q8-10, 19 in \tbl\ref{tab:s2questions}, one open-ended) .

\bnd{Q8}\it{We asked developers whether they currently rely on tools to
analyze opinions about APIs from developer forums.} 13.3\% responded with a
`Yes' and the rest (86.7\%) with a `No'.

\bnd{Q9}\it{We further probed the developers who responded with a `No'. We asked
them whether they feel the necessity of a tool to help them to analyze those
opinions.} The majority (62.5\%) of the developers were unsure (`I don't know').
9.7\% responded with a `Yes' and 27.8\% with a `No'.

\bnd{Q10}\it{We further probed the developers who responded with a `Yes'. We
asked them to write the name of the tool they currently use.} The developers
cited the following tools: \begin{inparaenum}
\item Google search,
\item Stack Overflow votes,
\item Stack Overflow mobile app, and
\item GitHub pulse.
\end{inparaenum}

\begin{figure*}
\centering
\includegraphics[scale=.86]{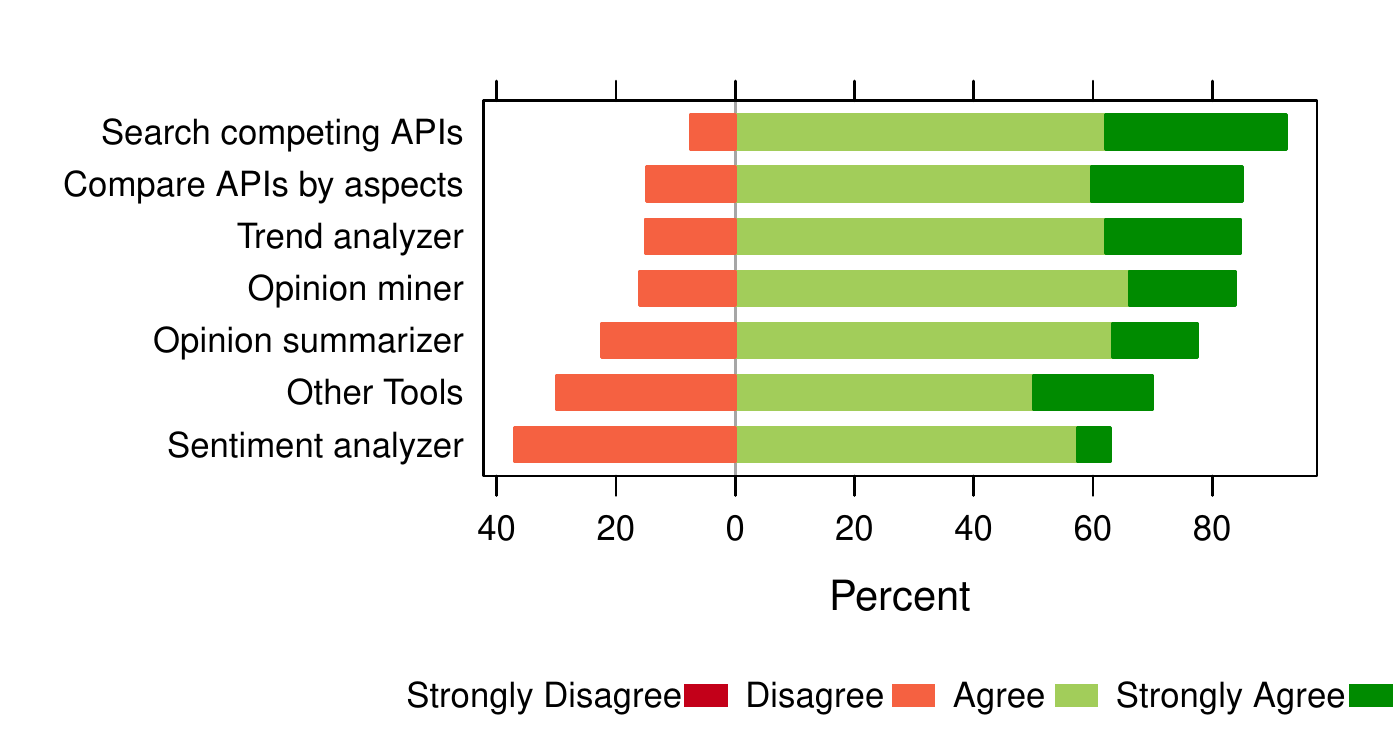}
\caption{ Tool support for analyzing API opinions.}
\label{fig:WhatToolsBetterSupportOpinionUnderstanding}
\end{figure*}
\bnd{Q19} was a
multiple choice question, with each choice referring to one specific tool.
The choice of the tools was inspired by research on sentiment analysis in
other domains~\cite{liu-sentimentanalysis-handbookchapter-2010}. To ensure that the participants understood what we
meant by each tool, we provided a one-line description to each choice. The choices were
as follows (in the order we placed those in the survey):

\begin{enumerate}[leftmargin=10pt,topsep=0pt]
\item \bf{Opinion miner}: for an API name, it provides only the positive and negative opinions collected from the forums.
\item \bf{Sentiment analyzer}: it automatically highlights the positive and
negative opinions about an API in the forum posts.
\item \bf{Opinion summarizer}: for an API name, it provides only a
summarized version of the opinions from the forums.
\item \bf{API comparator}: it compares two APIs based on opinions.
\item \bf{Trend analyzer}: it shows sentiment trends towards an
API.
\item \bf{Competing APIs}: it finds APIs co-mentioned positively or
negatively along an API of interest and compares those.
\end{enumerate}
The respondents' first choice was the `Competing APIs' tool (6),
followed by a `Trend Analyzer' to visualize trends of opinions about an API (5),
an `Opinion Miner' (1) and a `Summarizer' (3)
(see \fig\ref{fig:WhatToolsBetterSupportOpinionUnderstanding}).
The sentiment analyzer (2) is the least desired tool,
i.e., developers wanted tools that do not simply show sentiments, but also show the
context of the provided opinions.
Note that an opinion search and summarization engine in other domain (e.g,
camera reviews) not only show the mined opinions, but also offer insights by
summarizing and revealing trends. The engines facilitate the comparison among the
competing entities through different aspects of the entities. Such aspects
are also used to show the summarized viewpoints about the entities (ref
\fig\ref{fig:opiniontools-camera}). Therefore, it
was encouraging to see that developers largely agreed with the diverse summarization needs of API reviews and usage
information, which corroborate with the similar summarization needs in
other domains.

\begin{tcolorbox}[title=Tool Support for API Review Analysis (RQ2.1),
opacityback=0, standard jigsaw,]
To analyze opinions about APIs in the forums, developers leverage
traditional search features in the absence of a specialized tool. The developers
agree that a variety of opinion analysis tools can be useful in their exploration
of opinions about APIs, such
as, opinion miner and summarizer, a trend analyzer, and an API
comparator engine that can leverage those opinions to facilitate the comparison between
APIs.
\end{tcolorbox}
\subsection{Needs for API Review Summarization (RQ2.2)}
As we observed in the
previous section, a potential opinion summarization engine about APIs need to show both summarized viewpoints about an API
and support the comparison of APIs based on those viewpoints. To properly
understand how and whether such summarization can indeed help developers, we
probed the participants with a number of question.
Specifically, we present the analysis of their responses along
three themes (RQs 2.2a, b, and c in \secs\ref{sec:factors-summarization}-\ref{sec:summarization-types}.

\subsubsection{Factors Motivating Summarization Needs (RQ2.2.a)}\label{sec:factors-summarization}
We asked the developers three questions (Q20-22 in \tbl\ref{tab:s2questions}).

\bnd{Q20}\it{We asked developers how often they feel overwhelmed due to the
abundance of opinions about APIs in the developer forums.} 3.6\% of the
developers mentioned that they are `Always' overwhelmed,
48.2\% are `Sometimes' overwhelmed and 37.3\% are `Rarely'
overwhelmed. Only 10.8\% developers mentioned that they are `Never' overwhelmed
by the abundance of opinions about APIs in the developer forums.

\bnd{Q21}\it{We then asked them whether a summarization of those opinions would
help them make better decisions about APIs?} 38.6\% responded with a `Yes' and
13.3\% responded with a `No'. 48.2\% were unsure (`I don't know'). 

\begin{figure*}
\centering
\includegraphics[scale=.82]{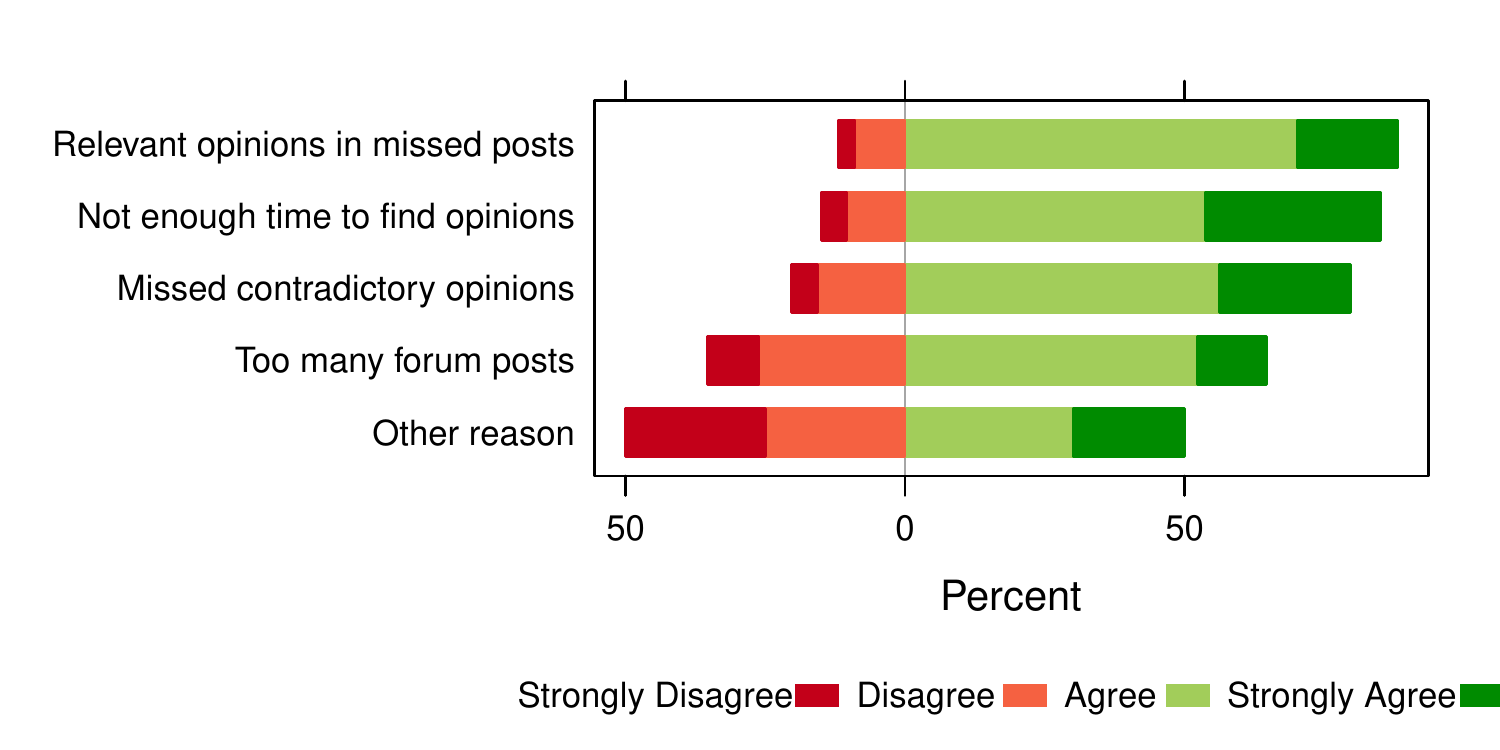}
\caption{Developer needs for opinion summarization about APIs.}
\label{fig:OpinionsNeedTobeSummarizedBecause}
\end{figure*}

\bnd{Q22}\it{We
examined the needs to opinion summarization using a five-point Likert-scale with seven options}\footnote{Before conducting the two surveys, the first author manually analyzed around 1,000 posts from Stack Overflow. The goal was
to understand the type of opinions developers share about APIs in the forum posts. The options were selected based on our observation of
API reviews in forum posts.}: \begin{inparaenum}
\item Too many forum posts with opinions,
\item Opinions can evolve over time in different posts,
\item Opinions can change over time,
\item The interesting opinion may not be in the posts that you have looked in,
\item Contradictory opinions about an API may be missed,
\item Not enough time to look at all opinions, and
\item Other reason.
\end{inparaenum} The same options were
also used in the pilot survey and we observed agreement from the developers
across the options. \rev{We remove the analysis of two options (``Opinions can evolve over time in different posts'' and ``Opinions can change over time''), because both 
may be perceived as similar by the participants without a clarification.} The analysis of the rest of the options
(Figure~\ref{fig:OpinionsNeedTobeSummarizedBecause}) shows that
nearly all developers agree that opinion summarization is much needed for several reasons.
The vast majority of the respondents think that an
interesting opinion about an API might be expressed in a
post that they have missed or have not looked at.
The need for opinion summarization is also motivated
by the presence of too many posts to explore.
Developers also believed that the lack of time to search for
all possible opinions and the possibility of missing an important
opinion are strong incentives for having opinions organized in a better way.

\begin{tcolorbox}[title=Needs for API Review Summarization (RQ2.2) \hrule
\it{Factors Motivating Summarization Needs (RQ2.2.a)},
opacityback=0, standard jigsaw,]
The developers reported that they feel overwhelmed due to the abundance of opinions
about APIs in the developer forums. The developers were in agreement that such
difficulty arises due to a variety of reasons, such as, relevant opinions about an API may be missed because those opinions were in posts not 
checked by the developer, the lack of enough time to find all such opinions, etc.
\end{tcolorbox}
\subsubsection{Summarization Preferences
(RQ2.2.b)}\label{sec:summarization-preferences}
We asked three questions (two open-ended) to understand the relative benefits
and problems developers may encounter during their usage of API review summaries
from developer forums (Q13,14, and 23 in \tbl\ref{tab:s2questions}).

\bnd{Q13}\it{We asked developers to write about the areas that can be
positively impacted by the summarization of opinions about APIs from developer
forums.} The following areas were considered to be benefited from the summaries: 
\begin{enumerate}[leftmargin=10pt]
\item The majority of the developers considered that \q{API selection}{30,25}
is an area that can reap the benefits from the summaries, \emt{It would make the initial review and whittling down of candidates quicker and easier.}. They
also considered that API review summarization can improve the
\q{Productivity}{12,11} of the developers by offering quick but informed insights about APIs, \emt{It would make the initial review and whittling down of candidates quicker and easier.} 
\item The developers expected the summaries to assist in their \q{API usage}{7,7},
such as, by showing reactions from other developers for a given code example
that can offer \q{situationally relevant}{3,3} insights about the code example
(e.g., if the code example does indeed work and solve the need as claimed). The
developers consider that they can use a summary as a form of documentation to
explore the relative strengths and weaknesses of an API for a given development
need. The developers do not expect the official documentation of an API to have
such insights, because \emt{You can see how the code really works, as opposed to how the documentation thinks it works}. Indeed, official documentation can be often
incomplete~\cite{Robillard-FieldStudyAPILearningObstacles-SpringerEmpirical2011a}.
Carroll et al.~\cite{Carroll-MinimalManual-JournalHCI1987a} advocated the needs
for \it{minimal API documentation}, where API usage scenarios should be
shown in a task-based view. In a way, our developers in the survey expect the
API usage scenarios combined with the opinions posted in the forum posts as
effective ingredients to their completion of the task at  hand. Hence, they
expect that the summaries of opinions about API can help them with the
efficient usage of an API during their development tasks, \emt{If summarization is beneficial to understanding API, then any problem
even remotely related to that use stands to achieve a network benefit.}
\begin{figure*}
\centering
\includegraphics[scale=.82]{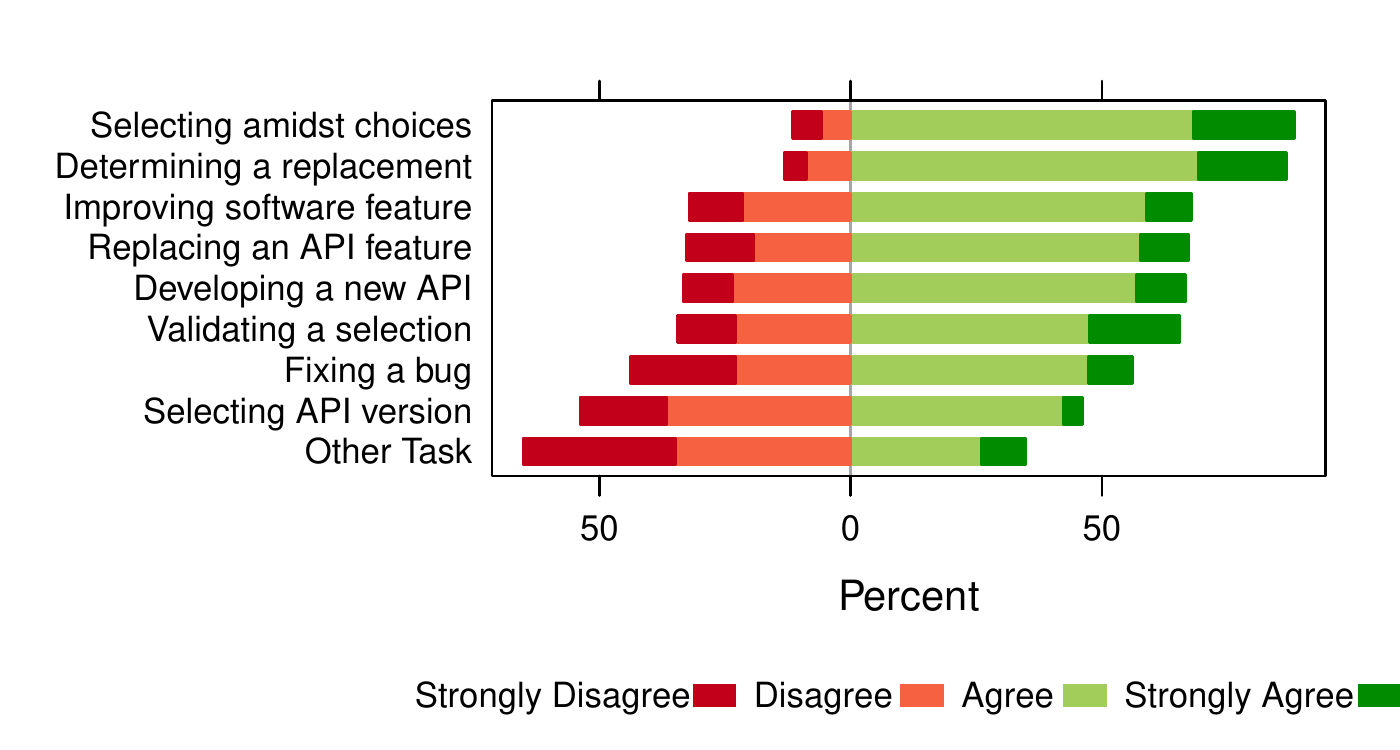}
\caption{ Impact of opinion summaries on decision making about APIs.}
\label{fig:OpinionSummarizationCanImproveDecisionMakingProcess}
\end{figure*}

\item As we noted earlier in this section, developers leverage the opinions about
diverse API aspects to compare APIs during their selection of an API among
choices. In the absence of any such automatic categorization available to
automatically show how an API operates based on a given aspect (e.g.,
performance), the developers had expectations that an opinion summarizer would
help them with such information. For example, the developers considered that
opinion summaries can also improve the analysis about
different API aspects, such as,
\q{Usability}{2,2}, \q{Community}{2,2}, \q{Performance}{3,3}, \q{Documentation}{6,5}, etc.
\item Finally, developers envisioned that opinion summaries can
improve the \q{Search}{6,6} for APIs and help them find better
\q{Expertise}{4,4} to learn and use APIs.
\end{enumerate}
\bnd{Q14}\it{We next asked developers to write about the areas that can be
negatively impacted by the summarization of opinions about APIs from developer
forums.} The purpose was to be aware of any drawback that can arise
due to the use of API opinion summaries. 
\begin{enumerate}[leftmargin=10pt]
  \item  Developers considered that opinion
summaries may \q{Miss Nuances}{18,12} in the API behavior that are subtle,
may not be as frequent, but that could be \emt{key design choices or the meta-information \ldots}. The
\q{reasoning}{7,6} about opinions can also suffer because of that.
\item While developers considered the selection of an API as an area that can be
positively affected by opinion summaries in general, they also raised
the concern that summaries may negatively impact \q{API selection}{6,6} when
more subtle API aspects are not as widely discussed as other aspects, because \emt{Popularity and fashion trends may mask more objective quality criteria
}. Summaries may become \q{API Barrier}{3,3} when new APIs with
less number of opinions are not ranked higher in the summaries, \emt{Just relying on the summarization for deciding for or against an API will probably not be enough and may lead to decisions that are less than optimal.}  
\end{enumerate}

\bnd{Q23}\it{We asked developers about how  opinion
summaries can support their tasks and decision making
process.} There were nine options. We used the same options we used to solicit the needs
for opinions about APIs in \fig\ref{fig:WhenDoYouSeekOpinionsAboutAPIs}.
This was due to the fact that the purpose of an opinion summarizer should be to facilitate the
easier and efficient consumption of opinions. In
\fig\ref{fig:OpinionSummarizationCanImproveDecisionMakingProcess}, we present
the responses of the developers. More than 70\% of the
developers believe that opinion summarization can help
them with two decisions: 1) determining a replacement of an API and
2) selecting the right API among choices. Developers agree that
summaries can also help them improve a software
feature, replace an API feature, and validate the choice of API.
Fixing a bug, while receiving 38.6\% of the
agreement, might not be well supported by summaries since this task requires
certain detailed information about an API that may not be present in
the opinions (e.g., its version, the code itself, the feature). The
developers mostly disagreed with the use of opinions to select an API version
(28.9\% agreement only).

\begin{tcolorbox}[title=Needs for API Review Summarization (RQ2.2) \hrule
\it{Summarization Preferences (RQ2.2.b)},
opacityback=0, standard jigsaw,]
The summarization of opinions about APIs can be
effective to support the selection of an API
among choices, the learning of API usage cases, etc.
Developers expect a gain in productivity in supporting such needs by the opinion
summarizer. However, developers expressed their cautions with a potential
opinion summarizer for APIs, when for example, subtle nuances of an
API can be missed in the summary. Another concern is that summaries can make it difficult for a new API to be adopted. For example,
developers may keep using the existing APIs, because they are possibly reviewed
more than the new APIs.
\end{tcolorbox}

\subsubsection{Summarization Types
(RQ2.2.c)}\label{sec:summarization-types}
We asked developers three questions (Q11,12,15 in \tbl\ref{tab:s2questions}) to
learn about their preferences on how they would like to have opinion summaries
about APIs to be produced.

\bnd{Q11}\it{We asked developers to write about the different factors in an API
that play a role during their decision to choose the API.} The purpose was to
learn what specific API aspects developers consider as important.
Intuitively, developers would like to see opinions about those aspects in the forum posts to
make a decision on the API. The developers mentioned about the following API
aspects as factors contributing to their decisions to choose an API: 
\begin{enumerate}[leftmargin=10pt]
\item\bf{\q{Usability}{87,56}} 
\begin{enumerate}
  \item Simple. \emt{Simplicity, compliance with standards}
  \item Ease of use. \emt{\ldots easy to replace, simple to use}
  \item Design. \emt{clearly designed \& documented usage protocol}
  \item Stable. \emt{\ldots, so it doesn't change every 10 days?}
\end{enumerate}
\item\bf{\q{Documentation}{47,43}} 
\begin{enumerate}
  \item Clear. \emt{Clear documentation \ldots}, 
  \item Consistent. \emt{\ldots, has to be intuitive (consistent naming)}
  \item Complete and correct with examples. \emt{Correctness, completeness, documentation \& examples}
  \item Availability. \emt{clear and complete description, and a good blogging is a big bonus}
\end{enumerate}

\item\bf{\q{Community}{31,22}} 
\begin{enumerate}
  \item Active Community. \emt{Code quality, maintainer activity, ratio of answered/unanswered questions about the API.}
  \item Mature. \emt{Strong Stack Overflow community (to show that the API is relatively mature).}
  \item Reputation. \emt{Reputation of the organization.}, \emt{Other works of its creators.}
\end{enumerate}
\item\bf{\q{Performance}{30,23}} \emt{Has it been used in production grade software before?}
\item\bf{\q{Compatibility}{18,14}} 
\begin{enumerate}
  \item Language. \emt{Implemented in the language I need it.}
  \item Other API. \emt{Uniformity with other API \ldots}
  \item Framework/Project. \emt{compatibility (will it work for project I'm working on)} 
\end{enumerate}
\item\bf{\q{Legal}{15,14}}\emt{\ldots the API is open source and maintained.}
\item\bf{\q{Portability}{5,5}} \emt{Thread safety, sync/async, cross-language and cross-OS support, etc.
}
\item\bf{\q{Security}{2,2}} \emt{Are security issues addressed quickly?}
\item\bf{\q{Bug}{1,1}} \emt{Features, ease of use, documentation, size and culture of community,
bugs, quality of implementation, performance, how well I understand its purpose,
whether it's a do-all framework or a targeted library, \ldots }
\end{enumerate}
The \q{Usage scenarios}{10,9} of an API, and the \q{Features}{4,4} it supports
are also
considered to be important factors to be part of the opinion summaries.

\bnd{Q12}\it{We asked developers to provide their analysis on how opinions
about APIs can be summarized when such opinions are scattered across multiple
forum posts.} The purpose was to know what techniques developers consider as
valuable to aggregate such opinions from diverse posts. While developers mention about using the \q{search}{12,12} features in Stack
overflow, they acknowledged that they never thought of this problem deeply. They acknowledge that search engines or Stack Overflow features 
are really not the summarizer they may need, \emt{Often there is a sidebar with similar questions/discussions, but no easy way to "summarize".}

To address their needs for such a summarizer, the idea of a dedicated API \q{Opinion Portal}{18,15} was discussed. In such a portal all opinions about an API can be aggregated
and presented in one place, \emt{aggregate similar information, collect `A vs B' discussions
} \ldots \emt{like an aggregated portal about an API with all organized or
non-organized pages}. Having a centralized portal for such information can be useful, because \emt{Then I can read through them all and synthesize a  solution.}. 
Developers advocated for machine learning approaches to develop such a portal \emt{It would be very interesting for this to be automated via machine learning.}
\begin{figure*}
\centering
\includegraphics[scale=.8]{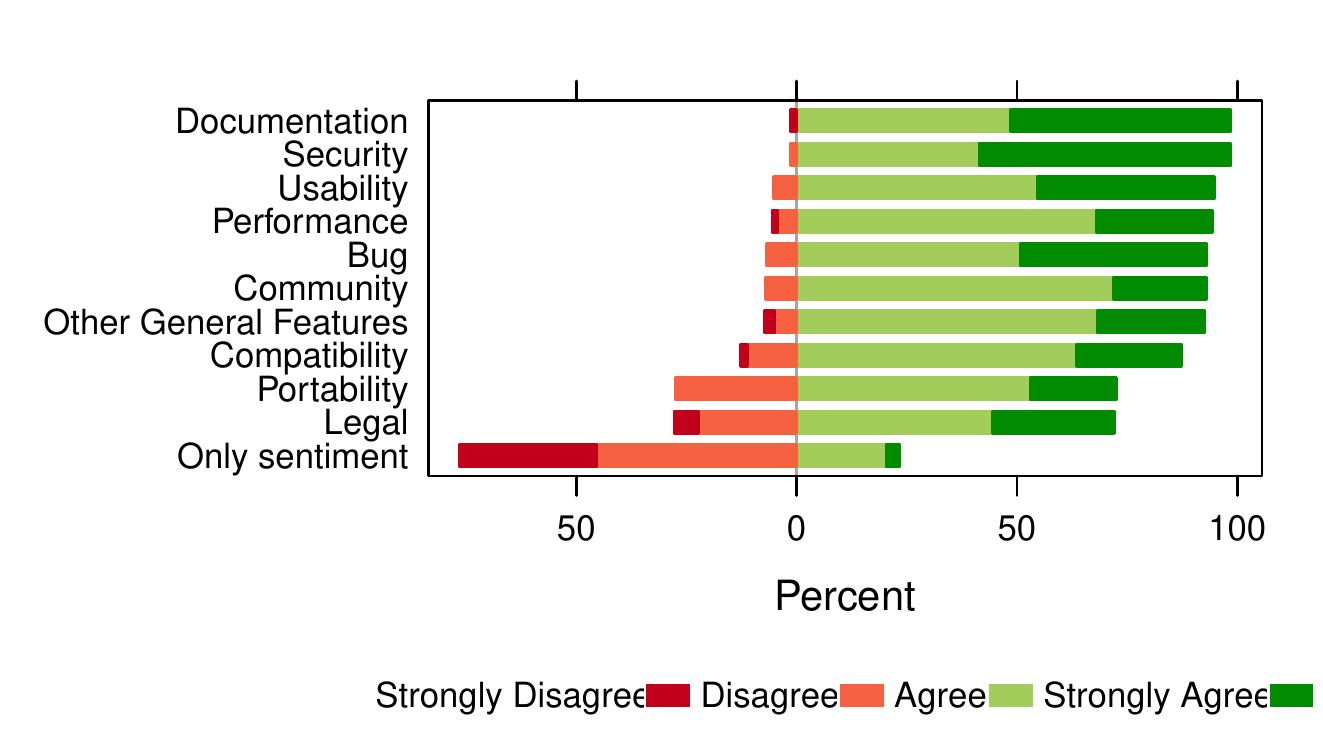}
\caption{Developers' preference to see opinions about API aspects.}
\label{fig:APIAspectImportantInOpinion}
\end{figure*}
%
Developers wished diverse presentation of opinions in the portal:
\begin{enumerate}[leftmargin=10pt]
\item \q{Categorized}{8,8} into different API aspects, such as \emt{Summarization by different criteria, categorization}, 

\item Distributed by \q{Sentiments}{8,5}, such as, star rating (similar
to other domains), \emt{do an x-out-of-5 rating for portability, stability, versatility and other attributes}.

\item Grouped by \q{Contrastive}{7,7} viewpoints to learn the strengths and weaknesses of an API feature from multiple forum posts, 
\emt{I make a research if the there are contradicting opinions on stack overflow / etc.}

\item Combined with code examples as a form of \q{API usage
scenarios}{7,6},  as well as integrated with the \q{API
Formal Documentation}{11,9}. 
\end{enumerate}
According to one developer \emt{Unified documentation on Stack Overflow (beta) looks like a positive step in the right direction.} 
Unfortunately, the Stack Overflow documentation site was shut down on August 8,
2017~\cite{website:sodoc-sunset-2017}, mainly due to a lower than the expected
number of visits to the documentation site. Therefore, we can draw insights both
from the problems faced by the Stack
Overflow documentation site and from the responses of the developers in our
survey to develop an
API portal that can offer better user experiences to the developers.

\bnd{Q15}\it{We asked developers about 11 API aspects, whether they would like
to explore opinions about APIs around those aspects.} The 11 aspects are:
\begin{inparaenum}
\item Performance,
\item Usability,
\item Portability,
\item Compatibility,
\item Security,
\item Bug,
\item Community,
\item Documentation,
\item Legal,
\item General Features,
\item Only sentiment.
\end{inparaenum} We note that each of these options are found in the open-ended
questions already. Thus, the response to this question can be used to further
ensure that we understood the responses of the developers. The results of the
relevant Likert-scale survey question are summarized in
\fig\ref{fig:APIAspectImportantInOpinion}. The vast majority of the developers agrees that
the presence of documentation, discussion of API's security features, mention
of any known bugs in an API, its compatibility, performance,  usability, and
the supporting user community are the main attributes that contribute to the opinion quality. Developers could not agree whether a useful opinion should include a
mention of any legal terms; while they agree that posts containing only one's
sentiment (e.g., ``I love API X'' or ``API Y sucks'') without any rationale are
not very useful. We note that there was a similar themed open-ended question in
Q11. However, the developers were not shown Q15 before their response of Q11.
Moreover, the two questions were placed in two separate sections, Q11 in
section 4 and Q15 in section 6. There was only one question (i.e., Q11) in
section 4. Therefore, the participants did not see Q15 or the options in Q15
during their response of Q11.


\begin{tcolorbox}[title=Needs for API Review Summarization (RQ2.2) \hrule
\it{Summarization Types (RQ2.2.c)},
opacityback=0, standard jigsaw,]
Developers expect to see opinions about diverse API aspects, such as, usability,
performance, security, compatibility, portability, etc. Developers wished
for a dedicated opinion portal engine for APIs where they can search for an API
and see all the opinions collected about the APIs from the developer forums.
\end{tcolorbox}


%% file: discussions.tex
\begin{table*}[!tbh]
  \centering
  \caption{Highlights of findings from Primary survey about developers' needs
  to seek opinions about APIs (RQ1). Subscript with a question number shows number of responses.}
  \begin{tabular}{r|p{16cm}}\toprule \multicolumn{1}{l}{\textbf{RQ1.1}} & \multicolumn{1}{l}{\textbf{Needs for seeking opinions about APIs from developer forums}} \\
    \midrule
    1$_{114}$     & \multicolumn{1}{l}{Do you visit developer forums to seek info about APIs?} \\
          &
          \ib{\begin{inparaenum}
          \item Yes 79.8\%,
          \item No 21.2\%
          \end{inparaenum}}
          \\
          \midrule
    2$_{93}$     & \multicolumn{1}{l}{List top developer forums you visited in the last two years} \\
          &
          \ib{
          \begin{inparaenum}
          \item Stack Overflow \& Exchange 56\%,
          \item Blogs \& mailing lists 18.1\%,
          \item GitHub 5.4\%
          \item Google dev 4.8\%,
          \item Apple dev 3.6\%
          \end{inparaenum}
          }
          \\
          \midrule
    4$_{83}$     & \multicolumn{1}{p{46.55em}}{What are your reasons for referring to opinions of other developers about APIs in online developer forums?} \\
          &
          \ib{
          \begin{inparaenum}
          \item Expertise 25.6\%,
          \item API usage 14\%,
          \item API selection 10.7\%,
          \item Opinion trustworthiness 10.7\%,
          \item Documentation/Community 9.1\%
          \end{inparaenum}
          }
          \\
          \midrule
    5$_{83}$     & \multicolumn{1}{p{46.55em}}{How do you seek information about APIs in a developer forum?} \\
          &
          \ib{
          \begin{inparaenum}
          \item Search 83.9\%,
          \item Sentiment statistics 3.6\%
          \item Similarity analysis 2.7\%,
          \item Expertise 1.8\%,
          \item Recency/Documentation 0.9\%
          \end{inparaenum}
          }
          \\
          \midrule
    6$_{83}$     & \multicolumn{1}{p{46.55em}}{What are your biggest challenges when seeking for opinions about an API in an online developer forum?} \\
          &
          \ib{
          \begin{inparaenum}
          \item Situational relevance 20.3\%,
          \item Trustworthiness 16.4\%
          \item Search 11.7\%,
          \item Recency 10.2\%,
          \item Community engagement 7\%
          \end{inparaenum}
          }
          \\
          \midrule
    16$_{83}$    & \multicolumn{1}{p{46.55em}}{Where do you seek help/opinions about APIs?} \\
          &
          \ib{
          \begin{inparaenum}
          \item Stack Overflow 34.7\%,
          \item Co-workers 26.4\%,
          \item Internal mailing lists 10\%,
          \item IRC Chats 9.2\%,
          \item Search/Diverse sources 19.7\%
          \end{inparaenum}
          }
          \\
          \midrule
    17$_{83}$    & \multicolumn{1}{p{46.55em}}{How often do you refer to developer forums to  get information about APIs?} \\
          &
          \ib{
          \begin{inparaenum}
          \item Every day 36.1\%,
          \item Two/three times a week 32.5\%,
          \item Once a month 16.9\%,
          \item Once a week 14.5\%
          \end{inparaenum}
          }
          \\
          \midrule
    18$_{83}$    & \multicolumn{1}{p{46.55em}}{ When do you seek opinions about APIs?} \\
          &
          \ib{
          \begin{inparaenum}
          \item Selection among choices 88\%,
          \item Determining replacement 80.7\%,
          \item Feature enhancement 69.9\%,
          \item Fixing a bug 68.7\%,
          \item Develop competing API 56.6\%,
          \item Replace a feature 49.4\%,
          \item Validate a selection 38.6\%,
          \item Select API version 27.7\%,
          \item Other task 15.7\%
          \end{inparaenum}
          }
          \\
          \midrule
    \multicolumn{1}{l}{\textbf{RQ1.2}} & \multicolumn{1}{l}{\textbf{Needs for opinion quality assessment}} \\
    \midrule
     3$_{93}$     & \multicolumn{1}{l}{Do you value the opinion of other developers in the developer forums when deciding on what API to use?} \\
          &
          \ib{
          \begin{inparaenum}
          \item Yes 89.2\%,
          \item No 10.8\%
          \end{inparaenum}
          }
          \\
          \midrule

    7$_{83}$     & \multicolumn{1}{p{46.55em}}{What factors in a forum post can help you determine the quality of a provided opinion about an API?} \\
          &
          \ib{
          \begin{inparaenum}
          \item Documentation 35.5\%,
          \item Reputation 23\%,
           \item Expertise 8.6\%,
          \item Situational relevance/Community 7.9\%,
          \item Trustworthiness 5.3\%
          \end{inparaenum}
          }
          \\
          \midrule
    24$_{10}$    & \multicolumn{1}{p{46.55em}}{Please explain why you don't value the opinion of other developers in the developer forums.} \\
          &
          \ib{\begin{inparaenum}
          \item Trustworthiness 23.5\%
          \item Situational relevance 11.8\%,
          \item Community 11.8\%,
          \item Expertise 11.8\%,
          \end{inparaenum}}
          \\
          \bottomrule
    \end{tabular}%
  \label{tab:s2rq1highlights}%
\end{table*}%
\section{Discussions} \label{sec:discussion}
In this section, we summarize the key points from our primary survey with
regards to the two research questions that we set forth to address
(\sec\ref{sec:summary}). We then analyze the findings from the two surveys along three demographics by: \begin{inparaenum}
\item profession in \sec\ref{sec:demographic-profession},
\item experience of the survey participants in \sec\ref{sec:demographic-experience}, and
\item the nine programming languages used to sample the primary survey participants in \sec\ref{sec:demographic-programming-langs}.
\end{inparaenum}

\begin{table*}[t]
  \centering
  \caption{Highlights of findings from Primary survey about developers' needs
  for tool support to analyze opinions about APIs (RQ2). Subscript with a question number shows number of responses.}
    \begin{tabular}{r|p{16cm}}\toprule
    \multicolumn{1}{l}{\textbf{RQ2.1}} & \multicolumn{1}{l}{\textbf{Tool support}} \\
    \midrule
    8$_{83}$     & \multicolumn{1}{l}{Do you rely on tools to help you understand opinions about APIs in online forum discussions?} \\
          &
          \ib{
          \begin{inparaenum}
          \item Yes 13.3\%,
		  \item No 86.7\%
          \end{inparaenum}
          }
          \\
    \midrule
    9$_{72}$     & \multicolumn{1}{p{49.95em}}{If you don't use a tool currently to explore the diverse opinions about APIs in developer forums, do you believe there is a need of such a tool to help you find the right viewpoints about an API quickly?} \\
          &
          \ib{
          \begin{inparaenum}
          \item Yes 9.7\%,
          \item No 27.8\%,
          \item I don't know 62.5\%
          \end{inparaenum}
          }
          \\
     \midrule
    10$_{11}$    & \multicolumn{1}{l}{You said yes to the previous question on using a tool to navigate forum posts. Please provide the name of the tool} \\
          &
          \ib{
          \begin{inparaenum}
          \item Google/Search 4,
          \item Stack Overflow votes/app/related post 3,
          \item GitHub issue/pulse 1,
          \item Safari 1,
          \item Reference documentation 1
          \end{inparaenum}
          }
          \\
     \midrule
    19$_{83}$    & \multicolumn{1}{l}{What tools can better support your understanding of API reviews in developer forums? } \\
          &
          \ib{
          \begin{inparaenum}
          \item Opinion mining 56.6\%,
          \item Sentiment analysis 41\%,
          \item Opinion summarization 45.8\%,
          \item API comparator 68.7\%,
          \item Trend analyzer 67.5\%,
          \item Co-mentioned competing APIs 73.5\%,
          \item Other Tools 8.4\%
          \end{inparaenum}
          }
          \\
     \midrule
    \multicolumn{1}{l}{\textbf{RQ2.2}} & \multicolumn{1}{l}{\textbf{Needs for Opinion Summarization}} \\
     \midrule
    11$_{83}$    & \multicolumn{1}{l}{What are the important factors of an API that play a role in your decision to choose an API?} \\
          &
          \ib{
          \begin{inparaenum}
          \item Usability 30.1\%,
          \item Documentation 16.3\%,
          \item Community 10.7\%,
          \item Performance 10.4\%,
          \item Compatibility 6.2\%,
          \item Legal 5.2\%,
          \end{inparaenum}
          }
          \\
     \midrule
    12$_{83}$    & \multicolumn{1}{p{49.95em}}{Considering that opinions and diverse viewpoints about an API can be scattered in different posts and threads of a developer\newline{}forum, what are the different ways opinions about APIs can be summarized from developer forums?} \\
          &
          \ib{
          \begin{inparaenum}
          \item Portal for opinion aggregation 12.7\%,
          \item Search 8.5\%
          \item Documentation 7.7\%,
          \item Sentiment statistics/Opinion categorization 5.6\%,
          \item Contrastive summaries/Usage scenario summaries 4.9\%
          \end{inparaenum}
          }
          \\
     \midrule
    13$_{83}$    & \multicolumn{1}{l}{What areas can be positively affected by the summarization of reviews about APIs from developer forums?} \\
          &
          \ib{
          \begin{inparaenum}
          \item API selection 26.1\%,
		  \item Productivity 10.4\%
		  \item API usage 6.1\%
		  \item Documentation 5.2\%
		  \item API popularity analysis 3.5\%
          \end{inparaenum}
          }
          \\
     \midrule
    14$_{83}$    & \multicolumn{1}{l}{What areas can be negatively affected by the summarization of reviews about APIs from developer forums?} \\
          &

          \ib{
          \begin{inparaenum}
          \item Opinion trustworthiness 19.3\%
          \item Missing nuances 16.5\%
		  \item Opinion reasoning 6.4\%
		  \item API selection 5.5\%
		  \item New API Entry 2.8\%
          \end{inparaenum}
          }
          \\
     \midrule
    15$_{83}$    & \multicolumn{1}{l}{An opinion is important if it contains discussion about the following API aspects?} \\
          &
          \ib{
          \begin{inparaenum}
          \item Usability 88\%
		  \item Documentation 85.3\%
		  \item Security 83.1\%
		  \item Performance 81.9\%
		  \item Bug 81.9\%
		  \item Compatibility 66.3\%,
		  \item Community 63.9\%,
		  \item Legal 47\%,
		  \item Portability 44.6\%,
		  \item General API Features 45.8\%,
		  \item Only Sentiment 18.1\%
          \end{inparaenum}
          }
          \\
     \midrule
    20$_{83}$    & \multicolumn{1}{l}{How often do you feel overwhelmed due to the abundance of opinions about an API?} \\
          &
          \ib{
          \begin{inparaenum}
          \item Sometimes 48.2\%
		  \item Rarely 37.3\%
		  \item Never 10.8\%
		  \item Always 3.6\%
          \end{inparaenum}
          }
          \\
     \midrule
    21$_{83}$   & \multicolumn{1}{l}{Would a summarization of opinions about APIs help you to make a better decision on which one to use? } \\
          &
          \ib{
          \begin{inparaenum}
          \item I don't know 48.2\%
		  \item Yes 38.6\%
		  \item No 13.3\%
          \end{inparaenum}
          }
          \\
     \midrule
    22$_{83}$    & \multicolumn{1}{l}{Opinions about an API need to be summarized because} \\
          &
          \ib{
          \begin{inparaenum}
          \item The interesting opinion may not be in the posts that you have
          looked in 71.1\%
		  \item Not enough time to look at all opinions 68.7\%
		  \item Contradictory opinions about an API may be missed 61.4\%
		  \item Too many forum posts with opinions 50.6\%
		  \item Other reason 12\%
          \end{inparaenum}
          }
          \\
     \midrule
    23$_{83}$    & \multicolumn{1}{l}{Opinion summarization can improve the following decision making processes} \\
          &
          \ib{
          \begin{inparaenum}
          \item Selection among choices 73.5\%,
		  \item Determining replacement of an API 71.1\%,
		  \item Validating an API selection 48.2\%,
		  \item Feature enhancement 45.8\% ,
		  \item Replacing an API feature 42.2\%,
		  \item Fixing a bug 38.6\%,
		  \item Selecting API version 28.9\%,
		  \item Other 9.6\%
          \end{inparaenum}
          }
          \\
    \bottomrule
    \end{tabular}%
  \label{tab:s2rq2highlights}%
\end{table*}%
\subsection{Summary}\label{sec:summary}
In \tbl\ref{tab:s2rq1highlights} and \tbl\ref{tab:s2rq2highlights}, we summarize the responses from our primary survey for RQ1 and RQ2, respectively. 
The insights are calculated using the same
rules that we used in \sec\ref{sec:pilot-survey}.

\rev{\addtocounter{o}{1}
\nd\it{\bf{Observation \arabic{o}.}} \bf{Developers leverage opinions about APIs to support development needs, such as API selection, usage, learning about 
edge cases, etc (\tbl\ref{tab:s2rq1highlights}.) 
Developers expect opinion summaries can facilitate those needs by offering an increase in productivity, e.g., save time by offering quick but synthesized insights (\tbl\ref{tab:s2rq2highlights})}
In the absence of a specific opinion summarization portal
available for APIs, though, we find that developers were unsure whether such
summaries can be feasible. However, they recommend to leverage machine
learning techniques to facilitate an opinion portal for APIs.}

\rev{\addtocounter{o}{1}
\nd\it{\bf{Observation \arabic{o}.}} A number of questions were paired as open-ended and closed
questions, with the open-ended questions being asked before the corresponding
closed-end question. For example, we asked developers about the reasons behind why
they seek opinions about APIs using two questions (Q4 and Q18). Another pair was Q11 and Q15, which we used to 
ask developers about their preference of specific API aspects/factors that they expect to see in the opinions about APIs.}

\rev{In the first pair (i.e., Q4 and Q18) the
responses to the open-ended question (Q4) show a slight variation from the
response to the closed question (Q18). In Q4, the main reason
mentioned was to build and look for expertise about APIs while
seeking for opinions. Q18 does not have that option. Nevertheless, the majority
of agreement for the options in Q18 show that those needs are also prevalent.
In the second pair (i.e., Q11 and Q15), the responses to both questions produced an almost similar set of
API aspects, such as, performance, usability, documentation, etc.}

\rev{\addtocounter{o}{1}
\nd\it{\bf{Observation \arabic{o}.}} In both surveys, two closed questions were paired together (Q18 and Q23 in the primary survey and Q4 and Q12 in
the pilot survey). The first question in each pair (i.e., Q18 in primary and Q4 in pilot surveys) aims at understanding the
needs of developers for seeking opinions about APIs. The second question in each pair (i.e., Q23 in primary survey and Q12 in pilot one)
aims to understand the needs for summarizing such opinions. The options for each of the four questions
remained the same (see \tbl\ref{tab:s2rq1highlights}). In both surveys, the highest ranked option was ``selection among choices''.
Therefore, developers seek opinions to decide on an API among choices and they believe that the summarization of opinions
can assist them in their selection process. The three selection-related options (i.e., selection among choices,
determining a replacement, and validating an API selection) are ranked as the top three in the pair of questions from the primary survey.
\bf{Therefore, the primary focus to aggregate and summarize opinions about APIs would be to assist developers in their selection of APIs and
any other tasks relevant to it.}}

\begin{table*}[t]
  \centering
  \caption{Highlights of Findings from Primary Survey RQ1 Closed Questions by Profession}
    \begin{tabular}{r|p{16cm}} \toprule
    \bf{No} & \bf{Question} \\ \midrule
    1     & Do you visit developer forums to seek info about APIs? Research Engineer\\ 
    &\it{\begin{inparaenum}
    \item Research Engineer: \ib{Yes 100\%*}, 
    \item Student: \ib{Yes 66.7\%}, 
    \item Software Developer: \ib{Yes 78.7\%*}, 
    \item Lead: \ib{Yes 100\%*}
    \end{inparaenum}}\\
    \midrule
    3     & Do you value the opinion of other developers in the developer forums when deciding on what API to use? \\
    & \it{\begin{inparaenum}
    \item Research Engineer: \ib{Yes 100\%*}, 
    \item Student: \ib{Yes 100\%}, 
    \item Software Developer: \ib{Yes 89.2\%*}, 
    \item Lead: \ib{Yes 100\%*}
    \end{inparaenum}
    }
    \\
    \midrule
    
    16    & Where do you seek help/opinions about APIs? \\
    &	 
    
    \begin{tabular}{lrrrrr}
          & \multicolumn{1}{l}{\textbf{Internal Mailing list}} & \multicolumn{1}{l}{\textbf{IRC chats}} & \multicolumn{1}{l}{\textbf{Co-workers}} & \multicolumn{1}{l}{\textbf{Stack Overflow}} & \multicolumn{1}{l}{\textbf{Search/Diverse Sources}} \\
    \textbf{Software Developer} & 24.3\% & 24.3\% & 71.6\% & \ib{89.2\%*} & 39.2\% \\
    \textbf{Other} & -     & -     & -     & \ib{33.3\%} & \ib{33.3\%} \\
    \textbf{Student} & 50.0\% & 50.0\% & 33.3\% & \ib{100.0\%} & 50.0\% \\
    \textbf{Research Engineer} & 33.3\% & 33.3\% & 33.3\% & \ib{100\%*} & 66.7\% \\
    \textbf{Lead} & 36.4\% & 18.2\% & 63.6\% & \ib{100\%*} & 63.6\% \\
    \end{tabular}%
    \\
   \midrule
    17    & How often do you refer to online forums (e.g., Stack Overflow) to get information about APIs? \\
    &	
    
        \begin{tabular}{lrrrrr}
          & \bf{Every day} & \bf{Two or three times a week} & \bf{Once a week} & \bf{One a month} & \bf{Never} \\
    \textbf{Software Developer} & \ib{33.3\%*} & 12.2\% & 28.4\% & 14.9\% & 11.2\% \\
    \textbf{Other} & 33.3\% & 0.0\% & 0.0\% & 0.0\% & \ib{66.7\%} \\
    \textbf{Student} & \ib{50.0\%} & 0.0\% & 0.0\% & \ib{50.0\%} & 0.0\% \\
    \textbf{Research Engineer} & \ib{33.3\%} & \ib{33.3\%} & \ib{33.3\%} & 0.0\% & 0.0\% \\
    \textbf{Lead} & 27.3\% & \ib{36.4\%*} & 18.2\% & 18.2\% & 0.0\% \\
    \end{tabular}%
      \\
    \midrule
    18    & When do you seek opinions about APIs?  \\
    &	 
    SAC = Selecting amidst choices, DAR = Determining a replacement, ISF = Improving software feature, RAF = Replacing software feature, DAN = Developing a new API, FAB = Fixing a bug, SAV = Selecting API version,
    VAS = Validating a selection
     \begin{tabular}{lrrrrrrrrr}\
          & {\textbf{SAC}} & {\textbf{DAR}} & {\textbf{ISF}} & {\textbf{RAF}} 
          & {\textbf{DAN}} & {\textbf{VAS}} & {\textbf{FAB}} 
          & {\textbf{SAV}} & {\textbf{Other}} \\
    \textbf{Software Developer} & \ib{81.1\%*} & 74.3\% & 60.8\% & 45.9\% & 51.4\% & 39.2\% & 59.5\% & 25.7\% & 17.6\% \\
    \textbf{Other} & 0.0\% & 33.3\% & \ib{33.3\%} & \ib{33.3\%} & \ib{33.3\%} & 0.0\% & \ib{33.3\%} & 0.0\% & 0.0\% \\
    \textbf{Student} & \ib{100.0\%} & \ib{100.0\%} & 50.0\% & 100.0\% & 100.0\% & 0.0\% & 50.0\% & 0.0\% & 0.0\% \\
    \textbf{Research Engineer} & 66.7\% & 33.3\% & 66.7\% & 0.0\% & 66.7\% & 0.0\% & \ib{100.0\%} & 0.0\% & 0.0\% \\
    \textbf{Lead} & \ib{81.8\%*} & 72.7\% & \ib{81.8\%*} & 36.4\% & 36.4\% & 27.3\% & 72.7\% & 36.4\% & 0.0\% \\
    \end{tabular}%
    \\
 
    \bottomrule
    \end{tabular}%
  \label{tab:rq1-sum-by-profession}%
\end{table*}%

\begin{table*}[t]
  \centering
  \caption{Highlights of Findings from Primary Survey RQ2 Closed Questions by Profession}
    \begin{tabular}{r|p{16cm}} \toprule
    \bf{No} & \bf{Question} \\ \midrule
    8     & Do you rely on tools to help you understand opinions about APIs in online forum discussions? \\
    &	\it{
    \begin{inparaenum}
    \item Research Engineer: \ib{No 66.7\%}, 
    \item Student: \ib{No 100\%}, 
    \item Software Developer: \ib{No 79.7\%*}, 
    \item Lead: \ib{No 72.7\%*}
    \end{inparaenum}
    }  \\
    \midrule
    9     & If you don't use a tool currently to explore the diverse opinions about APIs in developer forums, do you believe there is a need of such a tool to help you find the right viewpoints about an API quickly? \\
    &	\it{
    \begin{inparaenum}
    \item Research Engineer: \ib{No 33.3\%, I don't know 33.3\%} 
    \item Student: \ib{Yes 50\%, I don't know 50\%} 
    \item Software Developer: \ib{I don't know 48.6\%*}, 
    \item Lead: \ib{I don't know 54.5\%*}
    \end{inparaenum}
    }  \\
    \midrule
    15    & An opinion is important if it contains discussion about the following API aspects?  \\
    &	
    Per = Performance, Sec = Security, Use = Usability, Doc = Documentation, Comp = Compatibility, Comm = Community, Leg = Legal, Port = Portability, Sen = Only Sentiment, Feat = General Features
    
        \begin{tabular}{lrrrrrrrrrrr}
          & {\textbf{Per}} & {\textbf{Sec}} & {\textbf{Use}} & {\textbf{Doc}} 
          & {\textbf{Comp}} & {\textbf{Comm}} & {\textbf{Bug}} & {\textbf{Leg}} 
          & {\textbf{Por}} & {\textbf{Sen}} & {\textbf{Feat}} \\
    \textbf{Software Developer} & 75.7\% & 74.3\% & \ib{82.4\%*} & 75.7\% & 60.8\% & 56.8\% & 74.3\% & 41.9\% & 40.5\% & 13.5\% & 43.2\% \\
    \textbf{Other} & \ib{33.3\%} & \ib{33.3\%} & \ib{33.3\%} & 0.0\% & \ib{33.3\%} & 0.0\% & 0.0\% & 0.0\% & \ib{33.3\%} & 0.0\% & \ib{33.3\%} \\
    \textbf{Student} & 50.0\% & \ib{100.0\%} & \ib{100.0\%} & \ib{100.0\%} & 50.0\% & 50.0\% & 100.0\% & 0.0\% & 50.0\% & 50.0\% & 0.0\% \\
    \textbf{Research Engineer} & \ib{66.7\%} & \ib{66.7\%} & \ib{66.7\%} & \ib{66.7\%} & 33.3\% & \ib{66.7\%} & \ib{66.7\%} & \ib{66.7\%} & 0.0\% & 33.3\% & 33.3\% \\
    \textbf{Lead} & 72.7\% & 81.8\% & 63.6\% & \ib{100\%*} & 63.6\% & 72.7\% & 81.8\% & 54.5\% & 45.5\% & 27.3\% & 36.4\% \\
    \end{tabular}%
    
     \\
    \midrule
    19    & What tools can better support your understanding of API reviews in developer forums? \\
    &	OM =  Opinion Miner, SA = Sentiment Analyzer, OS = Opinion Summarizer, AC = API Comparator, TA = Trend Analyzer, CA = Competing APIs, OT = Other Tools 
    
    \begin{tabular}{lrrrrrrr}
          & \textbf{OM} & \textbf{SA} & \textbf{OS} & \textbf{AC} & \textbf{TA} & \textbf{CA} & \textbf{OT} \\
    \textbf{Software Developer} & 54.1\% & 39.2\% & 43.2\% & 62.2\% & 59.5\% & \ib{64.9\%*} & 6.8\% \\
    \textbf{Other} & 0.0\% & 0.0\% & 0.0\% & \ib{33.3\%} & \ib{33.3\%} & \ib{33.3\%} & 0.0\% \\
    \textbf{Student} & \ib{100.0\%} & 0.0\% & 50.0\% & 50.0\% & \ib{100.0\%} & \ib{100.0\%} & 0.0\% \\
    \textbf{Research Engineer} & 0.0\% & 33.3\% & 33.3\% & \ib{100.0\%} & 66.7\% & 66.7\% & 33.3\% \\
    \textbf{Lead} & 45.5\% & 36.4\% & 36.4\% & 54.5\% & 63.6\% & \ib{72.7\%*} & 9.1\% \\
    \end{tabular}%
     \\
    \midrule
    20    & How often do you feel overwhelmed due to the abundance of opinions about an API? \\
    &	
        \begin{tabular}{lrrrr}
          & \multicolumn{1}{l}{\textbf{Always}} & \multicolumn{1}{l}{\textbf{Sometimes}} & \multicolumn{1}{l}{\textbf{Rarely}} & \multicolumn{1}{l}{\textbf{Never}} \\
    \textbf{Software Developer} & 4.1\% & \ib{45.9\%*} & 31.1\% & 8.1\% \\
    \textbf{Other} & \ib{33.3\%} & 0.0\% & 0.0\% & 0.0\% \\
    \textbf{Student} & 0.0\% & \ib{100.0\%} & 0.0\% & 0.0\% \\
    \textbf{Research Engineer} & 0.0\% & \ib{33.3\%} & \ib{33.3\%} & \ib{33.3\%} \\
    \textbf{Lead} & 0.0\% & 27.3\% & \ib{63.6\%*} & 9.1\% \\
    \end{tabular}%
     \\
    \midrule
    21    & Would a summarization of opinions about APIs help you to make a better decision on which one to use? \\
    &	\it{
    \begin{inparaenum}
    \item Research Eng: No 66.7\%, 
    \item Student: Yes 50\%, don't know 50\%, 
    \item Software Dev: \ib{don't know 40.5\%*}, 
    \item Lead: \ib{don't know 63.6\%*}
    \end{inparaenum}
    }  \\
    \midrule
    22    & Opinions about an API need to be summarized because \\
    &	
        \begin{tabular}{lrrrrr}
          & {\textbf{Too many posts}} & {\textbf{Opinion in missed posts}} 
          & {\textbf{Missed contradictory}} & {\textbf{Not enough time}} 
          & {\textbf{Other reason}} \\
    \textbf{Software Developer} & 50.0\% & \ib{63.5\%*} & 54.1\% & 59.5\% & 12.2\% \\
    \textbf{Other} & \ib{33.3\%} & \ib{33.3\%} & \ib{33.3\%} & \ib{33.3\%} & 0.0\% \\
    \textbf{Student} & 50.0\% & \ib{100.0\%} & \ib{100.0\%} & \ib{100.0\%} & 0.0\% \\
    \textbf{Research Engineer} & 0.0\% & \ib{66.7\%} & \ib{66.7\%} & \ib{66.7\%} & 33.3\% \\
    \textbf{Lead} & 27.3\% & 63.6\% & 54.5\% & \ib{72.7\%*} & 0.0\% \\
    \end{tabular}%
     \\
    \midrule
    23    & Opinion summarization can improve the following decision making processes \\
    &	SAC = Selecting amidst choices, DAR = Determining a replacement, ISF = Improving software feature, RAF = Replacing software feature, DAN = Developing a new API, FAB = Fixing a bug, SAV = Selecting API version,
    VAS = Validating a selection
        \begin{tabular}{lrrrrrrrrr}
          & {\textbf{SAC}} & {\textbf{DAR}} & {\textbf{ISF}} & {\textbf{RAF}} & {\textbf{DAN}} & {\textbf{VAS}} 
          & {\textbf{FAB}} & {\textbf{SAV}} & {\textbf{Other}} \\
    \textbf{Software Developer} & \ib{63.5\%*} & 60.8\% & 41.9\% & 36.5\% & 37.8\% & 45.9\% & 33.8\% & 24.3\% & 10.8\% \\
    \textbf{Other} & \ib{33.3\%} & \ib{33.3\%} & 0.0\% & \ib{33.3\%} & 0.0\% & 0.0\% & 0.0\% & \ib{33.3\%} & 0.0\% \\
    \textbf{Student} & \ib{100.0\%} & \ib{100.0\%} & 0.0\% & 50.0\% & 0.0\% & 50.0\% & 50.0\% & 0.0\% & 0.0\% \\
    \textbf{Research Engineer} & \ib{66.7\%} & \ib{66.7\%} & \ib{66.7\%} & 33.3\% & \ib{66.7\%} & 33.3\% & 33.3\% & 0.0\% & 0.0\% \\
    \textbf{Lead} & \ib{81.8\%*} & \ib{81.8\%*} & 45.5\% & 45.5\% & 36.4\% & 36.4\% & 45.5\% & 45.5\% & 0.0\% \\
    \end{tabular}%
    \\
    \bottomrule
    \end{tabular}%
  \label{tab:rq2-sum-by-profession}%
\end{table*}%

\subsection{Analysis by Professions}\label{sec:demographic-profession}
\rev{In \tbls\ref{tab:rq1-sum-by-profession} and \ref{tab:rq2-sum-by-profession}, we summarize the results of closed questions from our final survey by the reported 
professions of survey participants. We report each question as follows:
\begin{enumerate}
  \item Multiple Choice Questions. If the question has three options (Yes, No, I don't know), we only show the result of the option(s) with the majority agreement. 
  \item Likert-scale Questions. We show all the options, by calculating agreements using the formula we introduce in \sec\ref{sec:pilot-survey} 
  (we used in \tbls\ref{tbl:s1questions},\ref{tab:s2rq1highlights}, and \ref{tab:s2rq2highlights}). 
  \item We highlight the value of an option for each profession as \it{bold italic}, if the option has the maximum value out of all options for the profession. 
  For example, for Q16 (where do you seek opinions about APIs?) in \tbl\ref{tab:rq1-sum-by-profession}, there are five options: \begin{inparaenum}
  \item Internal mailing lists,
  \item IRC chats,
  \item Co-workers,
  \item Stack Overflow,
  \item Search/Diverse sources
  \end{inparaenum}  
  The software developers in our survey show the most agreement (89.2\%) towards Stack Overflow. We thus make the value bold-italic among the five option for software developers.
  \item We place a star symbol beside a highlighted value of an option, if the value is statistically significant\footnote{We use Mann Whitney U test and a 95\% confidence
level (i.e., $p$ = 0.05).}   
\end{enumerate}}  

\rev{\addtocounter{o}{1}
\nd\it{\bf{Observation \arabic{o}.}} \bf{The necessity of seeking opinions about APIs is mostly prevalent and consistent across
the different reported professions in our surveys.} All the research engineers, students, and team leads, and 89.2\% of software developers who 
visit developer forums in our survey, also value the opinions about APIs in the forums. Unlike managers, technical leads are expected to work closely with the codebase
and overall system architecture design. The decision on an API during the design of a system can be beneficial for such team leads.
According to one team lead: \emt{The quality of APIs can vary considerably. Getting experience of others can save a
lot of time if you end up using a better API or end up skipping a bad one.} For the researchers, the motivation was to learn from the experts: \emt{
they have used the API, probably more extensively then I have, and may be experts in their subfield}.}

\rev{\addtocounter{o}{1}
\nd\it{\bf{Observation \arabic{o}.}} While all the team leads consult API information in the developer forums, most of them visit the forum two or
three times a week. In contrast, most of the developers who consult developer forums for API information, do so more every day.}
 
\rev{\addtocounter{o}{1}
\nd\it{\bf{Observation \arabic{o}.}} In both pilot and primary surveys, we asked developers about their preference for tools to better support their
understanding of opinions about APIs in the developer forums (Q19 in primary survey, Q7 in pilot survey). A 
summarization engine to compare APIs based on different features is considered as the most useful among the software developers, team leads and research engineers. 
In our pilot survey, the leads also preferred the same tool the most among all tools. In our pilot survey, 
we got responses from three managers, who showed equal preference (33.3\%) to all the tools except an opinion miner (66.7\%).}

\rev{\addtocounter{o}{1}
\nd\it{\bf{Observation \arabic{o}.}} There is a more clear distinction among the professions in their preference of API aspects that they prefer to explore in the opinions about APIs
(Q15 in primary survey). The team leads are most interested to find opinions about API documentation, while the other professions including software engineers are 
most interested to learn about the usability of the API. All the professions agreed the most that the summarizing of opinions can help in the selection of APIs.}  

\begin{table*}[t]
  \centering
  \caption{Highlights of Findings from Primary Survey RQ1 Closed Questions by Experience}
    \begin{tabular}{r|p{16cm}} \toprule
    \bf{No} & \bf{Question} \\ \midrule
    1     & Do you visit developer forums to seek info about APIs? Research Engineer\\ 
    &	\it{
    \begin{inparaenum}
    \item 10+: \ib{Yes 76.8\%*}, 
    \item 7-10: \ib{Yes 83.9\%*}, 
    \item 3-6: \ib{Yes 85.7\%*}
    \end{inparaenum}
    }  \\
    \midrule
    3     & Do you value the opinion of other developers in the developer forums when deciding on what API to use? \\
    &	\it{
    \begin{inparaenum}
    \item 10+: \ib{Yes 85.5\%*}, 
    \item 7-10: \ib{Yes 92.3\%*}, 
    \item 3-6: \ib{Yes 100\%*}
    \end{inparaenum}
    }   \\
    \midrule
    
    16    & Where do you seek help/opinions about APIs? \\
    &	 
    
    \begin{tabular}{lrrrrr}
          & {\textbf{Internal Mailing list}} & {\textbf{IRC chats}} & {\textbf{Co-workers}} 
          & {\textbf{Stack Overflow}} & {\textbf{Search/Diverse Sources}} \\
    \textbf{10+} & 21.8\% & 10.9\% & 67.3\% & \ib{85.5\%*} & 41.8\% \\
    \textbf{7--10} & 30.8\% & 34.6\% & 69.2\% & \ib{92.3\%*} & 46.2\% \\
    \textbf{3--6} & 33.3\% & 58.3\% & 66.7\% & \ib{100.0\%*} & 41.7\% \\
    \end{tabular}%
    \\
    \midrule
    17    & How often do you refer to online forums (e.g., Stack Overflow) to get information about APIs? \\
    &

    \begin{tabular}{lrrrrr}
          & {\textbf{Every day}} & {\textbf{Two or three times a week}} & {\textbf{Once a week}} & {\textbf{One a month}} & {\textbf{Never}} \\
    \textbf{10+} & \ib{32.7\%*} & 25.5\% & 12.7\% & 14.5\% & 0.0\% \\
    \textbf{7--10} & \ib{30.8\%*} & 30.8\% & 11.5\% & 19.2\% & 7.7\% \\
    \textbf{3--6} & 33.3\% & \ib{41.7\%*} & 16.7\% & 8.3\% & 0.0\% \\
    \end{tabular}%
      \\
    \midrule
    18    & When do you seek opinions about APIs?  \\
    &	 
    SAC = Selecting amidst choices, DAR = Determining a replacement, ISF = Improving software feature, RAF = Replacing software feature, DAN = Developing a new API, FAB = Fixing a bug, SAV = Selecting API version,
    VAS = Validating a selection
   
    \begin{tabular}{lrrrrrrrrr}
          & {\textbf{SAC}} & {\textbf{DAR}} & {\textbf{ISF}} & {\textbf{RAF}} & {\textbf{DAN}} & {\textbf{VAS}} 
          & {\textbf{FAB}} & {\textbf{SAV}} & {\textbf{Other}} \\
    \textbf{10+} & \ib{74.5\%*} & 65.5\% & 60.0\% & 38.2\% & 45.5\% & 34.5\% & 47.3\% & 25.5\% & 10.9\% \\
    \textbf{7--10} & \ib{80.8\%*} & 84.6\% & 61.5\% & 53.8\% & 57.7\% & 38.5\% & \ib{80.8\%*} & 26.9\% & 19.2\% \\
    \textbf{3--6} & \ib{91.7\%*} & 75.0\% & 75.0\% & 50.0\% & 58.3\% & 25.0\% & 83.3\% & 16.7\% & 16.7\% \\
    \end{tabular}%
    \\
 
    \bottomrule
    \end{tabular}%
  \label{tab:rq1-sum-by-exp}%
\end{table*}%

\begin{table*}[t]
  \centering
  \caption{Highlights of Findings from Primary Survey RQ2 Closed Questions by Experience}
    \begin{tabular}{r|p{16cm}} \toprule
    \bf{No} & \bf{Question} \\ \midrule
    8     & Do you rely on tools to help you understand opinions about APIs in online forum discussions? \\
    &	\it{
	\begin{inparaenum}
    \item 10+: \ib{No 70.9\%*}, 
    \item 7-10: \ib{No 84.6\%*}, 
    \item 3-6: \ib{No 91.7\%*}
    \end{inparaenum}    
    }  \\
    \midrule
    9     & If you don't use a tool currently to explore the diverse opinions about APIs in developer forums, do you believe there is a need of such a tool to help you find the right viewpoints about an API quickly? \\
    &	\it{
	\begin{inparaenum}
    \item 10+: \ib{I don't know 45.5\%*}, 
    \item 7-10: \ib{I don't know 53.8\%*}, 
    \item 3-6: \ib{I don't know 50\%*}
    \end{inparaenum}    
    }  \\
    \midrule
    15    & An opinion is important if it contains discussion about the following API aspects?  \\
    &	
    Per = Performance, Sec = Security, Use = Usability, Doc = Documentation, Comp = Compatibility, Comm = Community, Leg = Legal, Port = Portability, Sen = Only Sentiment, Feat = General Features
    
    \begin{tabular}{lrrrrrrrrrrr}
          & {\textbf{Per}} & {\textbf{Sec}} & {\textbf{Use}} & {\textbf{Doc}} & {\textbf{Comp}} & {\textbf{Comm}} 
          & {\textbf{Bug}} & {\textbf{Leg}} & {\textbf{Por}} & {\textbf{Sen}} & {\textbf{Feat}} \\
    \textbf{10+} & 69.1\% & 70.9\% & 70.9\% & \ib{72.7\%*} & 52.7\% & 58.2\% & \ib{72.7\%*} & 43.6\% & 41.8\% & 18.2\% & 38.2\% \\
    \textbf{7--10} & 69.2\% & 76.9\% & \ib{84.6\%*} & \ib{84.6\%*} & 61.5\% & 50.0\% & 65.4\% & 42.3\% & 38.5\% & 19.2\% & 13.3\% \\
    \textbf{3--6} & 100.0\% & 83.3\% & \ib{100.0\%*} & 75.0\% & 83.3\% & 66.7\% & 91.7\% & 33.3\% & 33.3\% & 0.0\% & 50.0\% \\
    \end{tabular}%
     \\
    \midrule
    19    & What tools can better support your understanding of API reviews in developer forums? \\
    &	OM =  Opinion Miner, SA = Sentiment Analyzer, OS = Opinion Summarizer, AC = API Comparator, TA = Trend Analyzer, CA = Competing APIs, OT = Other Tools 
    
    \begin{tabular}{lrrrrrrr}
          & \textbf{OM} & \textbf{SA} & \textbf{OS} & \textbf{AC} & \textbf{TA} & \textbf{CA} & \textbf{OT} \\
    \textbf{10+} & 49.1\% & 40.0\% & 38.3\% & 54.5\% & 50.9\% & \ib{63.6\%*} & 9.1\% \\
    \textbf{7--10} & 50.0\% & 30.8\% & 42.3\% & 65.4\% & \ib{69.2\%*} & \ib{69.2\%*} & 7.7\% \\
    \textbf{3--6} & 58.3\% & 33.3\% & 50.0\% & \ib{83.3\%*} & \ib{83.3\%*} & 66.7\% & 0.0\% \\
    \end{tabular}%
     \\
    \midrule
    20    & How often do you feel overwhelmed due to the abundance of opinions about an API? \\
    &	
    \begin{tabular}{lrrrr}
          & \textbf{Always} & \textbf{Sometimes} & \textbf{Rarely} & \textbf{Never} \\
    \textbf{10+} & 1.8\% & 34.5\% & \ib{40.0\%*} & 9.1\% \\
    \textbf{7--10} & 3.8\% & \ib{50.0\%*} & 26.9\% & 11.5\% \\
    \textbf{3--6} & 8.3\% & \ib{66.7\%*} & 16.7\% & 8.3\% \\
    \end{tabular}%
     \\
    \midrule
    21    & Would a summarization of opinions about APIs help you to make a better decision on which one to use? \\
    &	\it{
    \begin{inparaenum}
   \item 10+: \ib{I don't know 43.6\%*}, 
    \item 7-10: \ib{I don't know 46.2\%*}, 
    \item 3-6: \ib{Yes 58.3\%*}
    \end{inparaenum}
    }  \\
    \midrule
    22    & Opinions about an API need to be summarized because \\
    &	
    \begin{tabular}{lrrrrr}
          & \textbf{Too many posts} & \textbf{Opinion in missed posts} & \textbf{Missed contradictory opinion} & \textbf{Not enough time} & \textbf{Other reason} \\
    \textbf{10+} & 36.4\% & \ib{60.0\%*} & 54.5\% & 58.2\% & 10.9\% \\
    \textbf{7--10} & 50.0\% & \ib{65.4\%*} & 53.8\% & 61.5\% & 11.5\% \\
    \textbf{3--6} & \ib{75.0\%*} & \ib{75.0\%*} & 58.3\% & \ib{75.0\%*} & 8.3\% \\
    \end{tabular}%
     \\
    \midrule
    23    & Opinion summarization can improve the following decision making processes \\
    &	SAC = Selecting amidst choices, DAR = Determining a replacement, ISF = Improving software feature, RAF = Replacing software feature, DAN = Developing a new API, FAB = Fixing a bug, SAV = Selecting API version,
    VAS = Validating a selection
    \begin{tabular}{lrrrrrrrrr}
          & \textbf{SAC} & \textbf{DAR} & \textbf{ISF} & \textbf{RAF} & \textbf{DAN} & \textbf{VAS} & \textbf{FAB} & \textbf{SAV} & \textbf{Other} \\
    \textbf{10+} & 61.8\% & \ib{63.6\%*} & 41.8\% & 34.5\% & 40.0\% & 45.5\% & 30.9\% & 45.5\% & 5.5\% \\
    \textbf{7--10} & \ib{65.4\%*} & 61.5\% & 46.2\% & 38.5\% & 34.6\% & 46.2\% & 42.3\% & 30.8\% & 11.5\% \\
    \textbf{3--6} & \ib{83.3\%*} & 66.7\% & 25.0\% & 50.0\% & 25.0\% & 25.0\% & 33.3\% & 16.7\% & 16.7\% \\
    \end{tabular}%
    \\
    \bottomrule
    \end{tabular}%
  \label{tab:rq2-sum-by-exp}%
\end{table*}%

\subsection{Analysis by Experiences}\label{sec:demographic-experience}
\rev{In \tbls\ref{tab:rq1-sum-by-exp} and \ref{tab:rq2-sum-by-exp}, we summarize the results of closed questions from our final survey by the reported 
experiences of survey participants following the same reporting principles we discussed in \sec\ref{sec:demographic-profession}.} 

\rev{\addtocounter{o}{1}
\nd\it{\bf{Observation \arabic{o}.}} In both pilot and primary surveys, the developers with less experience show more interest to value the opinion
of other developers (Q2 in \tbl\ref{tab:rq1-sum-by-exp}). In both the surveys, the more experienced developers offer more uniform preferences towards
the different tools that could be developed to facilitate opinion analysis from developer forums (Q19 in \tbl\ref{tab:rq2-sum-by-exp}).}

\rev{\addtocounter{o}{1}
\nd\it{\bf{Observation \arabic{o}.}} The less experienced developers visit forums
more frequently (Q17 in primary survey): 75\%, 61.2\%, and 58.2\% developers with 3-6, 7-10, and 10+ years of experience, respectively visit developer forums at least two or three times a week.}


\rev{\addtocounter{o}{1}
\nd\it{\bf{Observation \arabic{o}.}}
Among the less experienced developers (3-6 years of experience), both Trend Analyzer and API Comparator were ranked over other tools 
when asked about their preference of tools to better support
their understanding of opinions about APIs (Q19 in primary survey).
The developers with 10+ years of experience were most interested to explore the Competing APIs tool. The preference towards a specific tool decreases
as developers become more experienced. For example, the developers with 10+ years of experience prefer the different tools with almost similar
preference.  The two tools (API Comparator and Competing APIs) are also ranked the highest (70.7\%)
by developers with 10+ years of experience in the pilot survey (Q7).} 

\rev{\addtocounter{o}{1}
\nd\it{\bf{Observation \arabic{o}.}} The developers with 10+ years of experience show almost equal preference to the different implicit API aspects about which they prefer to see or seek opinions. 
The less experienced developers also have more specific preference of 
API aspects about which they like to explore the opinions of other developers (Q15 in primary survey):
\begin{inparaenum}
\item More than 75\% agreement for the six API aspects by developers with 3-6 years of experience (maximum 100\% for Usability)
\item More than 75\% agreement for three API aspects by developers with 7-10 years of experience (maximum 84.6\% for both usability and documentation)
\item Maximum 72.7\% agreement for two API aspects by developers with 10+ years of experience (Documentation and Bug).
\end{inparaenum}} 

\begin{table*}[t]
  \centering
  \caption{Highlights of Findings from Primary Survey RQ1 Closed Questions by Programming Languages}
    \begin{tabular}{r|p{16cm}} \toprule
    \bf{No} & \bf{Question} \\ \midrule
    1     & Do you visit developer forums to seek info about APIs? Research Engineer\\ 
    &	\it{
    \begin{inparaenum}
    \item R: \ib{Yes 86.4\%*}, 
    \item Java: No 71.4\%, 
    \item Python: \ib{Yes 83.3\%*}
    \item Javascript: \ib{Yes 81.2\%*}
    \item C: \ib{Yes 75\%*}
    \item C++: \ib{Yes 83.3\%*}
    \item C\#: \ib{Yes 76.5\%*}
    \item Objective-C: \ib{Yes 70.0\%*}
    \item Ruby: \ib{Yes 100\%*}
    \end{inparaenum}
    }  \\
    \midrule
    3     & Do you value the opinion of other developers in the developer forums when deciding on what API to use? \\
    &	\it{
    \begin{inparaenum}
    \item R: \ib{Yes 78.9\%*} 
    \item Java: \ib{Yes 100\%*} 
    \item Python: \ib{Yes 83.3\%*}
    \item Javascript: \ib{Yes 84.6\%*}
    \item C: \ib{Yes 87.5\%*}
    \item C++: \ib{Yes 100.0\%*}
    \item C\#: \ib{Yes 92.3\%*}
    \item Objective-C: \ib{Yes 100\%*}
    \item Ruby: \ib{Yes 100\%*}
    \end{inparaenum}
    }   \\
    \midrule
    
    16    & Where do you seek help/opinions about APIs? \\
    &	 
    
      \begin{tabular}{lrrrrr}
          & \textbf{Internal Mailing list} & \textbf{IRC chats} & \textbf{Co-workers} & \textbf{Stack Overflow} & \textbf{Search/Diverse Sources} \\
    \textbf{R} & 26.3\% & 31.6\% & 63.2\% & \ib{78.9\%*} & 36.8\% \\
    \textbf{Java} & 40.0\% & 20.0\% & 80.0\% & \ib{100.0\%*} & 60.0\% \\
    \textbf{Python} & 16.7\% & 16.7\% & 66.7\% & \ib{83.3\%*} & 50.0\% \\
    \textbf{Javascript} & 30.8\% & 30.8\% & 53.8\% & \ib{84.6\%*} & 30.8\% \\
    \textbf{C} & 25.0\% & 18.8\% & 56.2\% & \ib{87.5\%*} & 43.8\% \\
    \textbf{C++} & 20.0\% & 20.0\% & 60.0\% & \ib{100.0\%*} & 50.0\% \\
    \textbf{C\#} & 30.8\% & 7.7\% & \ib{92.3\%*} & \ib{92.3\%*} & 30.8\% \\
    \textbf{Objective-C} & 71.4\% & 14.3\% & 42.9\% & \ib{100.0\%*} & 57.1\% \\
    \textbf{Ruby} & 25.0\% & 25.0\% & \ib{100.0\%*} & \ib{100.0\%*} & 25.0\% \\
    \end{tabular}%
    \\
    \midrule
    17    & How often do you refer to online forums (e.g., Stack Overflow) to get information about APIs? \\
    &	
    \begin{tabular}{lrrrrr}
          & \textbf{Every day} & \textbf{Two or three times a week} & \textbf{Once a week} & \textbf{One a month} & \textbf{Never} \\
    \textbf{R} & \ib{31.6\%*} & 15.8\% & 21.1\% & 10.5\% & 21.0\% \\
    \textbf{Java} & 40.0\% & 20.0\% & 0.0\% & 40.0\% & 0.0\% \\
    \textbf{Python} & 16.7\% & \ib{66.7\%*} & 0.0\% & 0.0\% & 16.6\% \\
    \textbf{Javascript} & \ib{30.8\%*} & \ib{30.8\%*} & 7.7\% & 15.4\% & 15.3\% \\
    \textbf{C} & 25.0\% & \ib{31.2\%*} & 18.8\% & 12.5\% & 12.5\% \\
    \textbf{C++} & 30.0\% & \ib{40.0\%*} & 10.0\% & 20.0\% & 0.0\% \\
    \textbf{C\#} & 38.5\% & 30.8\% & 15.4\% & 7.7\% & 7.6\% \\
    \textbf{Objective-C} & \ib{71.4\%*} & 0.0\% & 14.3\% & 14.3\% & 0.0\% \\
    \textbf{Ruby} & 0.0\% & 50.0\% & 0.0\% & 50.0\% & 0.0\% \\
    \end{tabular}%
      \\
    \midrule
    18    & When do you seek opinions about APIs?  \\
    &	 
    SAC = Selecting amidst choices, DAR = Determining a replacement, ISF = Improving software feature, RAF = Replacing software feature, DAN = Developing a new API, FAB = Fixing a bug, SAV = Selecting API version,
    VAS = Validating a selection
    \begin{tabular}{lrrrrrrrrr}
          & \textbf{SAC} & \textbf{DAR} & \textbf{ISF} & \textbf{RAF} & \textbf{DAN} & \textbf{VAS} & \textbf{FAB} & \textbf{SAV} & \textbf{Other} \\
    \textbf{R} & 68.4\% & \ib{73.7\%*} & 47.4\% & 31.6\% & 47.4\% & 42.1\% & 52.6\% & 31.6\% & 15.8\% \\
    \textbf{Java} & \ib{100.0\%*} & 80.0\% & 60.0\% & 20.0\% & 60.0\% & 60.0\% & 100.0\% & 20.0\% & 0.0\% \\
    \textbf{Python} & \ib{66.7\%*} & 66.7\% & \ib{66.7\%*} & 33.3\% & 50.0\% & 16.7\% & \ib{66.7\%*} & 33.3\% & 16.7\% \\
    \textbf{Javascript} & 61.5\% & \ib{76.9\%*} & \ib{76.9\%*} & 38.5\% & 30.8\% & 15.4\% & \ib{76.9\%*} & 15.4\% & 15.4\% \\
    \textbf{C} & \ib{75.0\%*} & 56.2\% & 56.2\% & 56.2\% & 56.2\% & 18.8\% & 50.0\% & 25.0\% & 12.5\% \\
    \textbf{C++} & \ib{100.0\%*} & \ib{100.0\%*} & 70.0\% & 70.0\% & 80.0\% & 60.0\% & 40.0\% & 20.0\% & 20.0\% \\
    \textbf{C\#} & \ib{92.3\%*} & 61.5\% & 69.2\% & 46.2\% & 53.8\% & 53.8\% & 69.2\% & 23.1\% & 7.7\% \\
    \textbf{Objective-C} & \ib{85.7\%*} & 71.4\% & 71.4\% & 42.9\% & 28.6\% & 28.6\% & 71.4\% & 42.9\% & 14.3\% \\
    \textbf{Ruby} & \ib{75.0\%*} & 75.0\% & 50.0\% & 50.0\% & 50.0\% & 0.0\% & 50.0\% & 0.0\% & 25.0\% \\
    \end{tabular}%
    \\
    \bottomrule
    \end{tabular}%
  \label{tab:rq1-sum-by-lang}%
\end{table*}%

\begin{table*}[tbp]
\vspace{-5mm}
  \centering
  \caption{Highlights of Findings from Primary Survey RQ2 Closed Questions  by Programming Languages}
  \vspace{-2mm}
    \begin{tabular}{r|p{16cm}} \toprule
    \bf{No} & \bf{Question} \\ \midrule
    8     & Do you rely on tools to help you understand opinions about APIs in online forum discussions? \\
    &	\it{
	\begin{inparaenum}
    \item R: \ib{No 57.9\%*}
    \item Java: \ib{No 100\%*}
    \item Python: \ib{No 100\%*}
    \item Javascript: \ib{No 76.9\%*}
    \item C: \ib{No 75\%*}
    \item C++: \ib{No 90.0\%*}
    \item C\#: \ib{No 84.6\%*}
    \item Objective-C: \ib{No 71.4\%*}
    \item Ruby: \ib{No 100\%*}    
    \end{inparaenum}    
    }  \\
    \midrule
    9     & If you don't use a tool currently to explore the diverse opinions about APIs in developer forums, do you believe there is a need of such a tool to help you find the right viewpoints about an API quickly? \\
    &	\it{
	\begin{inparaenum}
    \item R: No 26.3\%, 
    \item Java: don't know 60.0\%, 
    \item Python: \ib{don't know 83.3\%*}
    \item Javascript: \ib{don't know 69.2\%*}
    \item C: \ib{No 50\%, don't know 50\%}
    \item C++: \ib{don't know 80.0\%*}
    \item C\#: don't know 53.8\%*
    \item Objective-C: \ib{don't know 57.1\%*}
    \item Ruby: don't know 50\%, No 50\%    
    \end{inparaenum}    
    }  \\
    \midrule
    15    & An opinion is important if it contains discussion about the following API aspects?  \\
    &	
    Per = Performance, Sec = Security, Use = Usability, Doc = Documentation, Comp = Compatibility, Comm = Community, Leg = Legal, Port = Portability, Sen = Only Sentiment, Feat = General Features
    
    \begin{tabular}{lrrrrrrrrrrr}
          & \textbf{Per} & \textbf{Sec} & \textbf{Use} & \textbf{Doc} & \textbf{Comp} & \textbf{Comm} & \textbf{Bug} & \textbf{Leg} & \textbf{Por} & \textbf{Sen} & \textbf{Feat} \\
    \textbf{R} & 68.4\% & 68.4\% & \ib{78.9\%*} & 73.7\% & 42.1\% & 52.6\% & 78.9\% & 42.1\% & 36.8\% & 21.1\% & 4.8\% \\
    \textbf{Java} & \ib{100.0\%*} & \ib{100.0\%*} & \ib{100.0\%*} & 80.0\% & \ib{100.0\%*} & 60.0\% & 80.0\% & 60.0\% & 60.0\% & 0.0\% & 60.0\% \\
    \textbf{Python} & 66.7\% & \ib{83.3\%*} & \ib{83.3\%*} & 66.7\% & 66.7\% & 50.0\% & 66.7\% & 50.0\% & 33.3\% & 16.7\% & 16.7\% \\
    \textbf{Javascript} & \ib{76.9\%*} & 69.2\% & 69.2\% & 69.2\% & 61.5\% & 46.2\% & 53.8\% & 30.8\% & 23.1\% & 7.7\% & 46.2\% \\
    \textbf{C} & 56.2\% & 62.5\% & 56.2\% & \ib{68.8\%*} & 62.5\% & 56.2\% & 68.8\% & 43.8\% & 50.0\% & 0.0\% & 37.5\% \\
    \textbf{C++} & 80.0\% & 70.0\% & 90.0\% & \ib{100.0\%*} & 70.0\% & 60.0\% & 80.0\% & 50.0\% & 70.0\% & 40.0\% & 50.0\% \\
    \textbf{C\#} & 84.6\% & \ib{92.3\%*} & \ib{92.3\%*} & 84.6\% & 61.5\% & 61.5\% & 84.6\% & 38.5\% & 30.8\% & 30.8\% & 76.9\% \\
    \textbf{Objective-C} & 71.4\% & 71.4\% & \ib{85.7\%*} & \ib{85.7\%*} & 71.4\% & 71.4\% & 85.7\% & 57.1\% & 42.9\% & 14.3\% & 28.6\% \\
    \textbf{Ruby} & \ib{75.0\%*} & \ib{75.0\%*} & \ib{75.0\%*} & 50.0\% & 0.0\% & \ib{75.0\%*} & 50.0\% & 0.0\% & 0.0\% & 0.0\% & 25.0\% \\
    \end{tabular}%
     \\
    \midrule
    19    & What tools can better support your understanding of API reviews in developer forums? \\
    &	OM =  Opinion Miner, SA = Sentiment Analyzer, OS = Opinion Summarizer, AC = API Comparator, TA = Trend Analyzer, CA = Competing APIs, OT = Other Tools 
    
    \begin{tabular}{lrrrrrrr}
          & \textbf{OM} & \textbf{SA} & \textbf{OS} & \textbf{AC} & \textbf{TA} & \textbf{CA} & \textbf{OT} \\
    \textbf{R} & 31.6\% & 10.5\% & 31.6\% & 36.8\% & 47.4\% & \ib{52.6\%*} & 0.0\% \\
    \textbf{Java} & 60.0\% & 40.0\% & 60.0\% & \ib{80.0\%*} & \ib{80.0\%*} & \ib{80.0\%*} & 20.0\% \\
    \textbf{Python} & \ib{66.7\%*} & 16.7\% & 16.7\% & 50.0\% & 33.3\% & 50.0\% & 0.0\% \\
    \textbf{Javascript} & 61.5\% & 38.5\% & 46.2\% & 53.8\% & \ib{69.2\%*} & \ib{69.2\%*} & 7.7\% \\
    \textbf{C} & 37.5\% & 37.7\% & 18.8\% & 56.2\% & 43.8\% & \ib{62.5\%*} & 12.5\% \\
    \textbf{C++} & 70.0\% & 50.0\% & 60.0\% & \ib{80.0\%*} & 70.0\% & \ib{80.0\%*} & 0.0\% \\
    \textbf{C\#} & 61.5\% & 69.2\% & 61.5\% & 69.2\% & \ib{76.9\%*} & 61.5\% & 15.4\% \\
    \textbf{Objective-C} & 42.9\% & 57.1\% & 57.1\% & \ib{100.0\%*} & 85.7\% & 85.7\% & 14.3\% \\
    \textbf{Ruby} & 50.0\% & 0.0\% & 25.0\% & \ib{75.0\%} & 50.0\% & \ib{75.0\%} & 0.0\% \\
    \end{tabular}%
     \\
    \midrule
    20    & How often do you feel overwhelmed due to the abundance of opinions about an API? \\
    &	
    \begin{tabular}{lrrrr}
          & \textbf{Always} & \textbf{Sometimes} & \textbf{Rarely} & \textbf{Never} \\
    \textbf{R} & 5.3\% & \ib{31.6\%} & 31.6\% & 10.5\% \\
    \textbf{Java} & 20.0\% & \ib{60.0\%} & 20.0\% & 0.0\% \\
    \textbf{Python} & 0.0\% & \ib{50.0\%} & 33.3\% & 16.7\% \\
    \textbf{Javascript} & 15.3\% & \ib{46.2\%} & 30.8\% & 7.7\% \\
    \textbf{C} & 12.5\% & 12.5\% & \ib{56.2\%} & 18.8\% \\
    \textbf{C++} & 0.0\% & \ib{70.0\%*} & 30.0\% & 0.0\% \\
    \textbf{C\#} & 7.7\% & \ib{53.8\%} & 30.8\% & 7.7\% \\
    \textbf{Objective-C} & 0.0\% & \ib{57.1\%} & 14.3\% & 28.6\% \\
    \textbf{Ruby} & 0.0\% & \ib{50.0\%} & 25.0\% & 25.0\% \\
    \end{tabular}%
     \\
    \midrule
    21    & Would a summarization of opinions about APIs help you to make a better decision on which one to use? \\
    &	\it{
    \begin{inparaenum}
    \item R: \ib{I don't know 36.8\%}
    \item Java: \ib{I don't know 40.0\%, Yes 40.0\%} 
    \item Python: \ib{I don't know 50.0\%}
    \item Javascript: \ib{I don't know 46.2\%}
    \item C: \ib{No 50.0\%*}
    \item C++: \ib{Yes 70.0\%*}
    \item C\#: \ib{I don't know 46.2\%*, Yes 46.2\%}
    \item Objective-C: \ib{Yes 57.1\%}
    \item Ruby: \ib{I don't know 75\%}    
    \end{inparaenum}        }  \\
    \midrule
    22    & Opinions about an API need to be summarized because \\
    &	
    \begin{tabular}{lrrrrr}
          & \textbf{Too many posts} & \textbf{Opinion in missed posts} & \textbf{Missed contradictory opinion} & \textbf{Not enough time} & \textbf{Other reason} \\
    \textbf{R} & 36.8\% & 57.9\% & 42.1\% & \ib{63.2\%*} & 5.3\% \\
    \textbf{Java} & 80.0\% & \ib{100.0\%*} & \ib{100.0\%*} & 60.0\% & 20.0\% \\
    \textbf{Python} & 50.0\% & \ib{66.7\%*} & 50.0\% & 50.0\% & 0.0\% \\
    \textbf{Javascript} & 46.2\% & \ib{61.5\%*} & 46.2\% & \ib{61.5\%*} & 15.4\% \\
    \textbf{C} & 25.0\% & 56.2\% & 50.0\% & 56.2\% & 6.2\% \\
    \textbf{C++} & 50.0\% & 60.0\% & \ib{70.0\%*} & \ib{70.0\%*} & 20.0\% \\
    \textbf{C\#} & 53.8\% & \ib{61.5\%*} & \ib{61.5\%*} & 61.5\% & 15.4\% \\
    \textbf{Objective-C} & 71.4\% & \ib{85.7\%*} & 57.1\% & \ib{85.7\%*}& 14.3\% \\
    \textbf{Ruby} & 25.0\% & \ib{50.0\%} & \ib{50.0\%} & 25.0\% & 0.0\% \\
    \end{tabular}%
     \\
    \midrule
    23    & Opinion summarization can improve the following decision making processes \\
    &	SAC = Selecting amidst choices, DAR = Determining a replacement, ISF = Improving software feature, RAF = Replacing software feature, DAN = Developing a new API, FAB = Fixing a bug, SAV = Selecting API version,
    VAS = Validating a selection
    \begin{tabular}{lrrrrrrrrr}
          & \textbf{SAC} & \textbf{DAR} & \textbf{ISF} & \textbf{RAF} & \textbf{DAN} & \textbf{VAS} & \textbf{FAB} & \textbf{SAV} & \textbf{Other} \\
    \textbf{R} & 36.8\% &\ib{42.1\%*} & 15.8\% & 26.3\% & 21.1\% & 21.1\% & 21.1\% & 26.3\% & 0.0\% \\
    \textbf{Java} & \ib{80.0\%*} & 60.0\% & 0.0\% & 20.0\% & 40.0\% & 60.0\% & 80.0\% & 20.0\% & 20.0\% \\
    \textbf{Python} & \ib{66.7\%*} & \ib{66.7\%*} & 50.0\% & 50.0\% & \ib{66.7\%*} & 50.0\% & 50.0\% & 33.3\% & 0.0\% \\
    \textbf{Javascript} & \ib{69.2\%*} & 84.6\% & \ib{69.2\%*} & 46.2\% & 23.1\% & 46.2\% & 38.5\% & 30.8\% & 7.7\% \\
    \textbf{C} & 50.0\% & \ib{56.2\%*} & 50.0\% & 31.2\% & 25.0\% & 25.0\% & 31.2\% & 25.0\% & 0.0\% \\
    \textbf{C++} & \ib{90.0\%*} & 80.0\% & 40.0\% & 50.0\% & 50.0\% & 80.0\% & 20.0\% & 10.0\% & 20.0\% \\
    \textbf{C\#} & \ib{84.6\%*} & 61.5\% & 53.8\% & 46.2\% & 53.8\% & 61.5\% & 46.2\% & 30.8\% & 23.1\% \\
    \textbf{Objective-C} & \ib{85.7\%*} & \ib{85.7\%*} & 42.9\% & 42.9\% & 57.1\% & 57.1\% & 28.6\% & 28.6\% & 14.3\% \\
    \textbf{Ruby} & \ib{75.0\%} & 50.0\% & 25.0\% & 25.0\% & 25.0\% & 0.0\% & 25.0\% & 25.0\% & 0.0\% \\
    \end{tabular}%
    \\
    \bottomrule
    \end{tabular}%
  \label{tab:rq2-sum-by-lang}%
\end{table*}%

\subsection{Analysis by Programming Languages}\label{sec:demographic-programming-langs}
\rev{In \tbls\ref{tab:rq1-sum-by-lang} and \ref{tab:rq2-sum-by-lang}, we summarize the results of closed questions from our final survey by the nine 
targeted programming languages from where our survey sample was collected. We follow the same reporting principles that we discussed in \sec\ref{sec:demographic-profession}.} 

\rev{\addtocounter{o}{1}
\nd\it{\bf{Observation \arabic{o}.}} Among the nine programming languages that we targeted in our primary survey to sample the participants of our primary survey,
at least 70\% respondents from each language reported that they visit developer forums to seek information about APIs.
The opinions posted about APIs in the forums are valued by those developers (100\% of all Java developers, followed by at least
78\% of the respondents from other languages).} 

\rev{\addtocounter{o}{1}
\nd\it{\bf{Observation \arabic{o}.}} All the Java developers also report that they do not have any tools that
they can currently use to analyze those opinions. The lack of such tool support is also prevalent among developers
across other languages.} 

\rev{\addtocounter{o}{1}
\nd\it{\bf{Observation \arabic{o}.}} The developers across all the programming languages (except Python and C\#) preferred most the tools that can offer summarized comparisons
between APIs or show list of competing APIs.
For python, the most preferred tool was an opinion miner and for C\# it was a trend analyzer.} 

\rev{\addtocounter{o}{1}
\nd\it{\bf{Observation \arabic{o}.}} While developers prefer to explore opinions about 
diverse API aspects, the usability aspect was ranked the highest among the 
list of API aspects across all the languages except C, C++ and Javascript. For C and C++, the most pressing aspects are
documentation and for Javascript it was performance.} 

%
%

%% file: research-journey.tex
\begin{table*}[htbp]
  \centering
  \caption{The major findings from the survey results. CID = ID for Core Finding (to be used for reference in the paper)}
    \begin{tabular}{cl|p{11cm}|r}\toprule
    \multicolumn{1}{l}{\textbf{RQ ID}} & \textbf{Theme} & \textbf{Core Findings} & \multicolumn{1}{l}{\textbf{CID}} \\
    \midrule
    \multirow{2}[0]{*}{1.1} & \multirow{2}[0]{*}{Needs for seeking opinions about APIs} & Developers seek opinions to support diverse development needs & 1 \\
    \cmidrule{3-4}
          &       & Developers consider the combination of code examples and opinions about an API as a source of API documentation & 2 \\
    \cmidrule{3-4}
          & & Developers currently use search engines or Stack Overflow search which are not designed to seek and analyze opinions about APIs & 3 \\
          \midrule
    \multirow{2}[0]{*}{1.2} & \multirow{2}[0]{*}{Needs for opinion quality assessment} & Developers face challenges to determine the quality of provided opinions & 4 \\
    \cmidrule{3-4}
          &       & Developers wish opinions to be trustworthy, recent and combined with real-world facts & 5 \\
          \midrule
    \multirow{2}[0]{*}{2.1} & \multirow{2}[0]{*}{Needs for tool support} & Developers wish for diverse tool supports to analyze opinions, e.g., API comparator, trend analyzer, opinion summarizer, etc. & 6\\
          \midrule
    \multirow{3}[0]{*}{2.2} & \multirow{3}[0]{*}{Needs for summarization} & Developers are frequently overwhelmed with the huge volume of opinions about APIs in forums & 7 \\
    \cmidrule{3-4}
          &       & Developers wish to see opinions about APIs by diverse API aspects (e.g., performance) & 8 \\
    \cmidrule{3-4}
          &       & Developers wish to get quick, but complete and concise insights about API usage by analyzing both the provided code examples and reviews & 9 \\
          \bottomrule
    \end{tabular}%
  \label{tbl:core-findings}%
\end{table*}%
\begin{table*}[htbp]
  \centering
  \caption{The actionable findings used to elicit requirements to design an API opinion and usage summarization engine.
  R = Requirement, QID = Question ID from primary survey (Questions from Pilot survey are prefixed with a P). CID = ID of the Core findings from \tbl\ref{tbl:core-findings}}
    \begin{tabular}{r|p{7.5cm}|p{7.5cm}|p{1cm}}\toprule
    \textbf{CID} & \textbf{Actionable Findings/Elicited Design Requirements}  & \textbf{Tool Feature}& \textbf{QID} \\
    \midrule
    \multicolumn{4}{c}{\bf{\it{R1. Needs for Mining of Opinions and Usage Scenarios about APIs}}}\\
    \cmidrule{2-3}
    {1,3,6} & Opinions about an API need to be collected to assist in diverse development needs
     & Develop techniques to automatically mine opinions about APIs  & {18}\\
    {1-3} & Code examples with related opinions need to be collected to assist in API documentation and to show real-world usage experience
     & Develop techniques to automatically mine usage scenarios about APIs & 4, 6, 7, 12, 13\\

    \cmidrule{2-3}
    \multicolumn{4}{c}{\bf{\it{R2. Needs for Summarization of Opinions about APIs}}}\\
    \cmidrule{2-3}
    {7,8} & Categorize opinions about an API by aspects, e.g., performance & Aspect based summarization of opinions  & 11,15\\
    {6} & Show progression of popularity (i.e., trends) about an API & Statistical Summarization & {19}\\
    6,9   & Need a dashboard to explore and compare competing APIs & (1) Rank competing APIs by popularity.
    (2) Rank competing APIs by aspects (e.g., popularity by performance)
    (3) Show APIs co-reviewed in the same posts& {13,19, 23}\\
    7, 8 & Show topics discussed in opinions and provide short summaries as paragraphs & Summarization by Topics, etc. & P17 \\
    \cmidrule{2-3}
    \multicolumn{4}{c}{\bf{\it{R3. Needs for Summarization of Usage Scenarios about APIs}}}\\
    \cmidrule{2-3}

    1,9   & Show situationally relevant usage examples together & Concept-Based summarization  & 12,24\\
    1,9   & Integrate usage scenarios into API documentation & Type-Based summarization & {12}\\
    6,7   & Provide situationally relevant information to an API usage  & (1) Show APIs used together, (2) Show API elements used together  & 19,7\\
    \cmidrule{2-3}
    \multicolumn{4}{c}{\bf{\it{R4. Needs for Quality Assurance about API Opinions}}}\\
    \cmidrule{2-3}
    3,4   & Provide situational relevance to an opinion by showing the context   & Trace opinions to Stack Overflow posts  & 6,7\\
    3,4   & Show most recent opinions before the older opinions &  Rank opinions and usage scenarios by recency  & {7}\\
    3,4   & Show ratings to a code example to inform developers of its quality &  Show star ratings for each code example based on the reactions of developers
    towards the post containing the example  & {7}\\
    {6} & Show contrastive viewpoints about API features & Contrastive Viewpoint Summarization  & {12}\\
    \cmidrule{2-3}
    \multicolumn{4}{c}{\bf{\it{R5. Needs for a portal for APIs}}}\\
    \cmidrule{2-3}
    1-9   & Need a dedicated portal for APIs to explore opinions and usage about APIs
    & A search and opinion summarization engine for APIs to explore and compare APIs  & 12\\
    \bottomrule
    \end{tabular}%
  \label{tbl:actionable-findings}%
\end{table*}%
\begin{figure*}
\centering
\hspace*{-.5cm}%
\includegraphics[scale=.70]{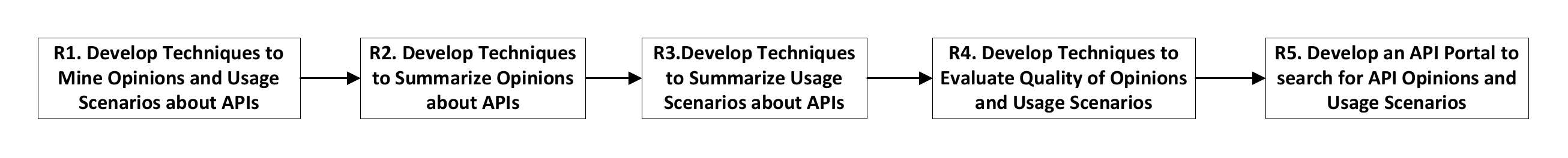}
	\caption{The major research steps undertaken to incorporate the requirements from the two surveys into our proof of concept tool, Opiner}
 \label{fig:researchJourneyToolDevelopment}
\end{figure*}
\section{Research Journey}\label{sec:research-journey}
We noted in \sec\ref{sec:introduction} (\fig\ref{fig:researchJourneySteps}) that the findings from the surveys led us to develop techniques and tools
to assist developers in their analysis of opinions and usage about APIs from Stack Overflow. In this section, we break down the journey into two major phases
that we undertook after the surveys:
\begin{enumerate}
  \item Identification of core and actionable insights from the survey results that can be used to guide the design of tools to assist developers in their
  analysis of opinions and usage of APIs from developer forums (see \sec\ref{sec:core-findings}).
  \item Development and evaluation of techniques and our tool (Opiner) based on the insights (see \sec\ref{sec:implications}).
\end{enumerate}

\subsection{Core and Actionable Findings}\label{sec:core-findings}
In \tbl\ref{tbl:core-findings}, we summarize the core findings obtained from the surveys. Each finding
is provided a unique ID (denoted by CID). The first two columns show the research
questions, the third column presents the core findings. In \tbl\ref{tbl:actionable-findings}, we
identify requirements for future tool designs to support each core finding. Each requirement is mapped to the CIDs
from  \tbl\ref{tbl:core-findings} (first column). The second column shows the requirements, i.e., actionable
findings. We cluster the requirements into five categories (R1-R5). 
The third column lists the questions from
the primary survey that we used to identify those requirements. The last column presents the features implemented in
our tool Opiner to address those requirements. In \fig\ref{fig:researchJourneyToolDevelopment}, 
we show how requirements are implemented into our tool. We now discuss the findings below along with the elicited requirements.

\subsubsection{Needs for Mining API Opinions \& Usage (R1)}
The developers in our surveys seek and analyze opinions about APIs to support diverse development needs and
they consider the combination of opinions and code examples in the forum posts as a form of API documentation (CID 1, 2 in \tbl\ref{tbl:core-findings}).
To search for opinions and usage discussions about APIs, developers use search engines
or the tags in Stack Overflow. They raise the concern that this approach of using search engines (e.g., find sentiments about an API)
can be sub-optimal when they only explore the top results (CID 3 in \tbl\ref{tbl:core-findings}).
The developers reported to look for sentiments and situational relevance in the search results (Q5). This exploration can be challenging, because
the results may not be \begin{inparaenum}[(1)]
\item \it{situationally relevant}, such as the API about which they would like to see opinions about may not be present in the search result. It could also be
that their development task is not properly supported by the code example found in the search result.
\item \it{trustworthy}, such as, sentiment towards an API found in the search result may not represent the overall sentiments expressed towards it (e.g., the
opinion in the search results may be biased).
\item \it{recent}, such as the opinion and code example may not be the most recent and thus the solution may not be applicable to the most recent version of the API.
\end{inparaenum}  In the absence of a dedicated engine for APIs, developers opt for the reformulation of their search query by
modifying search keywords. This approach is considered as challenging and time consuming. The developers wished for better search support to address their needs.

Intuitively, it is easier to find opinions/solutions for an API with regards to specific task (or situation), if they are not \it{scattered} in
millions of posts in the forums. As a first step towards facilitating such search, it is thus necessary to collect opinions and code examples about APIs from the millions of posts, where
their usage is discussed. During our tool design to assist developers in their exploration of opinions and usage about APIs from developer forums,
we formulate the following two requirements:
\begin{enumerate}
  \item \bf{Opinions about an API need to be collected}. 
  \rev{Sentiment and emotion mining in software engineering has so far focused on the usefulness of cross-domain sentiment detection tools for software engineering, 
  the development of sentiment detection tools for the domain of software engineering, and the relationship between team productivity with the sentiments expressed (see \sec\ref{sec:related-work-sentiment}).  
  We are aware of no technique that can mine opinions associated to an API from developer forums.} 
  We developed a framework to automatically mine opinions about APIs from developer forums.
  The framework currently supports the following major features:
  \begin{inparaenum}
  \item Detection of API names in the forum texts
  \item Detection of opinionated sentences in the forum texts, and
  \item Association of opinionated sentences to APIs.
  \end{inparaenum}  A detailed description and evaluation of the framework is the subject of our paper~\cite{Uddin-OpinionValue-TSE2018}.
  \item \bf{Code examples with reactions need to be collected together}. 
  \rev{A number of recent research efforts have been devoted to mine code examples about APIs from forums, such as detecting all the APIs used in a code example in a forum post~\cite{Subramanian-LiveAPIDocumentation-ICSE2014,Dagenais-RecoDocPaper-ICSE2012a}, etc. 
  We are aware of no technique that mines both a code example and the reactions towards the code example for an API from forum post.}   
  We developed another framework to automatically mine usage scenarios about APIs from developer forums. Each
  usage scenario about an API consists of three major components: \begin{inparaenum}
  \item The code example
  \item A short description in natural language about the code example, and
  \item The reactions of developers towards the forum post from where the code example is found.
  \end{inparaenum} A detailed description and evaluation of the framework is the subject of our technical report~\cite{Uddin-MiningUsageScenarios-ASE2018}.
\end{enumerate}

\subsubsection{Needs for Summarization of API Opinions (R2)}
In our primary survey, 89.2\% of the developers mentioned that they are overwhelmed by the huge volume of opinions posted
about APIs in forums (CID 7). While the
majority of the developers agreed that opinion summaries can help them evaluating APIs, the following summarization types were rated as (could be) useful (CID 8):
\begin{inparaenum}
\item categorization of opinions by API aspects,
\item show trends of API popularity,
\item show contrastive viewpoints,
\item show dashboards to compare APIs,
\item opinions summarized as topics; and
\item most importantly opinions summarized into a paragraph.
\end{inparaenum}
It would be interesting to further investigate the relative benefit of each
summarization type and compare such findings with other domains. For example,
Lerman et al.~\cite{Lerman-SentimentSummarizerEvaluation-EACL2009} found no clear winner for consumer product
summarization, while aspect-based summarization is predominantly present in camera and phone
reviews~\cite{liu-sentimentanalysis-handbookchapter-2010}, and topic-based
summarization has been studied in many domains (e.g., identification of popular topics in bug report detection~\cite{Nguyen-DuplicateBugReport-ASE2012}, software
traceability~\cite{AsuncionTylor-TopicModelingTraceabilityWithLDA-ICSE2010a},
etc.).

As a first step towards facilitating the summarization of opinions about APIs, we formulate the following design requirements:
\begin{enumerate}
  \item \bf{Categorize opinions by aspect.} The developers indicated that they would like to seek opinions about diverse API aspects, such as
  performance, usability, security, etc. We develop machine learning techniques to automatically categorize each opinionated sentences associated to an API into 11 different
  categories: \begin{inparaenum}[(i)]
  \item Performance
  \item Usability
  \item Security
  \item Compatibility
  \item Portability
  \item Legal
  \item Bug
  \item Community
  \item Documentation
  \item Other general features
  \item Only sentiment
  \end{inparaenum}. The first nine categories are frequently asked for in the responses of our surveys (e.g., Q11, Q15 in our primary survey).
  \item \bf{Statistical summarization.} Developers asked for automated analysis to see
  trends of usage of an API based on sentiment analysis.
  We have developed techniques to create time series of positivity and negativity towards an API by analyzing the sentiments expressed about
  the API in the different forum posts.
  \item \bf{API Comparator.} The most-asked tool was a dashboard to compare APIs and to see competing APIs given an API. We developed two techniques to facilitate
  comparison between APIs. \begin{inparaenum}
  \item We rank APIs by aspect to offer recommendation, such as based on the aspect performance the API Jackson is the most popular for JSON-based tasks in Java, but for
  usability aspect it is the Google GSON API. 
  \item We further apply collocation algorithm on the
  forum posts for each API mention and show which other APIs were positively or
  negative reviewed in the same forum post. This
  analysis can reveal other similar APIs to developers if they are not satisfied with their initially selected API.
  \end{inparaenum}
  \item \bf{Summarize opinions by topics and paragraphs.} In our pilot survey, we asked developers of their preference to summarize opinions by topics and in short
  paragraphs. The two options were ranked lower than the aspect-based categorization, but were still considered as useful. We investigated the following existing
  algorithms to summarize opinions along the two options. Each algorithm takes as input all the opinionated sentences (one bucket for positive and another for negative sentences).
  \begin{inparaenum}
  \item We applied the widely used topic modeling algorithm, LDA (Latent Dirichlet Allocation)~\cite{Blei-LDA-JournalMachineLearning2003} to find topics in the input and to
  cluster the opinionated sentences by topics.
  \item We applied four algorithms based on Extractive and Abstractive summarization of natural language texts to produce a summary of the input as a paragraph.
  \end{inparaenum}
\end{enumerate} The detail of the summarization algorithms and their evaluation was the subject of our recently published paper~\cite{Uddin-OpinerReviewAlgo-ASE2017}.

\subsubsection{Needs for API Usage Summarization (R3)}
Developers mentioned that opinions about
different API elements can help them better understand the benefits and shortcomings of the API.
In our primary survey, developers mentioned that they consider the combination of opinions and code examples posted in Stack Overflow as a form a API
documentation. Motivated by this finding, we developed a framework to mine usage scenarios about APIs (R1), where each usage scenario of an API consists of a code
example and the reactions (i.e., positive and negative opinions) towards the code example. In our survey, developers mentioned several quality criteria
for such usage scenarios while developing an API documentation based on the scenarios, such as offering clues to determine situational relevance of the usage,
clarity and correctness of the usage, etc. Indeed, summaries extracted
from Stack Overflow can be effective about the purpose and use of API
elements (classes, methods, etc.) by building opinion summarizer
tools that leverage informal documentation. This motivates
future extension of the recent research on augmenting insights about APIs from
the forums to the formal API
documentation~\cite{Treude-APIInsight-ICSE2016}. Such tools can be
integrated within IDEs to assist
developers during their APIs-based tasks.

As a first step towards producing summaries of usage scenarios of an API as a form of API documentation, we formulated
the following tool design requirements:
\begin{enumerate}
  \item \bf{Show situationally relevant usage scenarios together.} We developed an algorithm to find code examples of an API that are \it{conceptually similar}, i.e.,
  the tasks supported by the code examples may be closely related to each other. One such example would be, while using the Apache HttpClient API, developers
  first establish a connection to an HTTP server and then they send or receive messages using the connection~\cite{Uddin-TemporalApiUsage-ICSE2012}.
  \item \bf{Integrate usage scenarios into API documentation.} In our previous study, we found that
  the official API documentation can be often obsolete, incomplete, or outdated~\cite{Uddin-HowAPIDocumentationFails-IEEESW2015}. In our primary survey, developers
  asked for a unified documentation by combining API usage scenarios from developer forums with the official API documentation. One such integration would be
  the Live API documentation, as proposed by Subramanian et al.~\cite{Subramanian-LiveAPIDocumentation-ICSE2014}, i.e., link code examples from Stack Overflow into
  the Javadoc of each API class. We developed algorithms to produce type-based usage summaries of an API. In a type-based summary of an API, the usage scenarios
  of each type (e.g., a class) of the API is grouped and further summarized into situationally relevant clusters.
  \item \bf{Provide situationally relevant information to an API usage.} A development task often requires the usage of more than one API. While the above two summaries are
  focused on producing summaries for an API, we still need information on whether and how other APIs are co-used with our API of interest in the usage scenarios. We used collocation
  algorithms to find APIs that are co-used together, and other API types that are co-used together with an API type of our interest.
\end{enumerate} The development and evaluation of the usage summarization algorithm is the subject of our recent publication~\cite{Uddin-UsageSummarization-TSE2018}.

\subsubsection{Needs for Opinion Quality Analysis (R4)}
In our surveys,
the developers expressed their concerns about the trustworthiness of the
provided opinions, in particular those that contain strong bias.
Since by definition an ``opinion'' is  a ``personal view,
belief, judgment, attitude''~\cite{opinion}, all opinions are biased.
However, developers associate \it{biased} opinions with noise
defining them as the ones not being supported by facts (e.g., links to the
documentation, code snippets, appropriate rationale). While the detection of
spam (e.g., intentionally biased opinion) is an active research
area~\cite{liu-sentimentanalysis-handbookchapter-2010}, it was notable that
developers are mainly concerned about the \it{unintentional} bias that they
associate with the user experience. Developers are also concerned
about the overall writing quality.

As a first step towards assisting developers to analyze the quality of the provided opinions, we formulated the following design
requirements:
\begin{enumerate}
  \item   \bf{Contrastive viewpoint summarization.} We implement the contrastive opinion clustering algorithm proposed by Kim and Zhai~\cite{KimZhai-ContrastiveOpinionSummary-CIKM2009}
  to find pairs of opinions that offer opposing views about an API feature.
  \item \bf{Ranking by Recency.} Developers in our surveys were explicitly asked about the necessity for finding more recent opinions about APIs. The reason is that API versions evolve quickly and
  some old features can become obsolete or changed. We rank the opinionated sentences of an API by recency, i.e., the most recent opinion is placed at the top.
  \item \bf{Tracing Opinions to Posts.} Developers in our surveys highlight the necessity of investigating the context of a provided opinion, such as the features about which
  the opinion is provided, the particular configuration parameters used in the usage example, etc. To enable such exploration, we link each mined opinionated sentence about an
  API to the specific forum post from where the opinion was mined.
\end{enumerate} A number of additional research avenues can
be explored in this direction, such as, the design of a theoretical framework to
define the quality of the opinions about APIs, the development of a
new breed of recommender systems that can potentially warn users of any potential bias in an
opinion in the forums, etc.

\subsubsection{Needs for an API Portal (R5)}
In our primary survey, developers explicitly asked for a dedicated portal for APIs where they can search and analyze opinions and usage about APIs that are posted
in developer forums, such as Stack Overflow (Q12 in primary survey). Based on the themes and developer needs that emerged from the study,
we developed a prototype tool, named Opiner. The tool is developed as an opinion search engine where developers can search for an API by its name to
explore the positive and negative opinions provided for the API by the
developers in the forum posts.

\begin{figure*}[t]
\centering
\hspace*{-1.cm}%
\includegraphics[scale=.9]{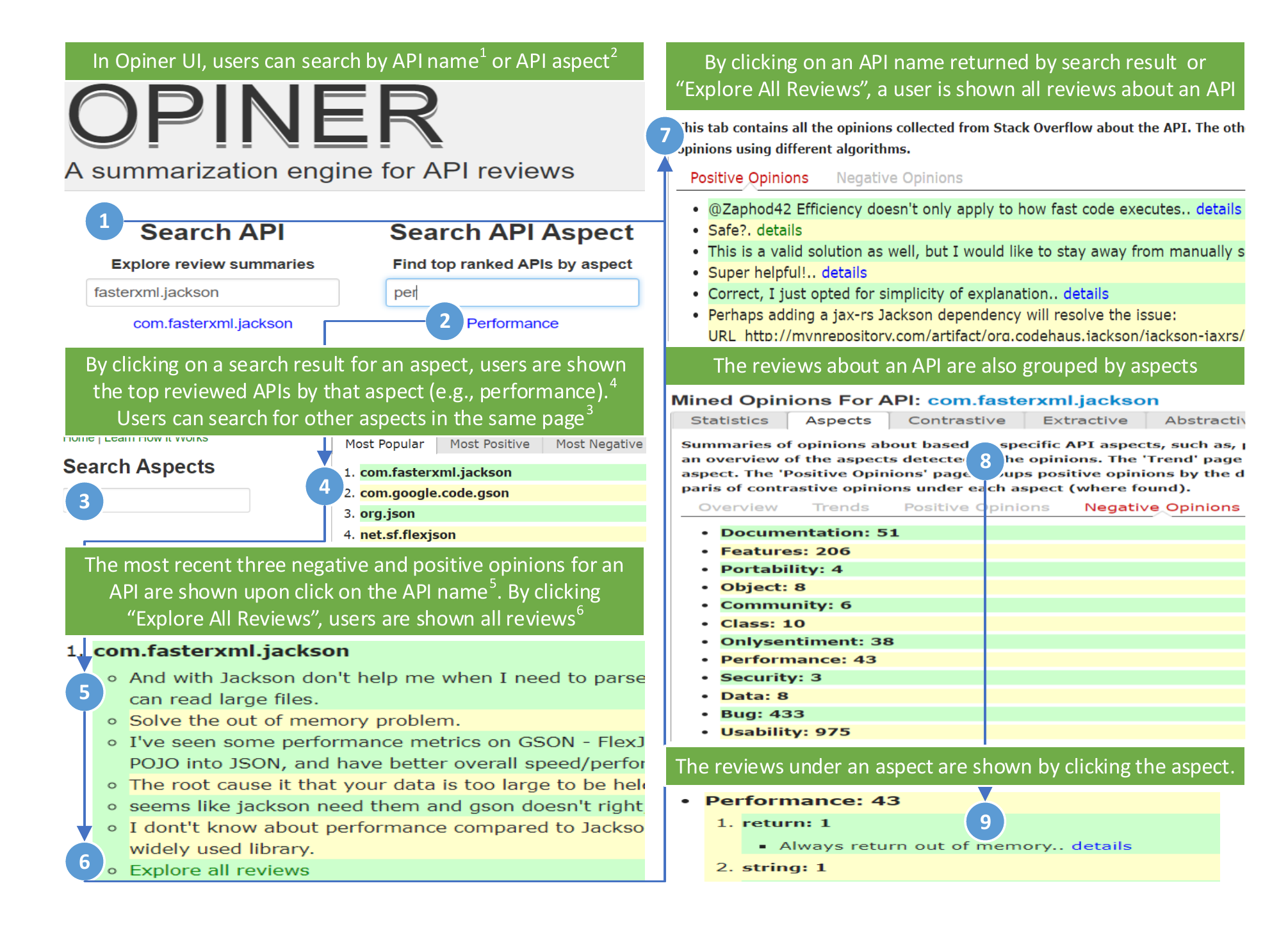}
\caption{Screenshots of Opiner opinion search engine for API reviews.}
\label{fig:opinerSC}
\end{figure*}
\begin{figure}[t]
  \centering
  \hspace*{-.8cm}%
  \includegraphics[scale=.72]{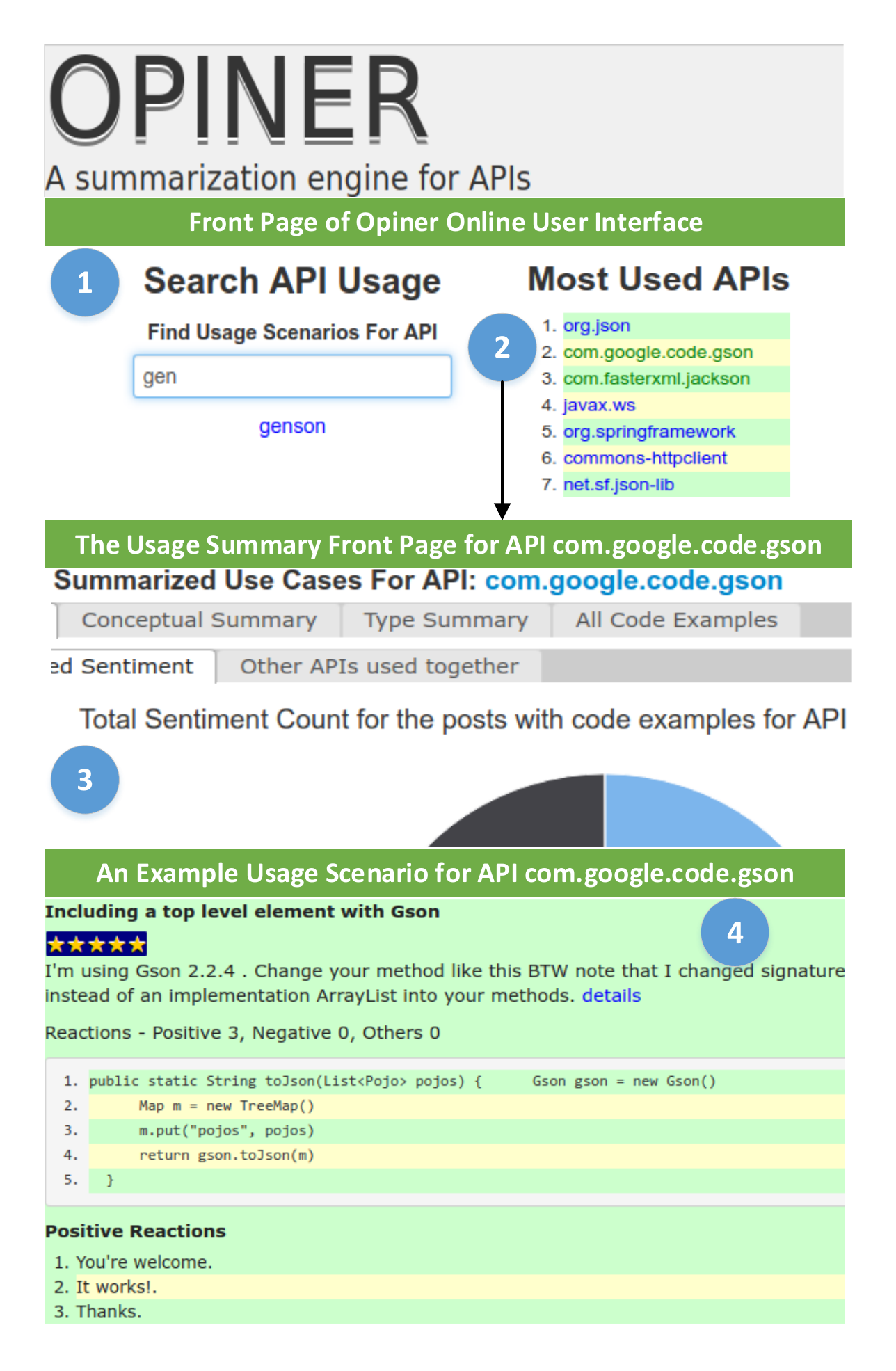}
  \hspace*{-.8cm}%
  \vspace{-4mm}
  \caption{Screenshots of the Opiner API usage summarizer}
  \label{fig:opiner-intro}
  \vspace{-4mm}
\end{figure}
\begin{figure}[t]
\centering
\vspace{-6mm}
\hspace*{-.8cm}%
\includegraphics[scale=1.2]{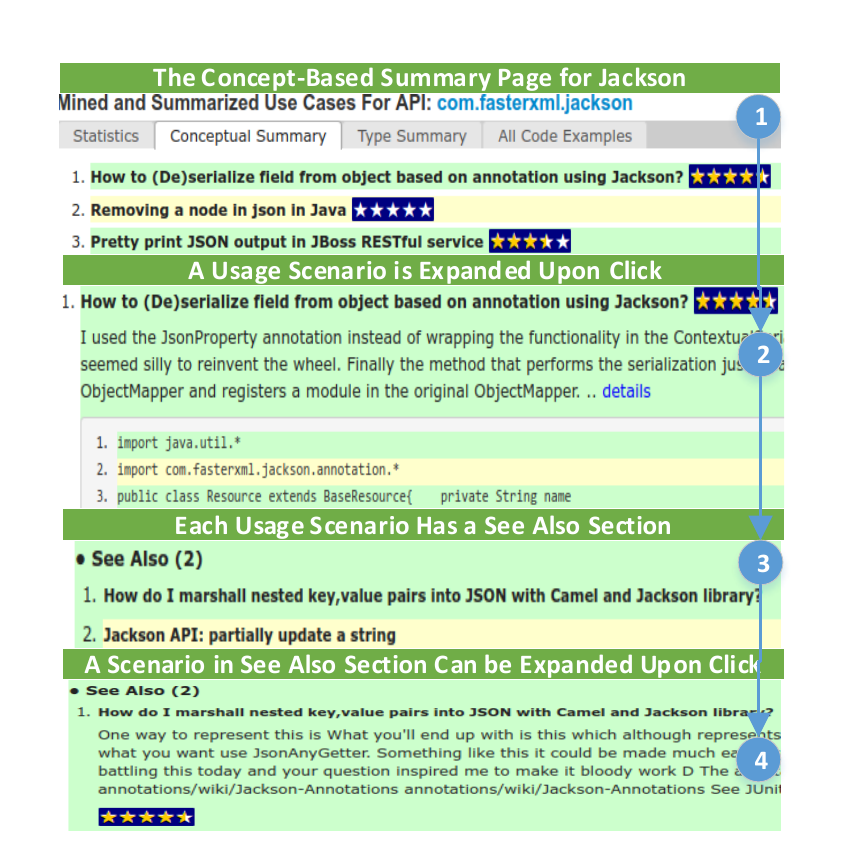}
\hspace*{-.8cm}%
\vspace{-4mm}
\caption{Concept based summary for API Jackson}
\label{fig:concept-summary}
\vspace{-4mm}
\end{figure}
The Opiner's infrastructure supports the implementation and deployment for all the above requirements (R1-R4).
Since the online deployment on October 30, 2017,
Opiner has been accessed and used by developers from 57 countries from all
the continents except Antarctica (as of July 29, 2018 by Google Analytics).

\nd\bf{$\bullet$ Opiner API Review Summarizer.} In \fig\ref{fig:opinerSC}, we show screenshots of the user interface of Opiner API review summarizer.
The UI of Opiner is a search engine, where users can search for APIs by their
names to look for the mined and categorized opinions about the API. There are
two search options in the front page of Opiner. A user can search for an API by
its name using \circled{1}. A user can also investigate the APIs by their
aspects using \circled{2}. Both of the search options provide auto-completion
and provide live recommendation while doing the search. When a user searches for
an API by name, such as, `jackson' in \circled{1}, the resulting query produces a
link to all the mined opinions of the API. When the user clicks the links
(i.e., the link with the same name of the API as `com.fasterxml.jackson'), all of the opinions
about the API are shown. The opinions are grouped as positive and negative (see
\circled{7}). The opinions are further categorized into
the API aspects by applying the aspect detectors on the detected opinions
\circled{8}. By clicking on each aspect, a user can see the opinions about the
API (see \circled{9}). Each opinion is linked to the corresponding post from
where the opinion was mined (using `details' link in \circled{9}). When a
user searches by an API aspect (e.g., performance as in \circled{2}), the user
is shown the top ranked APIs for the aspect (e.g., most reviewed,
most negatively reviewed in \circled{4}). For each
API in the Opiner page where the top ranked APIs are
shown, we show the most recent three positive and negative opinions about the API \circled{5}.
If the user is interested to further explore the reviews of an API from \circled{5}, he
can click the `Explore All reviews' which will take him to the page
in \circled{7}. The system architecture of Opiner API review summarizer is the subject of our tool demo paper~\cite{Uddin-OpinerReviewToolDemo-ASE2017}.

\nd\bf{$\bullet$ Opiner API Usage Scenario Summarizer.} In \fig\ref{fig:opiner-intro}, we show screenshots of Opiner
usage scenario summarizer.
The user can search an API by name to see the different usage summaries of the
API \circled{1}. The front page also shows the top 10 APIs for which the most number
of code examples were found \circled{2}. Upon clicking on the search result,
the user is navigated to the usage summary page of the API
\circled{3}, where the summaries are automatically mined and summarized from
Stack Overflow. A user can also click on each of the top 10 APIs listed in the
front page. An example usage scenario in Opiner is shown
in \circled{4}. The reactions included in a usage scenario can be simply a
``thank you'' note (when the code example serves the purpose) or more elaborated
(when the code example has certain limitations or specific usage requirements).
In \fig\ref{fig:concept-summary}, we show an overview of the concept-based API usage summary
for the API Jackson in Opiner. In Opiner, each concept consists of
one or more similar API usage scenarios. Each concept is titled as the title of the most
representative usage scenario (discussed below). In  \circled{1} of \fig\ref{fig:concept-summary}, we
show the most recent three concepts for API Jackson. The concepts are sorted by time of their most representative usage
scenario. The most recent concept is placed at the top of all concepts. Upon
clicking on a each concept title, the most representative scenario for the concept is
shown in \circled{2}. Each concept is provided a star
rating as the overall sentiments towards all the usage scenarios under the concept (see \circled{2}).
 Other relevant usage scenarios of the concept are grouped under a
`See Also' (see \circled{3}). Each usage scenario under the
`See Also' can be further explored (see \circled{4}). Each usage scenario is
linked to the corresponding post in Stack Overflow where the code example was
found (by clicking on the \it{details} word after the description text of a scenario).

The summaries page of an API in Opiner also contains a `Search' box, which developers can use to search for review and usage summaries of another API (e.g., a competing API).

\subsection{Effectiveness of Opiner}\label{sec:implications}\label{sec:implications}
We investigated the effectiveness of Opiner using both empirical evaluation and user studies. We conducted the empirical evaluation to
compute the performance of the mining techniques in Opiner (e.g., the precision of the sentiment detection and opinion association algorithms).
We observed a reasonable degree of precision in our mining techniques, such as more than 0.73 in our sentiment detection and .90 in our opinionated sentences to
API association algorithm. We conducted a total of six user studies to assess the usefulness of the opinion and usage summaries in Opiner. We found that developers were
able to select the right API with more accuracy while using Opiner and leveraging opinion summaries in Opiner. Additionally, we found that developers were
able to complete their coding tasks with more accuracy, in less time and using less effort while using the usage summaries in Opiner.
The details of the user studies and empirical evaluation are subject to our recent papers and
journals~\cite{Uddin-OpinerReviewAlgo-ASE2017,Uddin-OpinerReviewToolDemo-ASE2017,Uddin-MiningUsageScenarios-ASE2018,
Uddin-SummarizationUsageScenariosOpiner-TSE2018,Uddin-OpinionValue-TSE2018}. We summarize the user studies below.

\nd\bf{Effectiveness of Opiner Review Summaries.}
We conducted two user studies to assess the usefulness of the opinion summaries in Opiner and four studies were used to assess the
usefulness of the usage scenario summaries in Opiner.

In the first study, we compared our proposed summaries (aspect-based and statistical) against
the cross-domain review summarization techniques (summaries in paragraphs, topic-based). We compared the summaries using five development scenarios:
\begin{inparaenum}
\item Selection of APIs among choices,
\item Documentation of an API,
\item Presentation of an API to others to justify its selection to team members,
\item Staying aware of an API over time, and
\item Authoring a competing API to address the problems of an existing API.
\end{inparaenum} A total of 10 professional software developers were asked to rate the usefulness of the summaries for the five tasks. A number of criteria
were used to rate, such as usability of the API over other APIs, completeness of the summaries to produce a documentation, etc. The aspect-based summaries
were rated as the most useful (more than 85\% rating), followed by the statistical summaries (more than 70\%). The paragraph-based summaries were considered
as the least useful (less than 40\% rating).

We conducted the study on-site of a software company. A total of nine software developers from the company participated in the study.
The developers were given access to Opiner online tool. The developers were asked to complete the tasks using Opiner and Stack Overflow. Both tasks involved the
selection of an API among two competing APIs. For example, the first task asked the developers to pick one of the two APIs (GSON and org.json). The criteria
used to select were the usability and licensing restrictions of the two APIs. The developers were asked to consult Stack Overflow and the review summaries in
Opiner. The developers made the best decision while using Stack Overflow with Opiner, instead of while using Stack Overflow only. All the developers
considered Opiner to be usable and wished to use it in their daily development tasks.

\nd\bf{Effectiveness of Opiner Usage Summaries.}
We conducted four user studies to assess the
usefulness of the usage scenario summaries in Opiner. The first study involved the coding of four development tasks and the other involved surveys.
A total of 33 professional software developers and students participated in the four user studies (33 in the coding tasks, and 31 of the 34 in each of the surveys).

In the coding study, each participant was given four tasks, for which they wrote code. They used four different development resources (one each
for the four tasks): Stack Overflow, API official documentation, Opiner usage summaries, and everything including search engines. We observed an average accuracy of 0.62 in the provided solutions while the participants
used the Opiner usage summaries. The second best was the everything setting with an accuracy of 0.55, followed by an accuracy of 0.5 when the participants used only Stack Overflow.
The participants showed the least accuracy (0.46) while using the API official documentation. In subsequent surveys, more than 80\% of the participants agreed that
the usage summaries in Opiner can offer improvements over API official documentation and the developer forums. For example, the developers recommended that
the usage summaries should be integrated with the API official documentation.

%






%% file: threats.tex
\section{Threats to Validity} \label{sec:threats}
{We now discuss the threats to validity of our study by following the guidelines for empirical studies~\cite{Woh00}.}

\subsection{Construct Validity}
{\it{Construct validity threats} concern the relation between theory and observations. In our study, they could be due to the measurement of errors.
The accuracy of the open coding of the survey responses is subject to our
ability to correctly detect and label the categories. The exploratory
nature of such coding may have introduced the \it{researcher bias}. To mitigate
this, we coded 20\% of the cards for four questions independently and
measured the coder reliability on the next 20\% of the cards. We report the measures of agreement
in \tbl\ref{tab:agreementc1c2Fors2}. While we achieved a high level of agreement, we,
nevertheless, share the complete survey responses in an online
appendix~\cite{website:opinionsurvey-online-appendix}.}

{\it{Maturation threats} concerns the changes in a participant during the study due to the \it{passage of time}, such as change in development priorities or
environments (e.g., moving from open source-based APIs to proprietary APIs). Intuitively, we expect to see greater concentration of opinions about open-source APIs
in the forums. None of these concerns are applicable to
our surveys, because each survey was supposed to take not more than 30 minutes by each participant.}

\subsection{Internal Validity}
{Threats to internal validity refer to how well the research is conducted. In our case, it is about how well the design of the surveys
allow us to choose among alternative explanations of the
phenomenon. A high internal validity
in the design can let us choose one explanation (e.g., tool support to assist in opinion analysis) over another (e.g., no need for a tool)
with a high degree of confidence, because it avoids (potential) confounds.}

{In our primary survey, we sought to avoid confounding factors  by asking the participants open-ended questions. The questions with options (i.e., closed questions) were presented
to the participants only after they have responded to the open-ended questions. The two types of questions were divided into separate sections, i.e., the
participants could not see the closed-ended questions when they answered the open-ended questions. Despite this, we observed similar findings in the responses
of the questions. For example, one open-ended question was about the different factors in an API that can play a role in their decisions to choose the API (Q11). The paired
closed-ended question was Q15. In both the responses, we found that developers asked for similar API aspects about which they prefer to seek opinions, such as performance,
usability, etc.}

{Another threat could arise from the placement of the options in a multiple choice question. The threat may
influence the participants to agree/disagree with an option more than the other options, such as the option
that is placed at the top (i.e., rated first). We did not observe any such pattern in the responses. For example, the first option in our question about tool support in opinion analysis (Q19 in primary survey and Q7 in pilot survey). Both surveys followed the same ordering of the options for the question. Despite this,
we observed the lowest rank for the option ``sentiment miner'' in both surveys (second option) and highest rank for the option ``API comparator'' (fourth option).
In addition, there was no significant difference in the preference of the participants towards a specific tool over another (see \fig\ref{fig:WhatToolsBetterSupportOpinionUnderstanding}). Therefore, all such
tools were found as favorable by the participants, despite their placements.}

\subsection{External Validity}
{Threats to external validity compromise the confidence in stating whether the study results are applicable to other groups.}

\rev{While our sample size of 900 developers out of a population of 88K developers is small, we received 15.8\% response rate in our primary survey. 
Moreover, the qualitative assessment of the responses offered us interesting insights about the needs and challenges to seek and analyze opinions 
about APIs. However, it is desirable that future researches replicate our study on a larger and--or different population of developers to make our findings more generic/robust.}

{Due to the diversity of the domains where APIs can be used and developed, the provided opinions can be
contextual. Thus, the generalizability of the findings requires careful
assessment. Such diversity may introduce \it{sampling bias} if developers from a
given domain are under- or over-represented.} {One related threat could arise due to our sampling of
developers from GitHub (for pilot survey) and Stack Overflow (for primary survey). The sampling might
have missed developers who do not use GitHub or Stack Overflow. As we noted in \sec\ref{subsec:motivation-surveys}, we picked
developers from GitHub and Stack Overflow because of their popularity among the open-source developers. We observed
similar findings between our pilot and primary surveys. Nevertheless,
a replication of our study involving developers from other online forums could offer further validation towards the generalization of
our findings across the larger domains of software engineering.}

{Another potential threat of over-representation would be related to all the developers being
proficient in only one programming language (e.g., Javascript). This can happen because Javascript has been
the most popular language in Stack Overflow over the last six years (according to Stack Overflow survey of 2018~\cite{website:stackoverflow-survey}).
In our primary survey population of 88,021 Stack Overflow users, we observed that the Stack Overflow posts where the users participated in 2017, corresponded
to all the popular languages found in the Stack Overflow surveys. Moreover, while our sample attempted to assign each user to one of the top nine programming
languages, those users also participated in the discussions of APIs involving other programming languages. Therefore, their opinions
in the primary survey may be representative of the overall Stack Overflow population.}

{In our primary survey, we only collected responses from developers who visit developer forums to seek information about APIs. As we noted in \sec\ref{sec:methodology},
this decision was based on our observation from the pilot survey that developer forums are the primary resource for developers to see such information.
In the primary survey, we did not collect additional information from the participants who do not visit developer forums.
Further probing of such participants to understand why they do not use developer forums, could have offered us insights into the shortcomings
of developer forums to provide such needs. Intuitively, such insights about problems in developer forums
can be more concrete from a participant who actually uses developer forums. Therefore, to understand the shortcomings in developer forums, we probed the
participants with a number of questions, such as ``What are your biggest challenges while seeking opinions about APIs from developer forums?'' (Q6),
``What areas can be positively/negatively affected by summarization of reviews from developer forums?'' (Q13, 14), and ``What
factors in a forum post can help you determine the quality of the provided opinions?'' (Q7). The responses to those
questions showed us that developers face numerous challenges while seeking opinions about APIs from forum posts (such as
information overload, trustworthiness, etc.). We observed similar findings in our pilot survey. However, we could still have missed important insights from the participants
who do not visit developer forums. We leave the analysis of such participants for our future work.}

{Our survey samples are derived from GitHub and Stack Overflow developers.
We picked the pilot survey participants from a list of 4,500 GitHub users. 
We randomly selected the primary survey participants from a list of more than 88K Stack Overflow users, each 
of which also had an account in GitHub. We designed the primary survey by leveraging lessons learned from our pilot survey (see \sec\ref{sec:needs-for-primary-survey}). In our primary survey,
we attempted to fix each of the problems. For example, we applied \it{stratification} in our sampling to ensure that we involve developers from Stack Overflow
that are \it{proficient} in different programming languages. The stratification is necessary, because otherwise a random sample from the 88K Stack Overflow users
would have picked more developers from a language that is more popular (e.g., Javascript, Java, and Python) among the developers
(in terms of the number of users participating in the discussion of the posts related to the language). In addition, we attempted to include
the developers who can be considered as \it{experts} in those programming languages. Intuitively, expertise of a developer for a programming language
can be correlated to his reputation in Stack Overflow posts that are related to the language. The higher reputation score a user has, the more likely those reputation scores are provided by many different
users in Stack Overflow, and hence the higher likelihood that the user is considered as an expert within the community. An expert user is thus more likely to offer
more concrete information about the APIs used in the language, based on real-world experience.}

\subsection{Reliability Validity}
{Reliability threats concern the possibility of replicating this study. We attempted to provide all the necessary details to replicate the study.
The anonymized survey responses are provided in our online appendix~\cite{website:opinionsurvey-online-appendix}. The complete labels and list of quotes
used in the primary survey are also provided in the online appendix.}

%% file: summary.tex
\section{Conclusions and Future Work} \label{sec:conclusion}

Opinions can shape the perception and decisions of developers related to the
selection and usage of APIs.
The plethora of open-source APIs and the advent of developer forums have
influenced developers to publicly discuss their experiences and share opinions
about APIs.

To better understand the role of opinions and how
developers use them, we conducted a study based on two surveys of a total of 178 software developers.
We are aware of no such
previous surveys in the field of software engineering. The design of the two surveys and the survey responses form the first major contribution of
this paper.

The survey responses offer insights into how and why
developers seek opinions, the challenges they face, their assessment of the
quality of the available opinions, their needs for opinion summaries, and the
desired tool support to navigate through opinion-rich information. We observed the following major
findings in our analysis:
\begin{enumerate}
  \item Developers seek opinions about APIs to support diverse development needs, such as selection of an API among available choices. A primary source of such
  opinions is the online developer forums, such as Stack Overflow.
  \item Developers face several challenges associated with noise, trust, bias,
and API complexity when seeking opinions. High-quality opinions are typically viewed as clear, short and to the point, bias free,
and supported by facts.
\item Developers feel frustrated with the amount of available API opinions and desire a tool support to efficiently analyze the opinions, such as
API comparator, opinion summarizer, API sentiment trend analyzer, etc.
\end{enumerate} The findings and insights gained from this study
helped us to build a prototype tool, named Opiner. The Opiner framework can be used to
mine and summarize opinions about APIs in a fully automatic way. We observed promising results of leveraging
the Opiner API review summaries to support diverse development needs.
The detailed analysis of the survey responses and the findings form the second major contribution of this paper.

Our future work is broadly divided into two directions: 
\begin{inparaenum}

\nd\item \rev{\bf{Tool Support by Developer Experience.} As we noted in \sec\ref{sec:demographic-experience}, less experienced developers show more interest to value the opinions and 
were also more distinct in their preference of tools to support such analysis 
(API comparator and Trend Analyzer). We plan to investigate the particular characteristics of the less experienced developers which could motivate them more to 
value opinions of others and to use the tools. Such insights then can be used to motivate the design of tools and APIs by focusing on the demographic needs. 
It is thus desirable to replicate our study on a randomized sample of Stack Overflow users, because it may give us higher concentration of less experienced developers than the sample 
we have used (i.e., based on users with high reputation in Stack Overflow).}    

\nd\item \bf{Sentiments vs Emotions.} We
plan to investigate the role of finer grained emotions (e.g., anger, fear, etc.) in the daily development
activities involving APIs. In particular, we are interested in conducting more surveys and developing
tools to advance the knowledge on the impact of emotions in API usage analysis. 
\end{inparaenum}

%% file: appendix.tex
\begin{appendices}

\section{Open Coding of Final Survey Responses}\label{sec:app-open-coding-results}
\tbl\ref{tab:s2Categories} shows the distribution of the 
categories by the questions.
\begin{table*}[htbp]
  \centering
  \caption{The codes emerged during the open-coding of the responses (\#TC =
  Total codes).} \begin{tabular}{l|rr|rr|rr|rr|rr|rr|rr|rr|rr|r}\toprule
          & \multicolumn{2}{c}{\textbf{Q4}} & \multicolumn{2}{c}{\textbf{Q5}} & \multicolumn{2}{c}{\textbf{Q6}} & \multicolumn{2}{c}{\textbf{Q7}} & \multicolumn{2}{c}{\textbf{Q11}} & \multicolumn{2}{c}{\textbf{Q12}} & \multicolumn{2}{c}{\textbf{Q13}} & \multicolumn{2}{c}{\textbf{Q14}} & \multicolumn{2}{c}{\textbf{Q24}} &  \\
              \cmidrule{2-20}
          
          & \textbf{\#C} & \textbf{\#R} & \textbf{\#C} & \textbf{\#R} & \textbf{\#C} & \textbf{\#R} & \textbf{\#C} & \textbf{\#R} & \textbf{\#C} & \textbf{\#R} & \textbf{\#C} & \textbf{\#R} & \textbf{\#C} & \textbf{\#R} & \textbf{\#C} & \textbf{\#R} & \textbf{\#C} & \textbf{\#R} & \multicolumn{1}{l}{\textbf{\#TC}} \\
          \midrule
    \bf{Documentation} & {11} & {11} & {1} & {1} & {4} & {4} & {54} & {49} &
    {47} & {43} & {11} & {9} & {6} & {5} & {1} & {1} &       &       & 135 \\
    \cmidrule{2-20}
    \bf{Search} &       &       & {94} & {79} & {15} & {11} &       &       &      
    &       & {12} & {12} & {1} & {1} &       &       &       &       & 122 \\
    \cmidrule{2-20}
    \bf{Usability} & {1} & {1} &       &       & {2} & {2} &       &       &
    {87} & {56} & {2} & {2} & {2} & {2} &       &       &       &       & 94 \\
    \cmidrule{2-20}
    \bf{Community} & {11} & {9} &       &       & {9} & {7} & {12} & {10} & {31}
    & {22} & {6} & {3} & {2} & {2} &       &       & {2} & {1} & 71 \\
    \cmidrule{2-20}
    \bf{Trustworthiness} & {13} & {11} & {1} &       & {21} & {20} & {8}
    & {5} & {2} & {1} & {2} & {2} &       &       & {21} & {17} & {4} & {3} & 68 \\
    \cmidrule{2-20}
    \bf{Expertise} & {31} & {31} & {2} & {2} & {3} & {3} & {13} & {12} & {3} &
    {3} & {3} & {3} & {4} & {4} &       &       & {2} & {2} & 59 \\
    \cmidrule{2-20}
    \bf{API Selection} & {13} & {12} &       &       & {6} & {6} & {1} & {1} &      
    &       &       &       & {30} & {25} & {6} & {6} &       &       & 56 \\
    \cmidrule{2-20}
    \bf{API Usage} & {17} & {16} &       &       & {8} & {8} & {2} & {2} & {10}
    & {9} & {7} & {6} & {7} & {7} & {2} & {2} &       &       & 53 \\
    \cmidrule{2-20}
    \bf{Situational Relevance} &       &       &       &       & {26} & {22} &
    {12} & {9} & {8} & {7} & {3} & {3} & {3} & {3} &       &       & {2} & {2} & 52 \\
    \cmidrule{2-20}
    \bf{Reputation} &       &       & {1} & {1} & {1} & {1} & {35} & {33} & {5}
    & {4} &       &       &       &       &       &       &       &       & 42 \\
    \cmidrule{2-20}
    \bf{Performance} & {8} & {8} &       &       &       &       &       &      
    & {30} & {23} &       &       & {3} & {3} &       &       &       &       & 41 \\
    \cmidrule{2-20}
    \bf{Recency} &       &       & {1} & {1} & {13} & {13} & {6} & {6} &       &    
    & {1} & {1} & {3} & {3} & {3} & {3} &       &       & 27 \\
    \cmidrule{2-20}
    \bf{Productivity} & {9} & {8} &       &       & {1} & {1} &       &       &     
    &       &       &       & {12} & {11} &       &       &       &       & 22 \\
    \cmidrule{2-20}
    \bf{Compatibility} &       &       &       &       & {4} & {4} &       &      
    & {18} & {14} &       &       &       &       &       &       &       &       & 22 \\
    \cmidrule{2-20}
    \bf{Aggregation Portal} &       &       &       &       &       &       &      
    & &       &       & {18} & {15} & {1} & {1} &       &       &       &       & 19 \\
    \cmidrule{2-20}
    \bf{Missing Nuances} &       &       &       &       &       &       &      
    & &       &       &       &       &       &       & {18} & {12} &       &       & 18 \\
    \cmidrule{2-20}
    \bf{Legal} &       &       &       &       &       &       &       &       &
    {15} & {14} &       &       &       &       &       &       &       &       & 15 \\
    \cmidrule{2-20}
    \bf{Sentiment Statistics} &       &       & {4} & {4} &       &       &      
    & &       &       & {8} & {5} &       &       &       &       &       &       & 12 \\
    \cmidrule{2-20}
    \bf{General Insight} &       &       &       &       & {1} & {1} &       &      
    & {4} & {4} & {1} & {1} & {2} & {2} & {4} & {4} &       &       & 12 \\
    \cmidrule{2-20}
    \bf{API Maturity} &       &       &       &       &       &       &       &      
    & {6} & {5} & {3} & {2} &       &       &       &       &       &       & 9 \\
    \cmidrule{2-20}
    \bf{API Adoption} &       &       &       &       &       &       & {2} &
    {1} & {4} & {4} & {3} & {3} &       &       &       &       &       &       & 9 \\
    \cmidrule{2-20}
    \bf{Portability} &       &       &       &       &       &       &       &      
    & {5} & {5} &       &       & {4} & {4} &       &       &       &       & 9 \\
    \cmidrule{2-20}
    \bf{Opinion Categorization} &       &       &       &       &       &      
    & &       &       &       & {8} & {8} & {1} & {1} &       &       &       &       & 9 \\
    \cmidrule{2-20}
    \bf{Opinion Reasoning} &       &       &       &       &       &       &      
    & & &       &       &       & {1} & {1} & {7} & {6} &       &       & 8 \\
    \cmidrule{2-20}
    \bf{Contrastive Summary} &       &       &       &       &       &       &      
    &       &       &       & {7} & {7} &       &       &       &       &       &       & 7 \\
    \cmidrule{2-20}
    \bf{Info Overload} &       &       &       &       & {3} & {3} &       &      
    & &       & {1} & {1} &       &       & {2} & {2} &       &       & 6 \\
    \cmidrule{2-20}
    \bf{Similarity} &       &       & {3} & {3} &       &       &       &      
    & &       & {2} & {2} &       &       &       &       &       &       & 5 \\
    \cmidrule{2-20}
    \bf{Security} &       &       &       &       &       &       &       &      
    & {2} & {2} & {1} & {1} &       &       &       &       &       &       & 3 \\
    \cmidrule{2-20}
    \bf{Learning Barrier} &       &       &       &       &       &       &      
    & &       &       &       &       &       &       & {3} & {3} &       &       & 3 \\
    \cmidrule{2-20}
    \bf{Bug}   &       &       &       &       &       &       &       &       &
    {1} & {1} &       &       & {1} & {1} &       &       &       &       & 2 \\
    \cmidrule{2-20}
    \bf{Extractive Summary} &       &       &       &       &       &       &      
    &       &       &       & {2} & {2} &       &       &       &       &       &       & 2 \\
    \cmidrule{2-20}
    \bf{API Improvement} &       &       &       &       &       &       &      
    & &       &       &       &       & {1} & {1} & {1} & {1} &       &       & 2 \\
    \cmidrule{2-20}
    \bf{FAQ}   &       &       &       &       &       &       &       &       &    
    &       & {1} & {1} &       &       &       &       &       &       & 1 \\
    \cmidrule{2-20}
    \bf{Problem Summaries} &       &       &       &       &       &       &      
    & &       &       & {1} & {1} &       &       &       &       &       &       & 1 \\
    \cmidrule{2-20}
    \bf{Machine Learning} &       &       &       &       &       &       &      
    & &       &       & {1} & {1} &       &       &       &       &       &       & 1 \\
    \cmidrule{2-20}
    \bf{Wrong Info} &       &       &       &       &       &       &       &      
    &       &       &       &       &       &       & {1} & {1} &       &       & 1 \\
    \cmidrule{2-20}
    \bf{Lack of Info} &       &       &       &       & {1} & {1} &       &      
    & &       &       &       &       &       &       &       &       &       & 1 \\
    \midrule
    Not Sure &       &       &       &       &       &       &       &       &       &       & {10} & {9} & {23} & {22} & {26} & {25} & {1} & {1} & 59 \\
    \cmidrule{2-20}
    Irrelevant & {7} & {4} & {5} & {5} & {10} & {8} & {7} & {4} & {11} & {7} & {25} & {18} & {8} & {7} & {14} & {13} & {6} & {3} & 87 \\
    \midrule
    \bf{Total} & {121} & {83} & {112} & {83} & {128} & {83} & {152} & {83} &
    {289} & {83} & {139} & {83} & {115} & {83} & {109} & {83} & {17} & {10} & 1019 \\
    \midrule
    \bf{\# Categories} & \multicolumn{2}{c|}{9} & \multicolumn{2}{c|}{8} &
    \multicolumn{2}{c|}{17} & \multicolumn{2}{c|}{10} & \multicolumn{2}{c|}{17}
    & \multicolumn{2}{c|}{21} & \multicolumn{2}{c|}{19} &
    \multicolumn{2}{c|}{14} & \multicolumn{2}{c|}{5} &   {37} \\
    \midrule
    \multicolumn{20}{l}{Notes: \#C = the number of code, \#R = the number of
    respondents, Q4 = Reasons for referring to opinions of other
    developers about APIs? Q5 = How do you} \\                                                             
    \multicolumn{20}{l}{seek information about APIs in a
    developer forum?, Q6 = Challenges when seeking for
    opinions about an API? Q7 = Factors in a forum post to
    determine the} \\
    \multicolumn{20}{l}{the quality of a provided opinion about an API?, Q11
    = Factors of an API that play a role in your decision to choose an API? Q12
    = Different ways opinions}
    \\
    \multicolumn{20}{l}{about APIs can be summarized from developer
    forums? Q13=Areas positively affected by the summarization of reviews
about APIs from developer forums?}
    \\
    \multicolumn{20}{l}{Q14=Areas negatively affected by the summarization of
    reviews about APIs from developer forums? Q24= Reasons to not value the
    opinion of other devs.}
    \\
    \bottomrule
    \end{tabular}%
  \label{tab:s2Categories}%
\end{table*}%

\end{appendices}